\documentclass[iop]{emulateapj}
\pdfoutput=1
\usepackage{amsmath}
\usepackage{subfigure}
\usepackage{txfonts}
\usepackage{graphicx}
\usepackage{rotating}
\usepackage{lscape}
\usepackage{hyperref}
\usepackage{color,hyperref}
\definecolor{linkcolor}{rgb}{0,0,0.5}
\definecolor{notecolor}{rgb}{0.8,0,0}
\hypersetup{colorlinks=true, linkcolor=linkcolor, citecolor=linkcolor,
  filecolor=linkcolor, urlcolor=linkcolor}

\def\co{$^{12}$CO}
\def\13co{$^{13}$CO}
\def\c18o{C$^{18}$O}
\def\msun{$M_{\odot}$}
\def\vol{cm$^{-3}$}
\def\cm2{cm$^{-2}$}
\def\cmt{cm$^{-3}$}
\def\vel{km s$^{-1}$}
\def\um{$\mu$m}
\def \coor{$\alpha_{2000}=18^{\mathrm{h}}08^{\mathrm{m}}52\fs80, \delta_{2000}=-18^{\circ}16\arcmin22\farcs8$}
\def\higal{Hi-GAL}

\def\vel{km s$^{-1}$}

\slugcomment{In preparation. Revised Version: \today}

\shorttitle{Liu et al.}

\begin{document}

\title{  Interactions of the Infrared bubble N4 with the surroundings}

\author{Hong-Li Liu\altaffilmark{1,2,8$\dag$}, Jin-Zeng Li\altaffilmark{1},Yuefang Wu\altaffilmark{3},
Jing-Hua Yuan\altaffilmark{1},Tie Liu\altaffilmark{4}, G. Dubner\altaffilmark{5}, S. Paron\altaffilmark{5,6},
 M. E. Ortega\altaffilmark{5}, Sergio Molinari\altaffilmark{7}, Maohai Huang\altaffilmark{1},
 Annie Zavagno\altaffilmark{8}, Manash R. Samal\altaffilmark{8},
 Ya-Fang Huang\altaffilmark{1}, Si-Ju Zhang\altaffilmark{1,2}}

\altaffiltext{1}{National Astronomical Observatories, Chinese Academy of Sciences, 20A Datun Road, Chaoyang District, Beijing 100012, China}
\altaffiltext{2}{University of Chinese Academy of Sciences, 100049 Beijing, China}
\altaffiltext{3}{Department of Astronomy, Peking University, 100871 Beijing, China}
\altaffiltext{4}{Korea Astronomy and Space Science Institute 776, Daedeokdae-ro, Yuseong-gu, Daejeon, Republic of Korea 305-348}
\altaffiltext{5}{1Instituto de Astronom\'\i a y F\'\i sica del Espacio (IAFE, CONICET-UBA), CC 67, Suc. 28,
1428 Buenos Aires, Argentina}
\altaffiltext{6}{2FADU - Universidad de Buenos Aires, Ciudad Universitaria, Buenos Aires, Argentina}
\altaffiltext{7}{Istituto di Astrofisica e Planetologia Spaziali -- IAPS, Istituto Nazionale di Astrofisica -- INAF, via Fosso del Cavaliere 100, 00133 Roma, Italy}
\altaffiltext{8}{Aix Marseille Universit¨¦, CNRS, LAM (Laboratoire d'Astrophysique de Marseille) UMR 7326, 13388, Marseille, France}

\altaffiltext{$\dag$}{hlliu@nao.cas.cn}

\begin{abstract}
    The physical mechanisms that induce the transformation of a certain mass of gas in new stars are
     far from being well understood. Infrared bubbles associated with
    {H\,{\sc ii}}  regions have been considered to be good samples for investigating triggered star formation.
    In this paper we report on the investigation of the
    dust properties of the infrared bubble N4 around the H\,{\sc ii}~region {G11.898+0.747}, analyzing
    its interaction  with its surroundings and star formation histories therein, with the aim of determining the
    possibility of star formation triggered by the expansion of the bubble. Using  \emph{Herschel} PACS and
    SPIRE images with a wide wavelength coverage, we reveal the dust properties over the entire bubble.
    Meanwhile, we are able to identify
    six dust clumps surrounding the bubble, with a mean size of 0.50 pc,
    temperature of about 22 K, mean column density of 1.7 $\times10^{22}$ \cm2, mean volume density
    of about  4.4 $\times10^{4}$ cm$^{-3}$, and a mean mass of 320 \msun. In addition, from PAH emission
    seen at 8 \um, free-free emission detected at 20~cm and a probability density function in special
    regions,    we could identify clear signatures  of the influence of the H\,{\sc ii} region on the surroundings.
    There are hints of star formation, though further investigation is required to demonstrate that N4 is the triggering source.

\end{abstract}


\section{Introduction}\label{s1}
Massive stars (of OB spectral type with masses greater than  $\sim10 M_{\odot}$) are thought to form in clusters within molecular
cloud complexes. Intense radiative and mechanical outputs from massive stars strongly affect
their parent molecular clouds  in two opposite ways: by disrupting molecular clouds and by triggering
star formation.  In effect, on the one hand, the natal molecular clouds can be accelerated beyond their escape velocities
by feedback from massive stars  through expanding H\,{\sc ii}~regions, outflows, stellar winds, and supernova explosions,
leading to the disruption of the clouds  which can stop the eventual star-forming processes.
On the other hand, this feedback is capable of sweeping up the surroundings and accumulating them
into condensations gravitationally bound, inducing  the  formation of new generations  of stars.

Over the past several decades,  numerous studies have focused on star formation triggered in the environs of H\,{\sc ii}~regions
\citep[e.g.,][]{deh05,deh06,zav07,deh08,deh09,ogu10,elm11,ken12,dal13,sam14,liu15,dal15}.
Two  different mechanisms have been proposed as models of  star formation  triggered in the surroundings of
 H\,{\sc ii} regions: the ``collect and collapse" process
\citep[C~\&~C,][]{elm77} and ``radiation-driven implosion" \citep[RDI,][]{ber89,lef94}.
In the C~\&~C process, the ultraviolet radiation from ionizing sources produces an ionization front (IF),
creating an expanding H\,{\sc ii}~region. The supersonic expansion of the H\,{\sc ii}~region drives the shock front
(SF) in front of the IF. Finally, a shell of circumstellar gas can be collected between the IF and the SF. In
due time, the shell may become denser
and collapse to form stars. Several numerical simulations \citep[e.g.,][]{hos05,hos06,dal07} have suggested
that expanding H\,{\sc ii}~regions can trigger star formation through the C \&~C process only if ambient
molecular material is massive enough. This process has been tentatively detected  to be at work in several well-known
H\,{\sc ii} regions such as Sh~104 \citep{deh03}, RCW~79 \citep{zav06}, and RCW~120 \citep{deh09}. By contrast,
in the RDI model,
the IF drives the SF into molecular clouds surrounding the H\,{\sc ii}~region, stimulating the collapse of pre-existing
subcritical clumps. Recent numerical simulations \citep[e.g.,][]{mia06,mia09,bis11} have
demonstrated that the RDI model can successfully interpret star formation
in bright-rimmed clouds (BRCs). This prediction seems to have been confirmed by
observations of BRCs \citep[e.g.,][]{rea09,liu12} indicating that star formation
concentrated along the central axis of the clouds  might be the consequence of the RDI process.

\citet{tho12} and \citet{ken12} speculated that around 14\%-30\% of the massive young stellar
objects (MYSOs) in the Milky Way might have been induced
by expanding bubbles/H\,{\sc ii} regions on the basis of the association of a large sample of
IR bubbles \citep{chu06,chu07} with Red MSX MYSOs \citep{urq08}.
However, the exact processes of triggered star formation are far from being understood.
It is hoped that detailed studies of the physical interaction of bubbles/H\,{\sc ii}~regions together
 with their environs  will lead to a better understanding of star formation associated
 with bubbles/H\,{\sc ii}~regions \citep{liu15}.

Far-infrared observations carried out by the \emph{Herschel Space Observatory} enable us to
get a better insight into the physical properties of the
bubble / H\,{\sc ii}~region, with unprecedented resolutions and sensitivities in that spectral range.
Thanks to the widespread wavelength coverage from 70 to 500 \um, the dust temperature
and column density distributions of all of the entire bubbles/H\,{\sc ii}~regions can be readily revealed
by fitting spectral energy distributions (SEDs) pixel by pixel. Additionally, \emph{Herschel} observations
can reveal star formation surrounding the system. The widespread coverage is adequate for better
constraining the SED profiles of YSOs at earlier phases and more accurately estimating their
physical parameters such as stellar mass and luminosity. Moreover, \emph{Herschel} observations coupled
with suitable molecular lines offering kinetic and dynamical information are very useful tools for
understanding the interactions of the bubbles/H\,{\sc ii}~regions with their surroundings. The present
paper presents a comprehensive study of the bubble N4 \citep{chu06} using a combination of
\emph{Herschel} and molecular line observations.

The bubble N4 appears as an almost complete ring in the IR regime (see Figure~\ref{fig-rgb1}), enclosing
the H\,{\sc ii}~region G11.898+0.747 \citep{loc89}. N4 is centered at \coor,
with a radius of $\sim 2\arcmin$, corresponding to $1.9\pm0.5$ pc at a distance of $3.2\pm0.9$ kpc \citep{chu06}.
By analyzing the spatial distribution of candidate YSOs around N4, \citet{wat10} found no evidence for
triggered star formation. However, they claimed that the presence of triggered star formation can not be ruled out
due to the lack of a complete sample of the YSO population as well as maps of the molecular gas and ionized gas in the environs.
In contrast, the expanding motion of N4 uncovered by the observations of $J=1-0$ of \co, \13co, and \c18o carried out
by \citet{li13} suggests
a signature of possible triggered star formation. Therefore, the issue of star formation
in N4 merits a more comprehensive investigation.

The purpose of this study is to analyze interactions of the bubble N4 with its surroundings
and star formation histories therein, and to explore the possibility of triggered star formation.
This paper is organized as follows:  the \emph{Herschel} and molecular line observations together with archival IR and
radio data are presented in Section 2, the results are described in Section 3, the discussion is
arranged in Section 4, and finally, we give a summary in Section 5.

\section{Observations and data reduction}\label{s2}
\subsection{\emph{Herschel} Observations}\label{s-ob1}
The bubble N4 was observed as part of the \higal\footnote{\higal,
the \emph{Herschel} infrared Galactic
Plane Survey, is an Open Time Key project on the \emph{Hershcel Space Observatory} (HSO) aiming
to map the entire Galactic Plane in five infrared bands. This survey covers a $|b|<1^{\circ}$ wide
strip of the Milky Way Galactic plane in the longitude range $-60^{\circ} \leq l \leq 60^{\circ}$}
survey. In this survey, the Photodetector Array Camera $\&$ Spectrometer \citep[PACS,][]{pog10} at 70
and 160~\um\  and the Spectral and Photometric Imaging Receiver \citep[SPIRE,][]{gri10} at 250, 350,
and 500~\um\  worked simultaneously in the parallel photometric mode. The observations were run in
two orthogonal scanning directions at a scan speed of 60 arcsec~s$^{-1}$.
The measured angular resolutions of these bands are 10$\farcs$7, 13$\farcs$9, 23$\farcs$9, 31$\farcs$3,
and 43$\farcs$.8 \citep{tra11}, respectively, corresponding to $0.12\pm0.05$ to $0.49\pm0.21$ pc at the distance to N4.
The detailed descriptions of the pre-processing of the data up to usable high-quality images
can be found in \citet{tra11}.

To search for early stages of star formation in N4,
we took 56 point sources from the Curvature Threshold Extractor package (CuTeX) catalog. They are encompassed
within the 2.5 times radii of the bubble, of detection in the PACS 70~\um\ band.
Details about the catalog and the photometric procedures considered in CuTeX can be found in \citet{mol11}.

\subsection{Molecular Line Observations}
\begin{deluxetable}{lllllll}
\tabletypesize{\scriptsize} \tablecolumns{7} \tablewidth{0pt}
\tablecaption{Summary of KVN single-dish observations \label{tbl-1}}
\tablehead{
\colhead{Line}   & \colhead{$\nu$}  &
\colhead{$\theta_{HPBW}$} &\colhead{$\eta$} & \colhead{$T_{sys}$}  &
\colhead{$T_{rms}$}& \colhead{$\delta V$} \\
 \colhead{ }   & \colhead{GHz}   & \colhead{arcsec} &
\colhead{ } & \colhead{K}  & \colhead{(K)}& \colhead{km s$^{-1}$} }
\startdata
H$_{2}$O $6_{1,6}-5_{2,3}$  &  22.235077  &  120   &    0.46   &    80   &   0.05  &  0.21   \\
CH$_{3}$OH $7_{0,7}-6_{1,6}$ &  44.069476  &  60.5  &    0.48   &    170  &   0.07  &  0.11   \\
HCO$^{+}$ (1-0)              &  89.188526  &  31.5  &    0.38   &    250  &   0.08  &  0.05   \\
o-H$_{2}$CO $2_{1,2}-1_{1,1}$  &  140.83952  &  23.4  &    0.33   &    360  &   0.09  &  0.04
\enddata
\end{deluxetable}

Single-dish observations of molecular lines were performed with the Korean VLBI Network
(KVN) 21-m radio telescope at Yonsei site, Seoul on 2014 December 28. The front end
of the single-dish telescope is equipped with the four receivers working at
22, 43, 86, and 129 GHz simultaneously.  A digital
filter bank (DFB) with a 64 MHz bandwidth split into 4096 channels was adopted, resulting
in velocity resolutions of 0.21~\vel, 0.11~\vel, 0.05~\vel, and 0.04~\vel,
respectively. The pointing and tracking accuracies were better than 3$\arcsec$ \citep{kim11}.

Since mapping observations are quite time-consuming, we only chose six dense clumps,
located on the border of N4, to make single-point observations. To investigate
the dynamical and kinetic properties (see Section \ref{s-mle} and \ref{s-sfc}) and star formation activity in the clumps,
H$_{2}$O (22 GHz), CH$_{3}$OH (44 GHz), HCO$^{+}$ (89 GHz), and o-H$_{2}$CO (141 GHz)
were observed simultaneously. The data were calibrated to antenna temperature
($T_{a}^{*}$) using the chopper wheel method.
A summary of observations is given in Table \ref{tbl-1}, including the
half-power beam width ($\theta_{\mathrm{HPBW}}$), the beam efficiency ($\eta$), the typical systematic
temperature ($T_{\mathrm{sys}}$), the rms noise level ($T_{\mathrm{rms}}$) and the velocity resolution
($\delta V$). The data were analyzed and visualized with the software GILDAS \citep{gui00}.

\subsection{Archival data}\label{s-ad}
To study the physical characteristics of the bubble and its associated YSO candidates, auxiliary
data of images and point sources were complemented by the GLIMPSE \citep{ben03},
MIPSGAL  \citep{car09} and \emph{WISE} \citep{wri10} surveys.
The images of the Spitzer Infrared Array Camera (IRAC) at 3.6, 4.5, 5.8, and 8.0 \um, together with
the Multiband Imaging Photometer for \emph{Spitzer} (MIPS) at 24~\um\ were retrieved from the
\emph{Spitzer} Archive.\footnote{\url{http://sha.ipac.caltech.edu/applications/Spitzer/SHA/}}.
The angular resolutions of the images in the IRAC bands are $< 2\arcsec$ and it is
$\sim 6\arcsec$.0 in the MIPS 24~\um\ band. In addition,
the \emph{WISE} survey has mapped the entire sky in four infrared
bands centered at 3.4, 4.6, 12, and 22~\um, with a
resolution of 6$\farcs$1, 6$\farcs$4, 6$\farcs$5, and 12$\farcs$0, respectively \citep{wri10}.
 Point sources were obtained from the
\emph{AllWISE} Source and MIPSGAL Catalogs.\footnote{\url{http://irsa.ipac.caltech.edu/frontpage/}}
Both catalogs contain the photometries of point sources cross matched with the Two Micron All Sky
\citep[2MASS,][]{skr06} survey.
The 20 cm radio continuum image was used from the
Multi-Array Galactic Plane Imaging Survey \citep[MAGPIS,][]{hel06} archive\footnote{\url{http://third.ucllnl.org/gps/}}
was used to analyze the properties of the H\,{\sc ii}~region associated with N4, . Its angular resolution is less than
6$\farcs$0 and the sensitivity is better than 0.15 mJy beam$^{-1}$.

\begin{figure}
    \centering
    \includegraphics[width=0.48\textwidth]{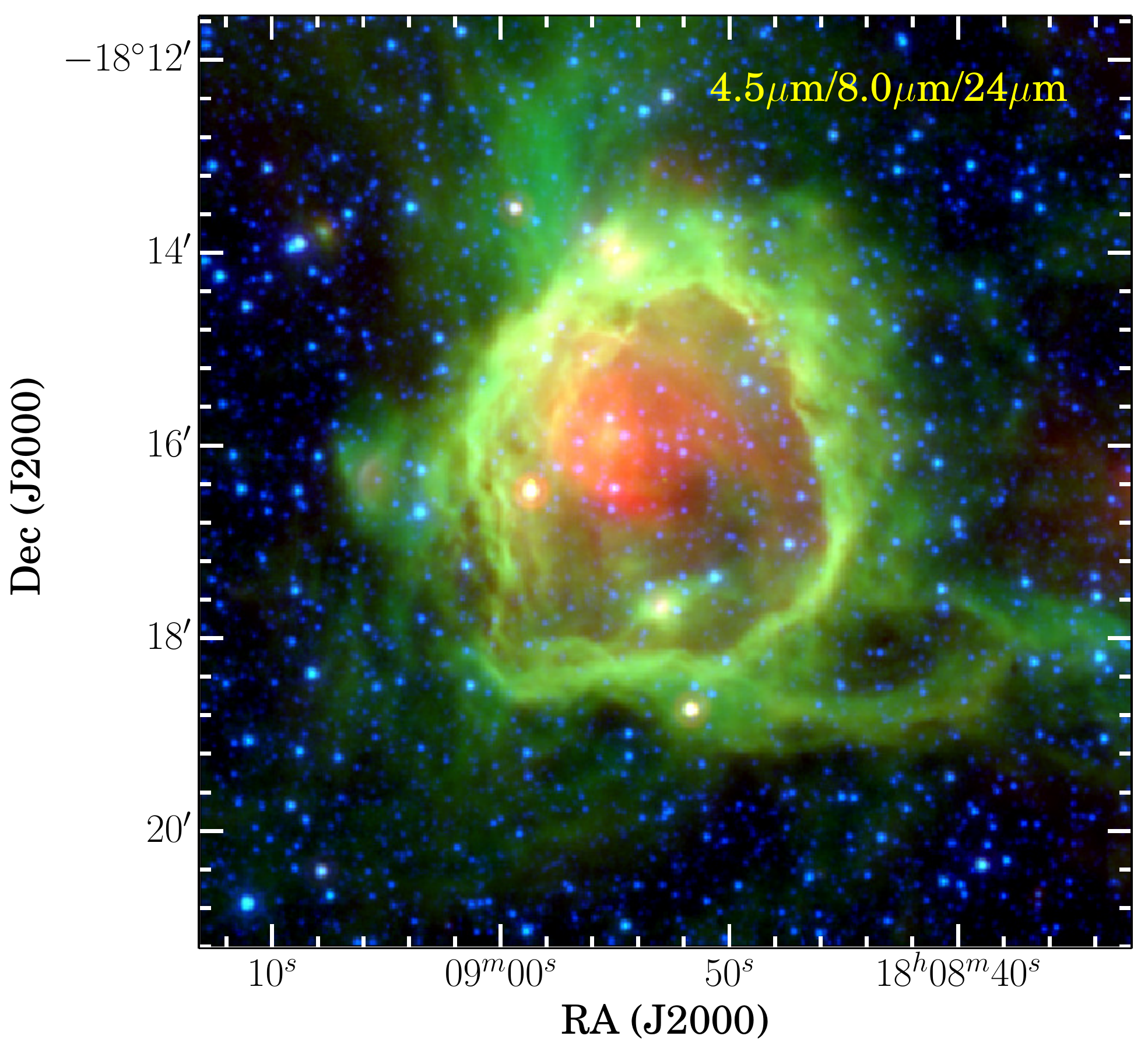}
    \caption{ Composite three-color image of N4.   The IRAC 4.5 \um, 8.0 \um, and 24~\um\
    are colored in blue, green, and red, respectively.
    The image is centered at \coor.}
    \label{fig-rgb1}
\end{figure}

\section{Results}\label{s3}

\subsection{Dust Emission} \label{s-dust}
\subsubsection{Dust Temperature and Column Density Distributions}
\emph{Herschel} observations with the widespread wavelength coverage can reveal the dust properties
of the entire bubble. To do this, a SED pixel-by-pixel fitting  was performed to obtain
the distributions of dust temperature ($T_{\mathrm{dust}}$) and H$_{2}$
column density ($N_{\mathrm{H_{2}}}$).
Assuming optically thin dust emission, a graybody function for a single temperature can be expressed
as follows:
    \begin{equation}\label{eq-1}
    I_{\nu} = \kappa_{\nu 0}(\nu/ \nu_{0})^{\beta}\ B_{\nu}(T_{\mathrm{d}})\ \mu \ m_\mathrm{H}\ N(\mathrm{H}_{2}),
    \end{equation}
where $I_{\nu}$ is the surface brightness, $B_{\nu}(T_{\mathrm{d}})$ is the blackbody function
for the dust temperature of $T_{\mathrm{d}}$,  the mean molecular weight $\mu$ is assumed to be
2.8 \citep[e.g.,][]{kau08,sad13}; and $m_\mathrm{H}$ is the mass of a hydrogen atom.
The dust opacity per unit mass of both dust and gas is defined as
$\kappa_{\nu}=\kappa_{\nu 0}(\nu/ \nu_{0})^{\beta}$, where $\kappa_{\nu 0}$ is assumed to be
0.1~cm$^{2}~$g$^{-1}$ at 1~THz \citep{bec90} under a gas-to-dust mass ratio of 100,
and $\beta$ is fixed to 2, which is a statistical value found in a large sample of
H\,{\sc ii}~regions \citep{and12}.

The \emph{Herschel} data of the four bands at 160, 250, 350, and 500~\um\  were
used for the SED fitting. Since emission at 70~\um\ may trace hotter components such as very
small grains (VSGs) and warmer material heated by protostars, it was not considered
in the SED fitting for a single temperature. The four images were
convolved to the same resolution 43$\farcs$8, and rebinned to the same pixel size 11$\farcs$5.
To minimize the contribution from the line of sight, we subtracted the backgrounds for each image using
a reference area away from the bubble. The corresponding backgrounds were estimated to be
634~MJy~sr$^{-1}$ for 160~\um, 387~MJy~sr$^{-1}$ for 250~\um, 205~MJy~sr$^{-1}$ for 350~\um\, and
79~MJy~sr$^{-1}$ for 500~\um. By treating $T_{\mathrm{dust}}$ and $N_{\mathrm{H_{2}}}$ as
free parameters, the SED fitting was performed using  the IDL program
MPFITFUN\footnote{\url{http://www.physics.wisc.edu/~craigm/idl/fitting.html}} \citep{mar09}.
In principle, the 870~\um\ image should have been considered in the SED fitting.
Although the 870~\um\ data constrain the dust temperature, as they are sensitive to colder
components than \emph{Herschel} data, they miss the bulk of extended emission,
which is the major interest in our current work. Thus, the image at 870~\um\ has been excluded in
the SED fitting.

\begin{figure*}
    \centering
    \includegraphics[width=0.8\textwidth]{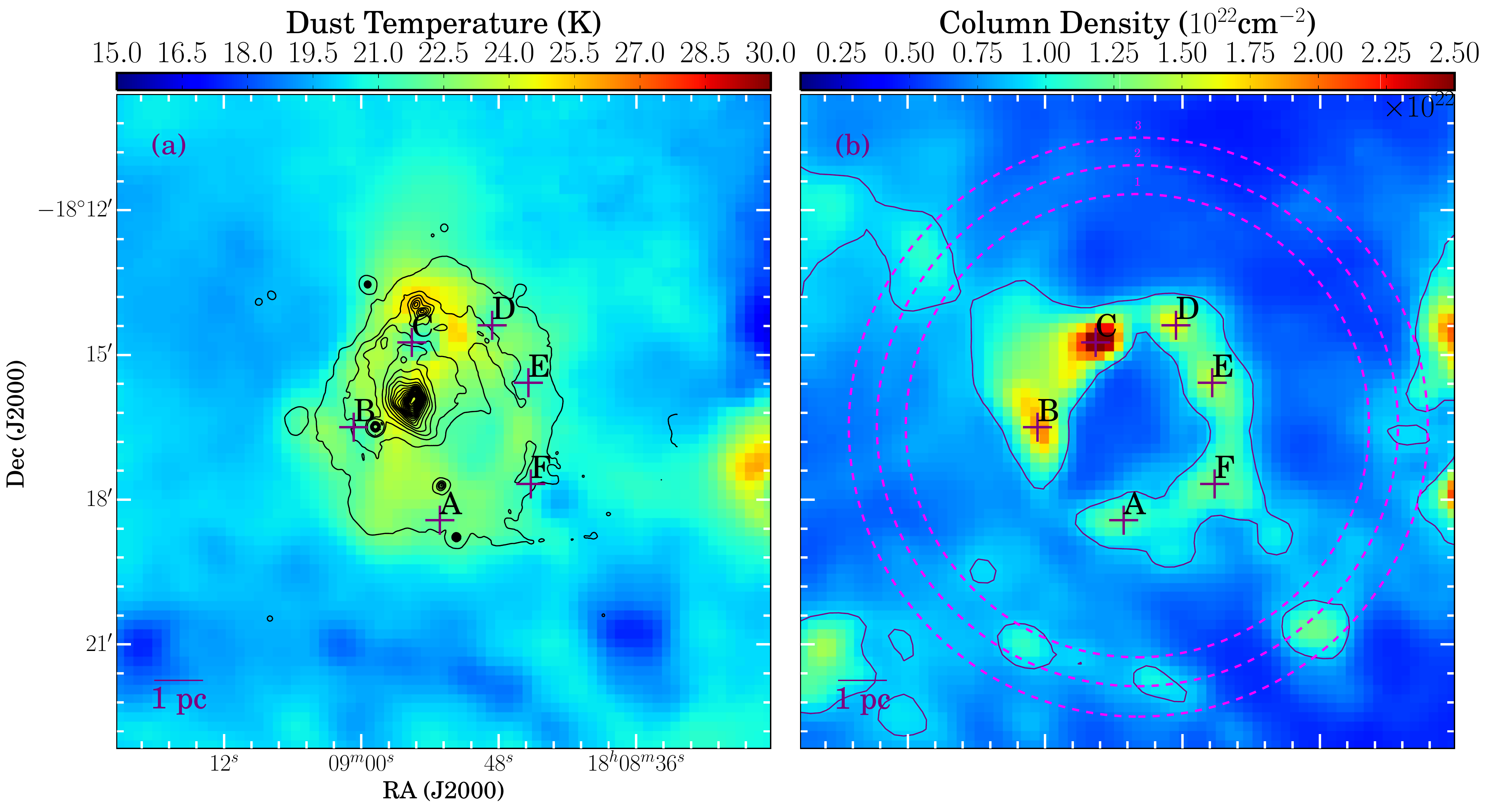}
    \caption{Maps of the dust temperature (a) and the $\mathrm{H}_{2}$ column density (b) of N4.
    The two maps were all built on the SED fitting pixel by pixel, centered at \coor. The crosses symbolize
    six dense clumps extracted from the $\mathrm{H}_{2}$ column density map. The purple contour shows
    a level of $9\times10^{21}$ cm$^{-2}$, which covers a region of $\sim 1.5$ radius region of the bubble.
    Column density PDFs within the three magenta dashed circles are taken into account (see Section \ref{s-feedback}).}.
    \label{fig-NT}
\end{figure*}

The resulting $T_{\mathrm{dust}}$ and $N_{\mathrm{H_{2}}}$ maps are shown in
Figure~\ref{fig-NT}, where the black contours stand for the emission at 24~\um and the
purple contour delineates a level of $N_{\mathrm{H_{2}}} = 9\times10^{21}$~\cm2, sculpting
a shell structure. Emission traced by 24~\um\ mainly comes from hot dust which can reach very
high temperatures after absorbing high-energy photons \citep[e.g.,][references therein]{deh10}.
The hot dust distribution shown in Figure~\ref{fig-NT}(a) indicates a dust heating from
the interior to the edge of N4 (also see Figure~\ref{fig-rgb1}), therefore, possible temperature gradients
along this direction would be expected. However, the dust distributions in the interior and on
the edge of N4 appear to be smooth. Given the resolution of 6$\arcsec$ at 24~\um, the smoothness could be
interpreted as the temperature gradients on small scales not resolved by the angular resolution
of 43$\farcs$8 of the dust temperature map. Additionally, we note that there is a cold region of $\sim 22$ K
within the bubble, which may be not real due to the low flux found there \citep{and12}.

On the large scale, the temperature distribution nearly matches with
24~\um\ emission (see Figure~\ref{fig-NT}(a)). That is, the temperatures
of cold dust in the bubble direction are hotter than those away the
bubble, suggesting that the H\,{\sc ii}~region enclosed by N4
has been heating its surroundings. On the small scale, there are three regions that are especially warmer
than other places in the $T_{\mathrm{dust}}$ map.
The first region, located near the bubble center, it
can be attributed to the heating of the central ionizing stars (see Section~\ref{s-HII}).
The second warm region, with which some YSO candidates are associated
(see Section~\ref{s-yso}), is to the north of N4. Hence, this warm region could be a result of the heating
from embedded protostars therein. However, the third warm region in the northern
border of N4 has no bright 24~\um\ counterpart or potential associated YSO candidate.
Figure~\ref{fig-NT}(b) shows that the column density of this region is lower than that of its neighbors.
This difference suggests that this warm region may be  heated premarily by the ionizing photons leaking
from the H\,{\sc ii}~region.

For the H$_{2}$ column density distribution in Figure~\ref{fig-NT}(b),
a shell morphology is inscribed as seen in IR bands.
In the shell, there is an anti-correlation between $N_{\mathrm{H_{2}}}$ and $T_{\mathrm{dust}}$.
This may be attributed to a  lower penetration of the external heating from the H\,{\sc ii} region into dense regions.
If the shell is bounded by a level of $9\times10^{21}$~cm$^{-2}$ as shown in Figure~\ref{fig-NT}(b),
which covers a region of $\sim 1.5$ radius of the bubble, the mean $N_{\mathrm{H_{2}}}$ column
density and total mass of the shell can be estimated to be $1.1\times10^{22}$ cm$^{-2}$
and $5.2\times10^{3}$~\msun, respectively. The total mass of the neutral material of the entire bubble is
estimated to be $\sim~5.5\times10^{3}$~\msun.

\subsubsection{Dense Clumps}\label{s-DC}
Six dense clumps were identified within the shell structure in the $\mathrm{H}_{2}$ column density image.
These sources were decomposed visually from the image by ellipses
bounded by the level of $9\times10^{21}$~cm$^{-2}$ enclosing
the majority of fluxes. The six clumps have different physical sizes and peak column densities
($N_{\mathrm{H_{2},peak}}$). Their average temperatures ($<T_{dust}>$)
 were obtained from the dust temperature image. We estimated the total mass (gas+dust)
 and the volume density for each clump.

\begin{deluxetable*}{llllllllll}[tbh!]
\tabletypesize{\scriptsize} \tablecolumns{12} \tablewidth{0pt}
\tablecaption{Characteristics of six dust clumps \label{tbl-2}}
\tablehead{
\colhead{Name}   & \colhead{R.A.}  & \colhead{Decl.} & \colhead{$r_{major}$} &
\colhead{$r_{minor}$} & \colhead{$r_{eff}\tablenotemark{a}$} &\colhead{$<T_{dust}>$} &
\colhead{$N_{\mathrm{H_{2},peak}}$}  & \colhead{$n_{\mathrm{H_{2}}}$}&
\colhead{$M_{(gas+dust)}$}  \\
\colhead{ }   & \colhead{(hh:mm:ss.ss)}   & \colhead{(dd:mm:ss.s)} &
\colhead{(arcsec)} & \colhead{(arcsec) } & \colhead{(pc)}  & \colhead{(K)}& \colhead{(10$^{22}$ \cm2)}
& \colhead{(10$^{4} $\vol)} & \colhead{(10$^{2}$ \msun)} }

\startdata

A  & 18:08:53.14  & -18:18:25.5   &    24.3  &    44.6   &    0.4 $\pm$ 0.1  &    22.3   $\pm$    0.2   &    1.0   $\pm$    0.1  &     4.0    $\pm$   1.4   &    0.8    $\pm$   0.5  \\
B  & 18:09:00.66  & -18:16:30.0   &    36.7  &    51.9   &    0.6 $\pm$ 0.2  &    22.1   $\pm$    0.5   &    1.9   $\pm$    0.3  &     5.0    $\pm$   2.1   &    3.5    $\pm$   2.5  \\
C  & 18:08:55.59  & -18:14:44.5   &    34.9  &    49.9   &    0.6 $\pm$ 0.2  &    23.1   $\pm$    0.9   &    2.8   $\pm$    0.5  &     7.8    $\pm$   3.6   &    4.5    $\pm$   3.4  \\
D  & 18:08:48.57  & -18:14:23.1   &    34.1  &    40.8   &    0.5 $\pm$ 0.1  &    22.3   $\pm$    0.7   &    1.6   $\pm$    0.2  &     5.2    $\pm$   2.1   &    1.9    $\pm$   1.3  \\
E  & 18:08:45.39  & -18:15:34.6   &    35.8  &    41.0   &    0.5 $\pm$ 0.2  &    21.7   $\pm$    0.7   &    1.5   $\pm$    0.1  &     4.7    $\pm$   1.8   &    1.9    $\pm$   1.3  \\
F  & 18:08:45.18  & -18:17:40.7   &    37.4  &    35.2   &    0.5 $\pm$ 0.1  &    23.6   $\pm$    0.7   &    1.3   $\pm$    0.1  &     4.4    $\pm$   1.6   &    1.4    $\pm$   0.9

\enddata
\tablenotetext{a}{$r_{eff}$ is the deconvolved effective radius.}
\end{deluxetable*}

Table \ref{tbl-2} gives a summary of the aforementioned parameters, where the main source of error
is the uncertainty of the distance to the bubble. The average size, dust temperature, column density,
number density, and mass of the six clumps are $\sim0.5$ pc, $\sim22$ K, $\sim1.7\times10^{22}$ cm$^{-2}$,
$\sim4.4\times10^{4}$ cm$^{-3}$, and $\sim3.2\times10^{2}$ \msun, respectively. The average density
$> 10^{4}$ cm$^{-3}$ is consistent with the detections of dense molecular lines within these clumps
(see Section \ref{s-mle}).

\subsection{Molecular Line Emission}\label{s-mle}
\begin{deluxetable*}{lllllllll}[tbh!]
\tabletypesize{\scriptsize} \tablecolumns{9} \tablewidth{0pt}
\tablecaption{Parameters of molecular lines of the six dust clumps \label{tbl-3}}
\tablehead{
\colhead{}  &  \multicolumn{3}{r}{ HCO$^{+}$} &\multicolumn{3}{r}{o-H$_{2}$CO} \\
\cline{2-5} \cline{6-9} \\
\colhead{Name}&\colhead{$V_{LSR}$}&\colhead{$FWHM$}& \colhead{$T_{mb}$}& \colhead{$\int T_{mb} \ \mathrm{d}V$}&
\colhead{$V_{LSR}$}&\colhead{$FWHM$}& \colhead{$T_{mb}$} & \colhead{$\int T_{mb} \ \mathrm{d}V$} \\
\colhead{ }&\colhead{(km s$^{-1}$)}&\colhead{(km s$^{-1}$)}&\colhead{(K)}&\colhead{(K km s$^{-1}$)}&\colhead{(km s$^{-1}$)}
&\colhead{(km s$^{-1}$)}&\colhead{(K)}&\colhead{(K km s$^{-1}$)}}
\startdata

  A  &   23.61    $\pm$  0.06   &   2.62   $\pm$   0.11   &   1.30  $\pm$    0.12  & 3.62  $\pm$     0.16  &   23.53  $\pm$    0.25  &    2.30  $\pm$    0.52   &  0.40  $\pm$    0.11 &  0.99  $\pm$    0.19    \\
  B  &   23.58    $\pm$  0.06   &   2.03   $\pm$   0.08   &   1.84  $\pm$    0.12  & 4.01  $\pm$     0.15  &   23.88  $\pm$    0.08  &    1.91  $\pm$    0.19   &  0.97  $\pm$    0.11 &  1.98  $\pm$    0.17    \\
 C-I &   23.56    $\pm$  0.21   &   2.54   $\pm$   0.21   &   0.99  $\pm$    0.11  & 3.84  $\pm$     0.31  &   24.51  $\pm$    0.38  &    4.32  $\pm$    0.70   &  0.53  $\pm$    0.12 &  2.42  $\pm$    0.41    \\
C-II &   26.68    $\pm$  0.06   &   1.79   $\pm$   0.21   &   2.91  $\pm$    0.11  & 4.74  $\pm$     0.26  &   26.87  $\pm$    0.04  &    1.16  $\pm$    0.11   &  1.78  $\pm$    0.12 &  2.19  $\pm$    0.29    \\
  D  &   25.00    $\pm$  0.06   &   3.06   $\pm$   0.10   &   1.47  $\pm$    0.11  & 4.77  $\pm$     0.17  &   25.11  $\pm$    0.12  &    2.00  $\pm$    0.35   &  0.67  $\pm$    0.11 &  1.43  $\pm$    0.19    \\
  E  &   25.28    $\pm$  0.06   &   2.46   $\pm$   0.06   &   2.33  $\pm$    0.11  & 6.12  $\pm$     0.17  &   25.47  $\pm$    0.09  &    2.23  $\pm$    0.28   &  0.96  $\pm$    0.12 &  2.26  $\pm$    0.21    \\
  F  &   24.86    $\pm$  0.09   &   4.02   $\pm$   0.24   &   0.85  $\pm$    0.12  & 3.64  $\pm$     0.23  &   25.36  $\pm$    0.20  &    2.91  $\pm$    0.51   &  0.49  $\pm$    0.11 &  1.53  $\pm$    0.22

\enddata
\end{deluxetable*}

\begin{figure}
  \centering
  \includegraphics[width=0.48\textwidth]{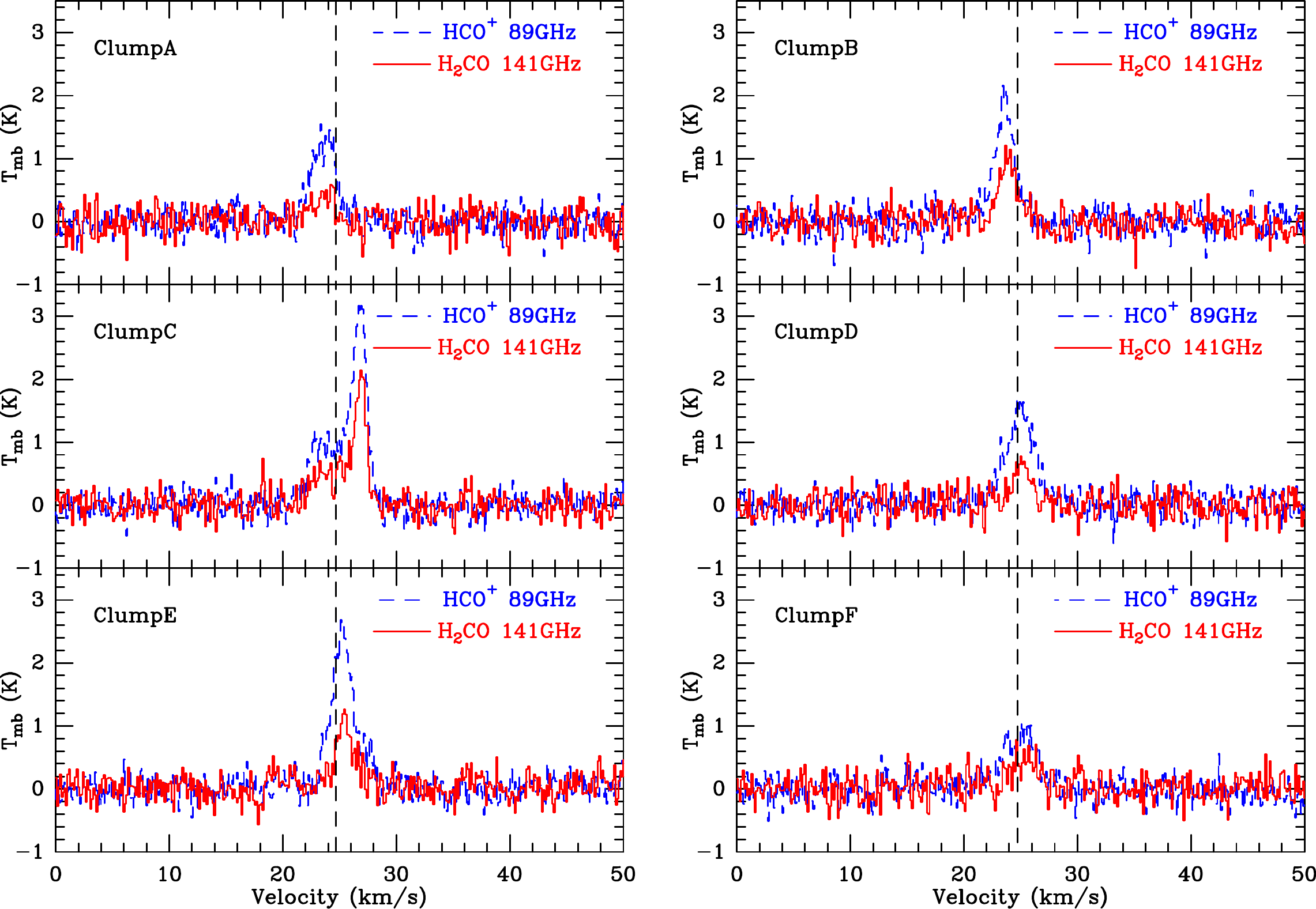}

\caption{HCO$^{+}$~(1-0) and  o-H$_{2}$CO $2_{1,2}-1_{1,1}$  spectra of the six clumps. The black dashed lines indicate a
systematic velocity of 24.7~\vel\ for the bubble. The systematic velocity is determined from the average spectra of
the HCO$^{+}$ and  H$_{2}$CO lines of the six clumps, which is coincident with the results
of \co, \13co, and \c18o\ (1-0) observations \citep{li13}. $T_{\mathrm{mb}}$ represents the main-beam temperature. The name of each clump is noted in the top left of each panel.}
\label{fig-coreSp}
\end{figure}

Figure~\ref{fig-coreSp} presents HCO$^{+}$ (1-0) and  o-H$_{2}$CO $2_{1,2}-1_{1,1}$ spectra of the six clumps. Since
the molecules H$_{2}$O and CH$_{3}$OH  were not detected, their spectra are not be
displayed. Emission of HCO$^{+}$ (1-0) and  o-H$_{2}$CO $2_{1,2}-1_{1,1}$ is strong enough to be detected in all clumps.
Their detections indicate high densities for the six clumps,
since the critical densities of both lines are $4.5\times10^{4}$ \vol, and $1.3\times10^{5}$ \vol\
for a kinetic temperature of 20 K \citep{shi15}.
Using the software GILDAS, we retrieved the observed parameters from the spectra of the six clumps, including
the peak velocity ($V_{LSR}$), line width (FWHM) and the main-beam temperature ($T_{mb}$), and the
velocity-integrated intensity ($\int T_{mb} \ \mathrm{d}V$)
of the different spectra (see Table \ref{tbl-3}).

It is interesting to analyze the centroid velocity variances of the six clumps. The spectra of HCO$^{+}$ and  o-H$_{2}$CO obviously show
the velocities blueshifted to a systematic velocity of 24.7~\vel in the southeastern (SE) part of the bubble
(i.e., Clumps A and B) and the redshifted velocities in the northwestern (NW) part (i.e., Clumps D and E).
The two remaining clumps show no evident velocity shifts. Observations of \emph{J=1-0} of \co, \13co, and \c18o
also have already revealed the same velocity difference between the SE and the NW  \citep[see Figure~3 of][]{li13}.
Due to the lack of blueshifted profiles in the front of the bubble and redshifted ones in the back cap
of the bubble in the CO observations, \citet{li13} suggested that the bubble N4 may be expanding along the SE-NW direction
with an inclination relative to the sky plane.
This scenario is compatible with \citet{bea10} interpretations of molecular gas studies around a number
of IR bubbles, where they conclude that the SFs driven by massive stars tend to create rings.
Therefore, the shell of dust and gas in the bubble N4 is presumably assembled by the expanding H\,{\sc ii}~region.

Assuming that the observed o-H$_{2}$CO line is optically thin and in local
thermodynamic equilibrium (LTE), we estimate the
column density of this molecular species toward each dense clump presented in Section \ref{s-DC} using \citep{goldsmith99}:
\begin{equation}
N =  \frac{8 \pi k \nu^{2}}{h c^{3} A_{ul}} \frac{Q_{rot}}{g_{u}}~{\rm exp}(\frac{E_{u}}{T_{rot}}) \int{T_{mb}~{\rm dv}},
\label{eqNh2co}
\end{equation}
where $A_{ul} = 5.4 \times 10^{-5}$ s$^{-1}$, $E_{u} = 21.9$ K, and $g_{u} = 15$. Assuming that the dust and the gas are coupled
at the same temperature, in LTE conditions we can approximate $T_{dust}=T_{kin}=T_{rot}$. Thus, for $T_{rot}$ we used the $T_{dust}$ value obtained for each clump (see Section \ref{s-DC}). Given a temperature of
$\sim22$~K averaged over the six clumps, the o-H$_{2}$CO partition function $Q_{rot}$ was estimated to be $\sim50$ by the extrapolation
to the relation between $Q_{rot}$ and $T_{rot}$ from the CDMS Catalog.\footnote{\url{www.astro.uni-koeln.de/cdms/catalog}}
The obtained column density values for each clump are presented in Table \ref{colabund}, which also includes
the o-H$_{2}$CO abundances ($X =$ N(o-H$_{2}$CO)/N(H$_{2}$)). The abundances were estimated using the H$_{2}$ column densities presented in Section \ref{s-DC}.  We note that  o-H$_{2}$CO emission is generally optically thick, therefore, the assumption of optically thin emission of o-H$_{2}$CO results in underestimated column density and abundance.

The obtained o-H$_{2}$CO column densities and abundances are quite similar to those obtained toward other photo-dominated
regions (PDRs), such as the Horsehead PDR \citep{guzman11}, and the Orion Bar \citep{vanderwiel,vanderwielE}.
Considering that the dust grains are cold ($T_{dust} \sim 22$ K) in the dense clumps around the N4 bubble and given the abundances
presented in Table \ref{colabund},
we suggest that in this region the formaldehyde may be formed mainly in the gas phase with a probable contribution
of photo-desorption of the grain mantles as was found in the Horsehead PDR \citep{guzman11}.

Observations of other o-H$_{2}$CO and p-H$_{2}$CO
lines toward this region would be very useful to give an important observational probe of the gas phase and grain surface
chemistry toward PDRs.

\begin{table}
\begin{center}
\caption{o-H$_{2}$CO LTE column densities and abundances.}
\label{colabund}
\begin{tabular}{lcc}
\tableline
\tableline
Clump            &  N(o-H$_{2}$CO)                 &  $X$  \\
                 &   ($\times10^{12}$ cm$^{-2}$)   &     (10$^{-10}$)        \\
\tableline
\noalign{\smallskip}
A                &   6.26                          &   6.2     \\
B                &   12.66                         &   6.6     \\
C                &   28.20                         &   9.9      \\
D                &    9.05                         &   5.6      \\
E                &   14.61                         &   9.7     \\
F                &   9.13                          &   7.0    \\
\tableline
\end{tabular}
\end{center}
\end{table}

\subsection{YSOs Associated with N4}\label{s-yso}

To understand more about star formation histories associated with the bubble, candidate YSOs
associated with N4 have been initially identified using the online SED fitting
tool\footnote{\url{http://caravan.astro.wisc.edu/protostars/}}
 of \citet{rob06,rob07}
and then further demonstrated and classified with color-color diagrams. The SED fitting tool
invokes a grid of 20,000 two-dimensional Monte Carlo radiation transfer models \citep[e.g.,][]{whi04}.
With 10 viewing angles for each model, there are actually 200,000 YSO SED models. This tool
works as a linear regression method to fit these models to the multi-wavelength photometry measurements
of a given source. The goodness/badness of each fit could be quantified by a specified value of the best $\chi^{2}$ ($\chi^{2}_{best}$), whereby YSO candidates could be robustly distinguished
from reddened photospheres of main-sequence and giant stars since YSOs require a thermal emission component
to reproduce the shapes of their mid-IR excesses. In addition, this method can make use of any available data
to constrain the shape of SED profiles as well as possible (see Appendix~\ref{s-app}). However, we have to acknowledge that
the disk and stellar parameters inferred from the SED models can be very uncertain
\citep[][and T.P. Robitaille 2015, private communication]{off12}.
Therefore, if these uncertain parameters are used to classify the identified YSO candidates with the scheme
of YSO classification of \citet{rob06}, the classification results must be unreliable. To avoid this uncertainty,
the color selection schemes of \citet{gut09} and \citet{koe12} have been adopted to group the YSO
candidates identified by the SED fitting into three ``Classes": Class~I, Class~II, and Transition
Disk (TD) YSOs. Class~I objects are deeply embedded protostars with a dominant infalling envelope,
Class~II objects are surrounded by a substantial accreting disk \citep{lad87}, and TD objects
are more evolved protostars where the inner parts of the disks have been cleared by photoevaporation
of the central stars or by planet-forming processes \citep{gut09,yua14,par15}. We believe that using
the infrared color schemes is
a better way to categorize the YSO candidates because the infrared color schemes are
a powerful proxy for measuring the excess emission \citep[e.g.,][]{all04,gut08,gut09,koe12}. Apart from
this, the candidate YSOs, which have been singled out by the SED fitting, can be further confirmed by
such color schemes.

Sixty potential YSOs within 5$\arcmin$ from the bubble center have been identified and classified.
Of these, there are 8 Class~I, 12 Class~II, and 40 TD objects. The classification results are
listed in Table \ref{tbl-5} and the detailed processes can be found in the Appendix~\ref{s-app}. We note that the
resulting 60 YSO candidates are incomplete but robust (see the Appendix~\ref{s-app}). Despite this incompleteness,
the resulting population with robust identifications is enough for simply learning about the
distributions of the YSOs associated with N4.

Figure~\ref{fig-N4rgbIII} shows the spatial distribution of the 60 YSO candidates.
One can see that the majority ($\sim82\%$) of YSOs are spatially correlated with the PDRs
as traced by 8 \um~emission (see Section~\ref{s-HII}), which indicates the good spatial association
of these potential YSOs with the bubble N4 and the strong impacts that the enclosed
H\,{\sc ii}~region is having on the surrounding star formation. In addition, there is an overdensity of
the number of YSO candidates residing on the verge of the bubble. This is also demonstrated by
the statistics of all candidates as a function of the normalized bubble radius
(see Figure~\ref{fig-ysoStat}). The statistics suggest that a large number ($\sim40\%$) of YSO candidates is
located within $1.0-1.5$ times the radius of the bubble. Such spatial distribution of YSOs in N4
agrees well with the statistical results based on the study of the association of a large sample
of YSOs with IR bubbles \citep{ken12,tho12}.

\begin{deluxetable}{lllllll}[tbh!]
\tabletypesize{\scriptsize} \tablecolumns{7} \tablewidth{0pt}
\tablecaption{Parameters of YSO candidates\label{tbl-5}}
\tablehead{
\colhead{No.} & \colhead{Name}  & \colhead{RA} & \colhead{Decl.} &
\colhead{$N$} & \colhead{$\chi^{2}_{\mathrm{best}}$}   & \colhead{Class}
}

\startdata

Y1   &  J180831.89-181639.1 & 272.133  & -18.278    &     7  &  13.4   &     II  \\
Y2   &  J180832.49-181616.6 & 272.135  & -18.271    &     7  &  35.7   &     TD  \\
Y3   &  J180834.88-181751.0 & 272.145  & -18.298    &     7  &  18.3   &     TD  \\
Y4   &  J180835.68-181814.8 & 272.149  & -18.304    &     7  &  11.0   &     TD  \\
Y5   &  J180839.36-181813.8 & 272.164  & -18.304    &     7  &  20.9   &     TD  \\
Y6   &  J180839.71-181841.5 & 272.165  & -18.312    &     7  &  47.7   &     TD  \\
Y7   &  J180840.11-181902.3 & 272.167  & -18.317    &     7  &  21.6   &     TD  \\
Y8   &  J180841.62-181600.0 & 272.173  & -18.267    &     7  &  12.9   &     TD  \\
Y9   &  J180842.08-181457.5 & 272.175  & -18.249    &     7  &   8.2   &     TD  \\
Y10  &  J180842.19-181329.1 & 272.176  & -18.225    &     7  &  34.6   &     TD  \\
Y11  &  J180842.69-181930.1 & 272.178  & -18.325    &     7  &  34.0   &     TD  \\
Y12  &  J180843.58-181441.8 & 272.182  & -18.245    &     7  &  38.3   &     TD  \\
Y13  &  J180844.11-181456.4 & 272.184  & -18.249    &     7  &  30.4   &     TD  \\
Y14  &  J180844.15-181401.9 & 272.184  & -18.234    &     7  &  43.4   &     TD  \\
Y15  &  J180844.97-181457.3 & 272.187  & -18.249    &     7  &  40.2   &     TD  \\
Y16  &  J180845.97-181612.4 & 272.192  & -18.270    &     8  &  32.2   &     II  \\
Y17  &  J180846.14-181844.0 & 272.192  & -18.312    &     7  &  32.0   &     TD  \\
Y18  &  J180847.47-181416.2 & 272.198  & -18.238    &     7  &  22.3   &     TD  \\
Y19  &  J180847.65-181450.0 & 272.199  & -18.247    &     7  &  32.9   &     TD  \\
Y20  &  J180847.67-181441.7 & 272.199  & -18.245    &     7  &  12.6   &     TD  \\
Y21  &  J180848.33-181357.2 & 272.201  & -18.233    &     7  &  33.9   &     TD  \\
Y22  &  J180849.08-181441.8 & 272.205  & -18.245    &     4  &   0.2   &      I  \\
Y23  &  J180849.78-181340.5 & 272.207  & -18.228    &     7  &   6.5   &     TD  \\
Y24  &  J180850.19-181816.2 & 272.209  & -18.305    &     7  &  27.3   &     II  \\
Y25  &  J180850.83-181836.6 & 272.212  & -18.310    &     7  &  14.0   &     II  \\
Y26  &  J180851.55-181831.9 & 272.215  & -18.309    &     7  &   7.6   &     TD  \\
Y27  &  J180852.42-181744.2 & 272.218  & -18.296    &     7  &  22.5   &     II  \\
Y28  &  J180852.59-181356.6 & 272.219  & -18.232    &     7  &   7.4   &     II  \\
Y29  &  J180852.97-181336.2 & 272.221  & -18.227    &     7  &  43.8   &     TD  \\
Y30  &  J180853.04-181742.2 & 272.221  & -18.295    &     8  &   5.5   &      I  \\
Y31  &  J180853.35-181130.9 & 272.222  & -18.192    &     7  &  41.9   &     TD  \\
Y32  &  J180853.65-181208.4 & 272.224  & -18.202    &     7  &  29.8   &     TD  \\
Y33  &  J180853.94-181422.4 & 272.225  & -18.240    &     5  &   0.3   &      I  \\
Y34  &  J180854.47-181406.5 & 272.227  & -18.235    &     4  &   0.1   &      I  \\
Y35  &  J180854.69-181230.5 & 272.228  & -18.208    &     7  &  39.4   &     TD  \\
Y36  &  J180854.87-181409.5 & 272.229  & -18.236    &     4  &   0.1   &      I  \\
Y37  &  J180855.80-181136.3 & 272.233  & -18.193    &     7  &  16.7   &     TD  \\
Y38  &  J180856.26-181505.7 & 272.234  & -18.252    &     7  &  26.5   &     II  \\
Y39  &  J180856.64-181420.4 & 272.236  & -18.239    &     7  &  12.4   &     II  \\
Y40  &  J180857.00-181354.7 & 272.238  & -18.232    &     4  &   0.1   &     II  \\
Y41  &  J180857.48-181853.3 & 272.240  & -18.315    &     7  &  25.0   &     TD  \\
Y42  &  J180858.07-181506.0 & 272.242  & -18.252    &     7  &  12.8   &     TD  \\
Y43  &  J180858.12-181807.2 & 272.242  & -18.302    &     7  &  26.4   &     TD  \\
Y44  &  J180858.75-181806.5 & 272.245  & -18.302    &     7  &  22.4   &     TD  \\
Y45  &  J180858.77-181629.7 & 272.245  & -18.275    &     11 &  25.1   &      I  \\
Y46  &  J180859.32-181316.1 & 272.247  & -18.221    &     7  &   9.8   &     II  \\
Y47  &  J180859.44-181332.8 & 272.248  & -18.226    &     7  &   0.9   &      I  \\
Y48  &  J180859.47-181855.8 & 272.248  & -18.316    &     7  &  14.1   &     TD  \\
Y49  &  J180900.26-181347.5 & 272.251  & -18.230    &     7  &  11.3   &     TD  \\
Y50  &  J180900.28-181921.8 & 272.251  & -18.323    &     7  &  31.4   &     TD  \\
Y51  &  J180900.48-181903.2 & 272.252  & -18.318    &     7  &  28.2   &     TD  \\
Y52  &  J180902.92-181316.2 & 272.262  & -18.221    &     7  &  41.9   &     TD  \\
Y53  &  J180904.56-181638.3 & 272.269  & -18.277    &     7  &   6.6   &     TD  \\
Y54  &  J180904.95-181806.2 & 272.271  & -18.302    &     7  &  23.3   &     TD  \\
Y55  &  J180905.60-181621.4 & 272.273  & -18.273    &     8  &   2.1   &     II  \\
Y56  &  J180906.19-181639.1 & 272.276  & -18.278    &     7  &   0.9   &     TD  \\
Y57  &  J180906.60-181847.9 & 272.278  & -18.313    &     7  &  19.7   &     TD  \\
Y58  &  J180908.60-181706.1 & 272.286  & -18.285    &     7  &  39.2   &     TD  \\
Y59  &   MG011.8545+00.7327 & 272.216  & -18.313    &     10  &   8.7   &     I  \\
Y60  &   MG011.9455+00.7481 & 272.248  & -18.226    &     9  &   6.6   &     II

\enddata
\tablenotetext{Notes}{Col.~1 is the number of the sources, Col.~2 is the source name retrieved
directly from the archived catalog, Cols.~3-4 are the coordinates, Col.~5 gives the count of data
points used in the SED fitting, Col.~6 provides the best square of the resulting SEDs, and Col.~7
the classification.}
\end{deluxetable}

\subsection{Properties of the Ionized Region}\label{s-HII}
\begin{figure}
  \centering
  \includegraphics[width=0.48\textwidth]{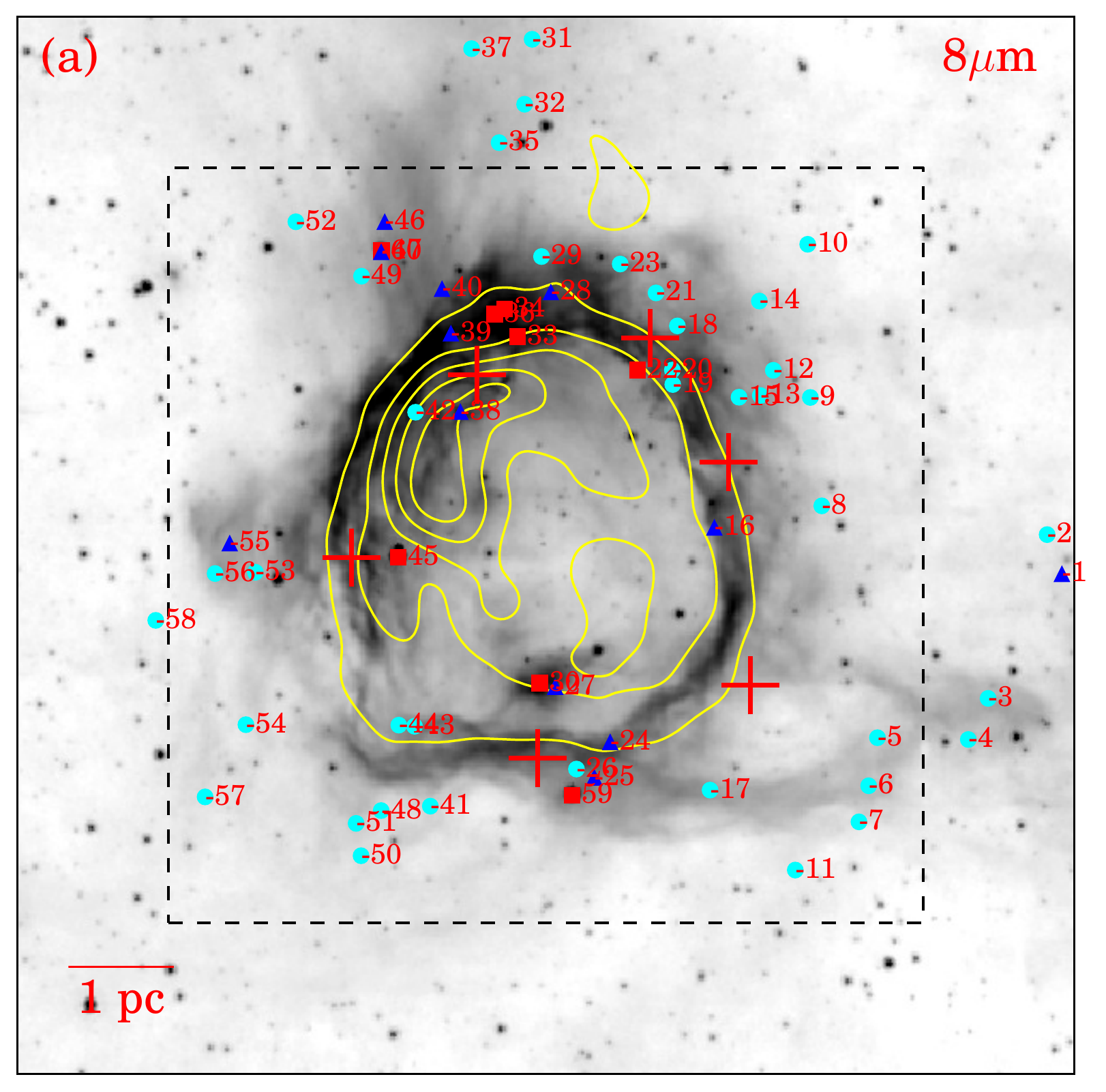}
  \includegraphics[width=0.48\textwidth]{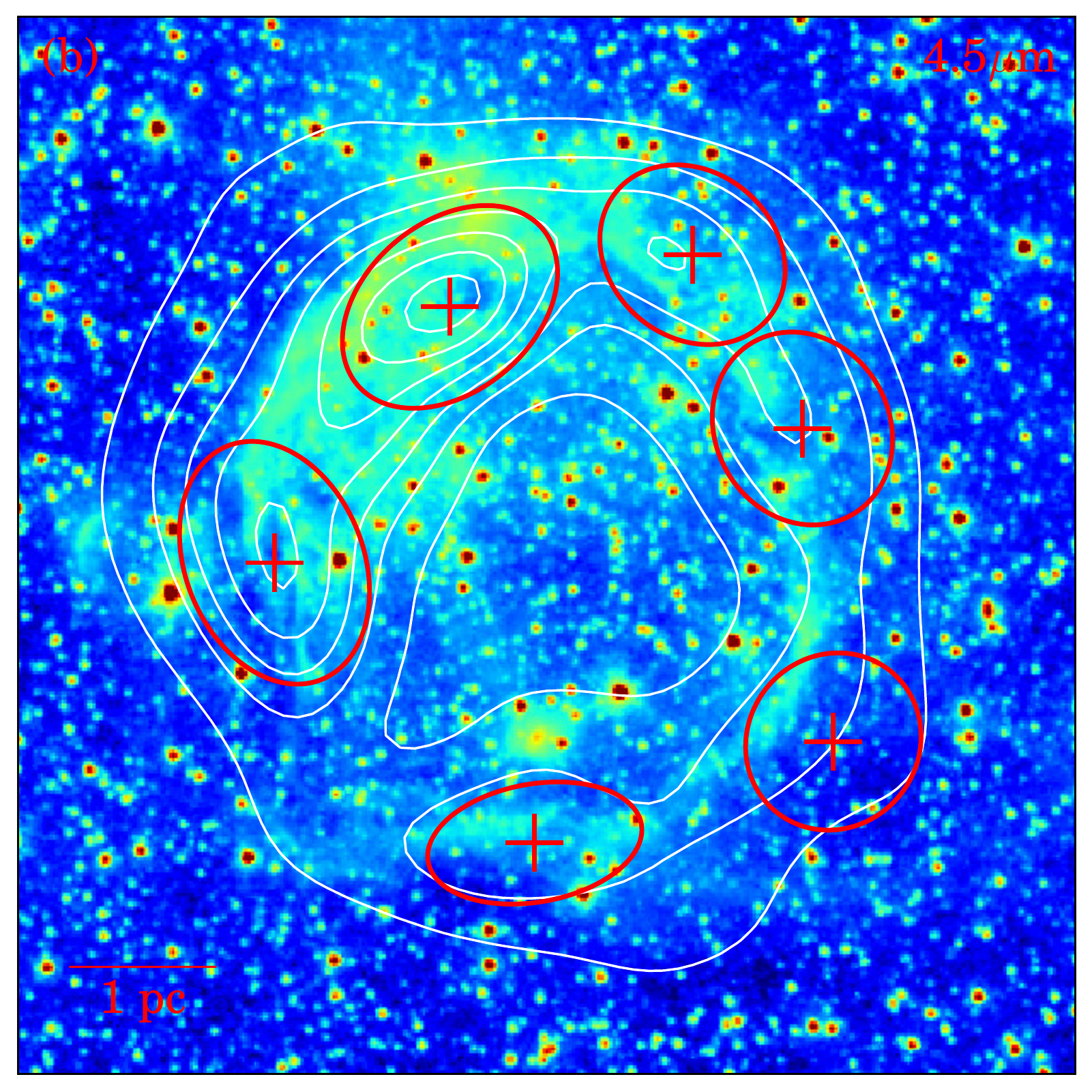}
\caption{(a) Image at 8~\um\ (grayscale) overlaid with free-free continuum
emission at 20~cm (contours). The yellow contours start from $5 \sigma$ (i.e.,
$1 \sigma=0.15$~mJy~Beam$^{-1}$) with a step of $8 \sigma$.
Class I, Class II, and transition disk YSO candidates are symbolized by square,
triangle, and circle symbols, respectively. The dashed rectangle is
 a selected region for a close-up view of 4.5~\um\ emission. (b) Image at 4.5~\um\
(colorscale) overlaid with dust emission at 250~\um\
(contours). The white contours start from 18~Jy~Beam$^{-1}$ with a step of
5.5~Jy~Beam$^{-1}$. The ellipses depict the
dust clumps extracted in Section~\ref{s-DC}, and the cross symbols represent
their peak positions. A scale bar of 1 pc is shown on the bottom left.}
\label{fig-N4rgbIII}
\end{figure}

\begin{figure}
  \centering
  \includegraphics[width=0.45\textwidth]{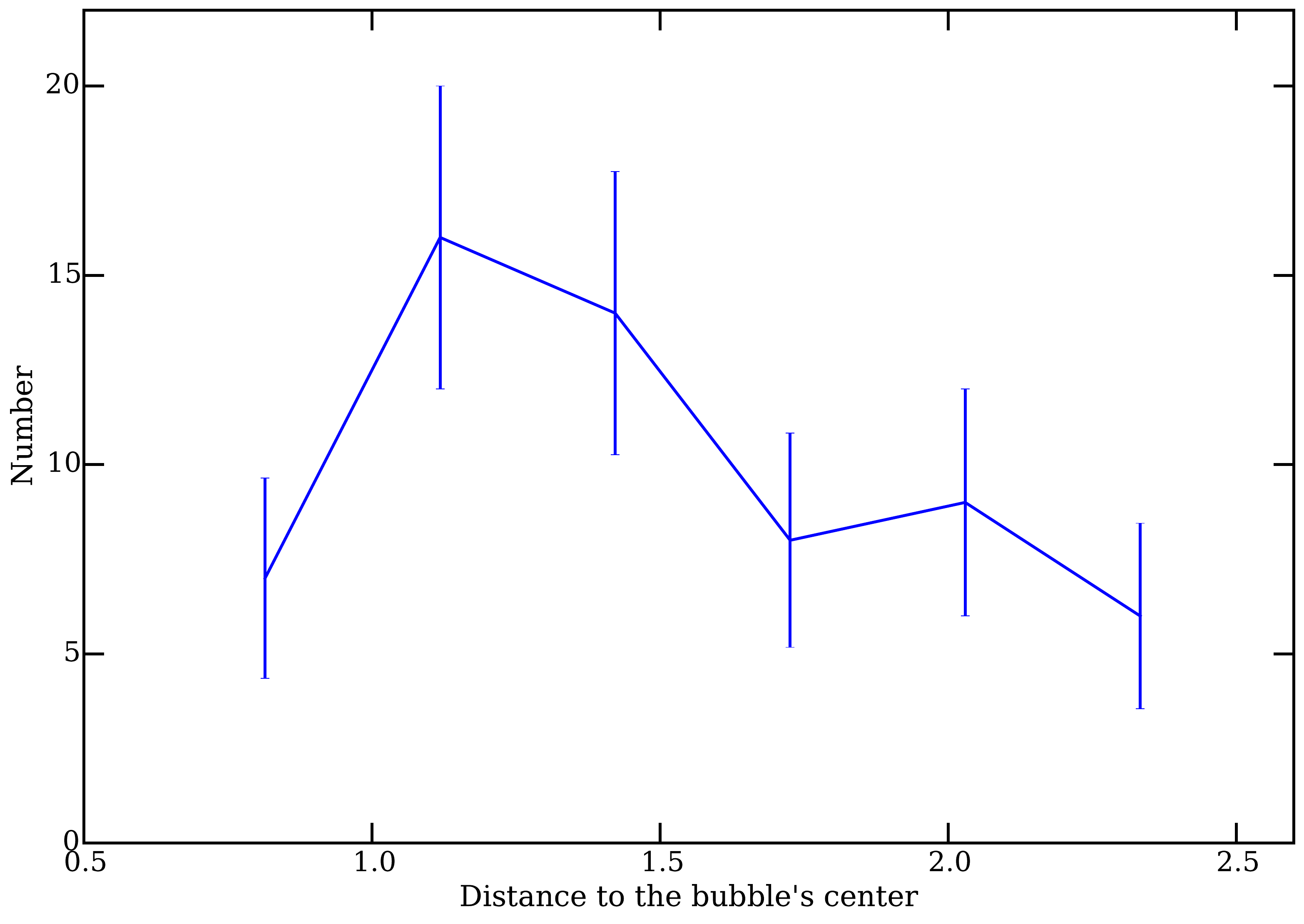}
\caption{Plot of the number counts of the 60 YSO candidates as a function of angular distance
from the center of the bubble N4. The distance is normalized by the bubble radius. Error
bars are determined via Poisson statistics.}
\label{fig-ysoStat}
\end{figure}

Figure~\ref{fig-N4rgbIII}~(a) presents the 8~\um\ image overlaid with radiation at
20 cm. The 8~\um\ emission predominantly comes from two prominent
polycyclic aromatic hydrocarbons (PAHs) at 7.7~\um\ and 8.6 \um. They are good tracers
of PDRs delineating the IF created by massive stars.
The 20 cm radiation represents free-free continuum emission from ionized gas.
As shown in Figure~\ref{fig-N4rgbIII}~(a), most of the ionized gas is surrounded
by the ring of PDRs, showing the strong effects of the H\,{\sc ii}~region on its
surroundings. Additionally, the ring of PDRs is also spatially well correlated
with the 250~\um\ emission (see Figure~\ref{fig-N4rgbIII}~(b)), which is
sensible to cold dust in thermal equilibrium.
As such, the cold dust wrapping ionized gas suggests a positive association
of the bubble N4 with the H\,{\sc ii}~region, enhancing their intensive mutual interaction.

Several properties
of the H\,{\sc ii}~region can be derived from 20 cm free-free continuum emission.
 \citet{chu78,win83,qui06} demonstrated the existence of a relationship between the H\,{\sc ii}
 region electron temperature, $T_{e}$, and the Galactocentric distance, $R_{\mathrm{Gal}}$.
 A detailed analysis of a sample of 76 H\,{\sc ii}
 regions with the highest quality data suggested that the electron temperature along
 the Galactic disk decreases approximately as a function of $R_{\mathrm{Gal}}$ \citep{qui06}:
    \begin{equation}\label{eq-4}
    T_{e} =  (5780\pm350) + (287\pm46)\ R_{\mathrm{Gal}}.
    \end{equation}
With an estimated Galactocentric distance of 5.5 kpc to N4, it yields an electron temperature of $7400\pm600$ K. If the H\,{\sc ii}~region
reaches the equilibrium at this temperature, the electron density, $n_{\mathrm{e}}$, the mass
of ionized gas, $M_{\mathrm{ion}}$, and the Lyman photons per second, $N_{\mathrm{Lym}}$, from the
exciting star(s), can be estimated
following \citet{kur94}:
    \begin{equation}\label{eq-5}
    n_{\mathrm{e}} = 2.878\times 10^{4}\ [(\frac{\theta}{\mathrm{arcsec}})^{-3}\ (\frac{D}{\mathrm{kpc}})^{-1} \ (\frac{\nu}{\mathrm{GHz}})^{0.1}\ (\frac{T_{\mathrm{e}}}{\mathrm{K}})^{0.35}\ (\frac{S_{\nu}}{\mathrm{Jy}})]^{0.5}\ \mathrm{cm}^{-3},
    \end{equation}

    \begin{equation}\label{eq-6}
    M_{\mathrm{ion}} = \frac{4}{3}\  \pi \ r^{3}_{\mathrm{H{\,\sc{\mathrm{ii}}}}}\ n_{e}\ \ m_\mathrm{\mathrm{p}},
    \end{equation}

    \begin{equation}\label{eq-7}
    N_{\mathrm{Ly}} = 7.588\times 10^{48}\ (\frac{T_{\mathrm{e}}}{\mathrm{K}})^{-0.5}\ (\frac{\nu}{\mathrm{GHz}})^{0.1}\ (\frac{S_{\nu}}{\mathrm{Jy}})\ (\frac{D}{\mathrm{kpc}})^{2} \ \mathrm{ph}\ \mathrm{s}^{-1},
    \end{equation}
where $S_{\nu}$ is the integrated flux density at the special frequency $\nu$ over the angular size $\theta$;
$r_{\mathrm{H{\,\sc{\mathrm{ii}}}}}$ is the radius of the H\,{\sc ii}~region, and $m_\mathrm{p}$ is the proton mass.
If the $\theta$ and $r_{\mathrm{H{\,\sc{\mathrm{ii}}}}}$ are considered to be approximately equal to those of the
bubble, a total of $32.8\pm0.01$ Jy at 20 cm measured by integrating over the 5 $\sigma$ contour
(see Figure~\ref{fig-N4rgbIII}) yields an estimated $n_{\mathrm{e}}$, $M_{\mathrm{ion}}$, and $N_{\mathrm{Lym}}$ of $75\pm15$ \cm2,
$50\pm38$ \msun, and $(1.5\pm1.1)\times10^{48}$ $\mathrm{ph}\ \mathrm{s}^{-1}$, respectively. The errors of
these parameters result mainly from the uncertainty in the distance of N4. According to \citet{mar05},
the estimated $N_{\mathrm{Lym}}$ (Log($N_{\mathrm{Lym}}$) $\simeq48.18$) indicates that a main O8.5V-O9V star was responsible for the ionization of N4.
These are lower limits given that any ionizing photons absorbed by dust or moving away from the bubble were not accounted for.

The central exciting stars of N4 may be located near the very bright 24~\um\ emission
inside the bubble. Based on the statistical analysis for the seven well-defined bubbles including
N4, \citet{deh10} argued that the hot dust grains seen at 24~\um\ inside the ionized region can be
heated by absorption of the Lyman continuum photons which are more numerous near exciting stars.
This indicates that the ionizing stars of N4 may reside near the very bright 24~\um\ emission
inside the bubble (see Figures~\ref{fig-rgb1} and \ref{fig-NT}(a)). This explanation can be further
supported by the positive association of intense free-free continuum radiation from ionized gas with
the bright 24~\um\ emission (see Figures~\ref{fig-NT}(a) and \ref{fig-N4rgbIII}(a)).
Observations of spectroscopy of ionizing stars \citep{mar10} are expected to reveal their natures.

\section{Discussion}\label{s4}
\subsection{Star Formation in Clumps}\label{s-sfc}
The good agreement of the PDRs with cold dust emission and the associated YSO candidates therein
are good indicators of the strong impact
of the H\,{\sc ii}~region on its surroundings and on star formation processes.
In what follows, we investigate the six identified dust
clumps in a context of star formation within them.

\begin{figure}
\begin{center}
    \includegraphics[width=0.45\textwidth]{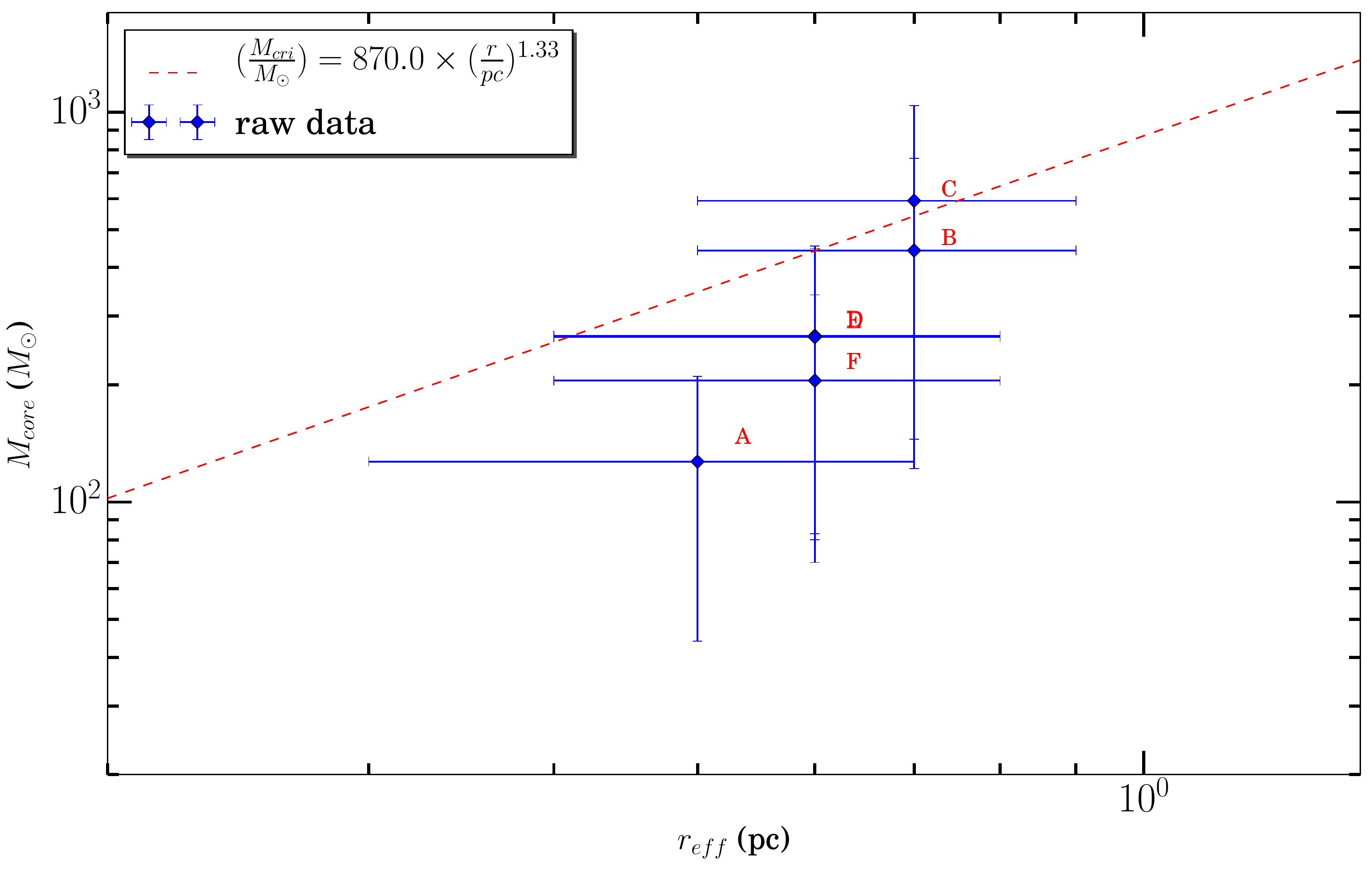}
    \caption{Mass-size relationship of the six clumps. The dashed line
    represents an empirical relation of $m(r) = 870$ \msun\ $(r/\mathrm{pc})^{1.33}$, which is
    expected to be as a threshold for forming high-mass stars \citep{kau10}.}
    \label{fig-kpmass}
\end{center}
\end{figure}

In the case of the Aquila Rift complex,  a column density threshold
of $N_{\mathrm{H_{2}}} > 7\times10^{21}$ cm$^{-2}$ was estimated for the formation of prestellar
cores \citep{and11}. For the six clumps in N4, the mean column density of $1.4\times10^{22}$ cm$^{-2}$
suggests that they may be capable of forming stars.
A mass-size relationship is generally adopted to predict whether the clumps will form low-mass stars or high-mass
stars \citep{kau10}. Based on the statistical analysis of nearby clouds without high-mass star formation
(i.e., Pipe Nebula, Taurus, Perseus, and Ophiuchus) and known samples with high-mass star formation
such as those of \citet{beu02,mue02,hil05}, and \citet{mot07}, a limiting mass-size
relation of $m(r)\geq 870$ \msun\ $(r/\mathrm{pc})^{1.33}$ was found to be an approximate threshold
for high-mass star formation \citep{kau10}. Figure~\ref{fig-kpmass} shows the mass versus size relation for the six clumps
in N4. Given uncertainties of the two parameters, the two clumps (Clumps B and C) are above
 the threshold, indicating that they may be forming massive stars, whereas the other four clumps
below the threshold may be inclined to form low-mass stars.
Indeed, several YSO candidates are close to all clumps but Clump F and six of these
(Y22, Y33, Y34, Y36, Y45 and Y59, see Figure~\ref{fig-N4rgbIII}) are classified as Class I YSOs, however,
there is no YSO candidate located at the peak positions of the six clumps. These facts
 suggest that the interiors of  six clumps may be in the process of forming stars but without active
ongoing star-forming activities.

No clear evidence for ongoing star formation has also been demonstrated in the
observations of the four molecular lines toward the peak positions of the six clumps.
The 22 GHz H$_{2}$O and 44 GHz Class I CH$_{3}$OH masers are known to occur in low- and/or high-mass
star formation regions \citep[e.g.,][]{for99,kur04}. These two types of masers are generally thought
to be associated with molecular outflows \citep[e.g.,][]{cod04,kur04}. Optically thick lines of HCO$^{+}$
and H$_{2}$CO are widely used to search for asymmetric
line profiles indicative of infall or outflow motions \citep[e.g.,][and references therein]{wu07,che10}.
Therefore, these four molecular lines are an important signpost of ongoing star formation. In the
bubble N4, it turns out that no detection of the 22 GHz H$_{2}$O and 44 GHz Class I CH$_{3}$OH masers
has been obtained toward the peak positions of all clumps, and the asymmetric red profiles
(i.e., a double peak spectral profile with red peak higher than blue peak)
have been detected only in Clump C by the HCO$^{+}$  and H$_{2}$CO lines (see Figure~\ref{fig-coreSp}).

As a result, no scenario of ongoing star formation is confirmed in the center of any of the six clumps
except for Clump C. On the one hand, this result may be a consequence of the poor angular
resolution of the present observations (23-120$\arcsec$, corresponding
to 0.26-1.35 pc). On the other hand, it can result from the incompleteness of the current
sample of YSO candidates. Therefore, we have correlated the six clumps with the YSOs in the
RMS survey database \citep{lum13}, the sources in the Methanol Multi-Beam survey
\citep{gre10}, and the compact H\,{\sc ii}~region catalog in the CORNISH survey \citep{pur13}.
It turns that there is only one source found in the RMS survey,
 which corresponds to the YSO candidate Y45 (see Figure~\ref{fig-N4rgbIII}). In addition,
 extended green objects (EGOs) identified based on their
extended 4.5~\um\ emission are thought to be massive YSO outflow candidates \citep{cyg08}.
Figure~\ref{fig-N4rgbIII}(b)
shows \emph{Spitzer} 4.5~\um\ emission overlaid with cold dust radiation at 250~\um. One can see
that six clumps except for Clumps C and D peak at dark 4.5~\um\ emission, and no brighter
extended 4.5~\um\ structures with respect to their backgrounds are detected for Clumps C and D,
indicating that there may be either no active ongoing star formation or deeply embedded protostars
at the centers of these clumps.

Due to the aforementioned lack of the H$_{2}$O, Class I CH$_{3}$OH maser detections, and other signs of ongoing
star formation, it is difficult to
confirm that the asymmetric red profiles are produced by outflow motions.
The fact that Clump C is located close to the strong ionization zone implies that it is fully
exposed to intense feedback from the main O8.5V-O9V star such as the compression of the
ionized region and stellar winds, forming the different two different velocity components and leading to the asymmetric
line profiles. Alternatively, this type of line profile may originate from the
clump rotation. Hence, it is worthwhile to determine the mechanism responsible for the asymmetric line profile in Clump C
 using a map observation with higher angular resolution.

\subsection{Feedback From Massive Stars}\label{s-feedback}
Massive stars are expected to affect their surroundings. Inside the bubble,
the hot dust traced by 24~\um\ emission has been presumably heated by absorption of
Lyman continuum photons whose free-free radiation is detected through radio continuum emission at 20 cm.
At the edge of the bubble, the shell of cold dust emission is associated with the PDR
that delimits the H\,{\sc ii}~region, indicating the bright illumination of the far
UV photons created by massive stars.
In the northeast of the bubble, brighter PAH emission is seen extending beyond the IF. This can be
attributed to far UV photons leaking from the H\,{\sc ii} region due to small-scale inhomogeneities in the IF and in the
surrounding medium \citep{zav07}. Likewise, the southwestern filament with bright PAH emission can also be attributed
to leaking radiation. All of the above observed facts demonstrate  the strong influence of the expanding H\,{\sc ii} region on
the surroundings. Therefore, the YSO candidates associated with them  must suffer
the physical action of the H\,{\sc ii} region. Moreover, 8~\um emission shows concave-convex characteristics, indicating
that the IF is locally distorted. The outline of the cold dust emission, and likewise that of the IF, shows that this type
of distortion occurs in the adjacent clouds. In this case such distortion could be attributed
to compressions of the expanding IF.
Such compression can be characterized by the probability density function (PDF) of the column density.

In turbulent simulations \citep{tre12a,tre12b}, it is argued that when the ionized gas pressure overweighs the ram pressure
of the turbulence, the ionization compression could produce PDF forms with two lognormal distributions:
    \begin{equation}\label{eq-pdf}
    p(\eta)=\frac{p_{0}}{\sqrt{2\pi\sigma_0^2}}\exp\left(\frac{-(\eta-\mu_0)^2}{2\sigma_0^2}\right)+
    \frac{p_{1}}{\sqrt{2\pi\sigma_1^2}}\exp\left(\frac{-(\eta-\mu_1)^2}{2\sigma_1^2}\right),
    \end{equation}
    where $\eta = \mathrm{ln}(N/\bar{N})$, $p_{i}$, $\mu_i$ and $\sigma_i$ are the integral, mean, and
    dispersion of each component, respectively. $\bar{N}$ represents the mean column density over a large region.
    The first lognormal form at low densities is generally
    thought to be a result of isothermal supersonic turbulence \citep{vaz94,vaz08,fed08,fed10}. The second
    lognormal distribution in the PDF at high densities is believed to be caused by the compression from the
    ionized gas pressure \citep{tre12a,tre12b,tre14}.
    Currently, this sort of PDF form has been observed in several
    massive star-forming regions, and interpreted as ionization compressions \citep[e.g.,][]{sch12,tre14}.

\begin{figure}
\begin{center}
    \includegraphics[width=0.45\textwidth]{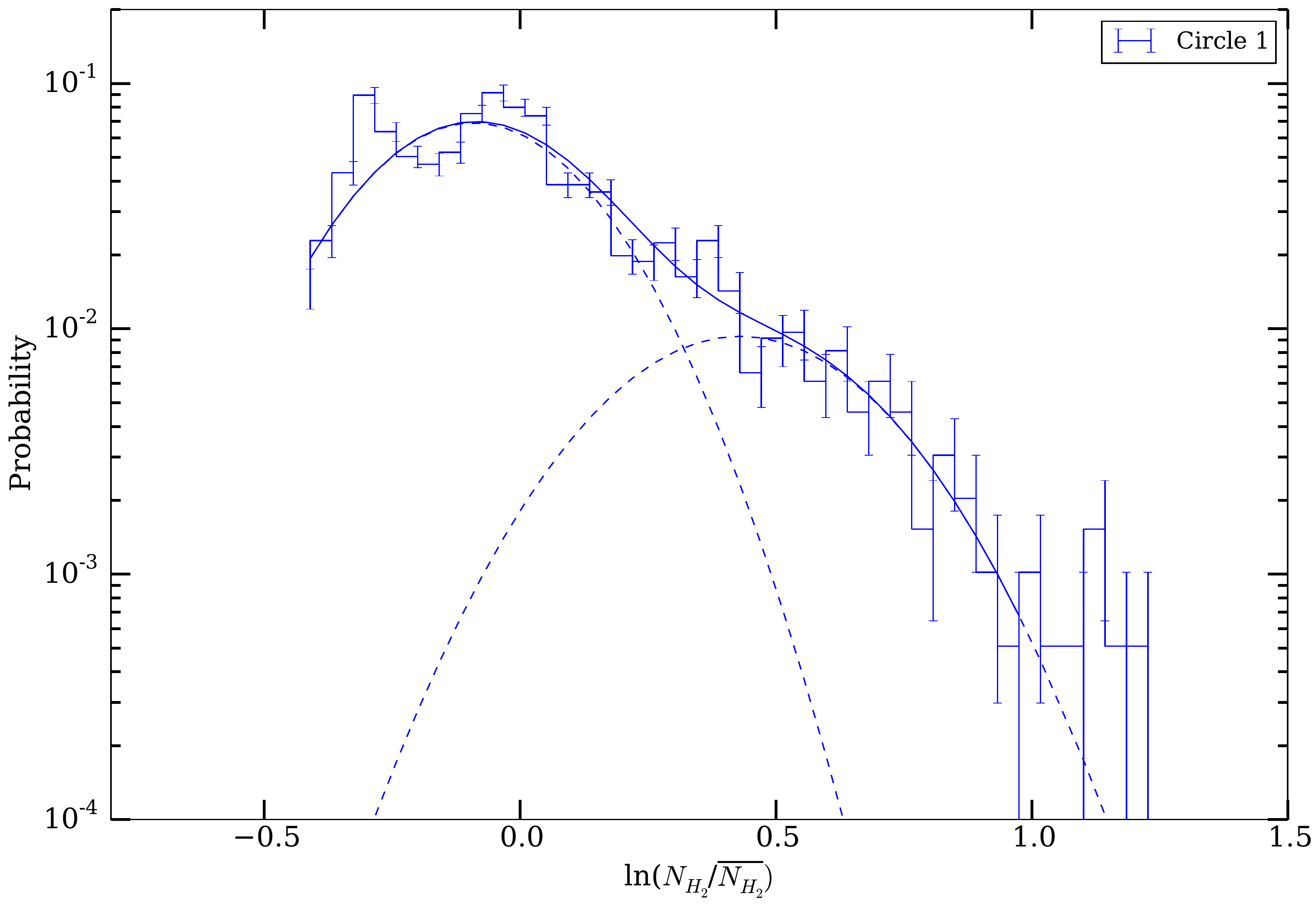}
    \includegraphics[width=0.45\textwidth]{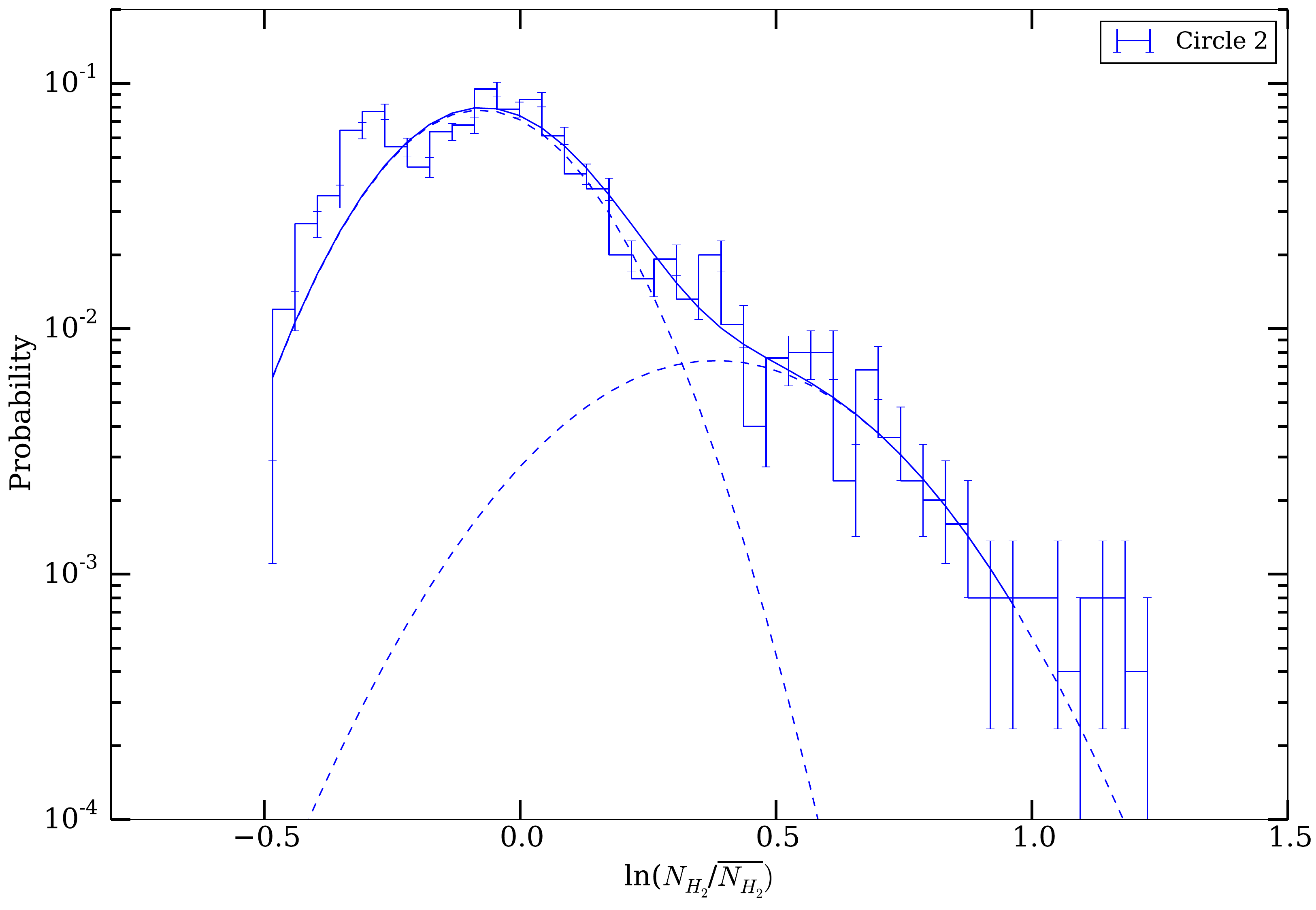}
    \includegraphics[width=0.45\textwidth]{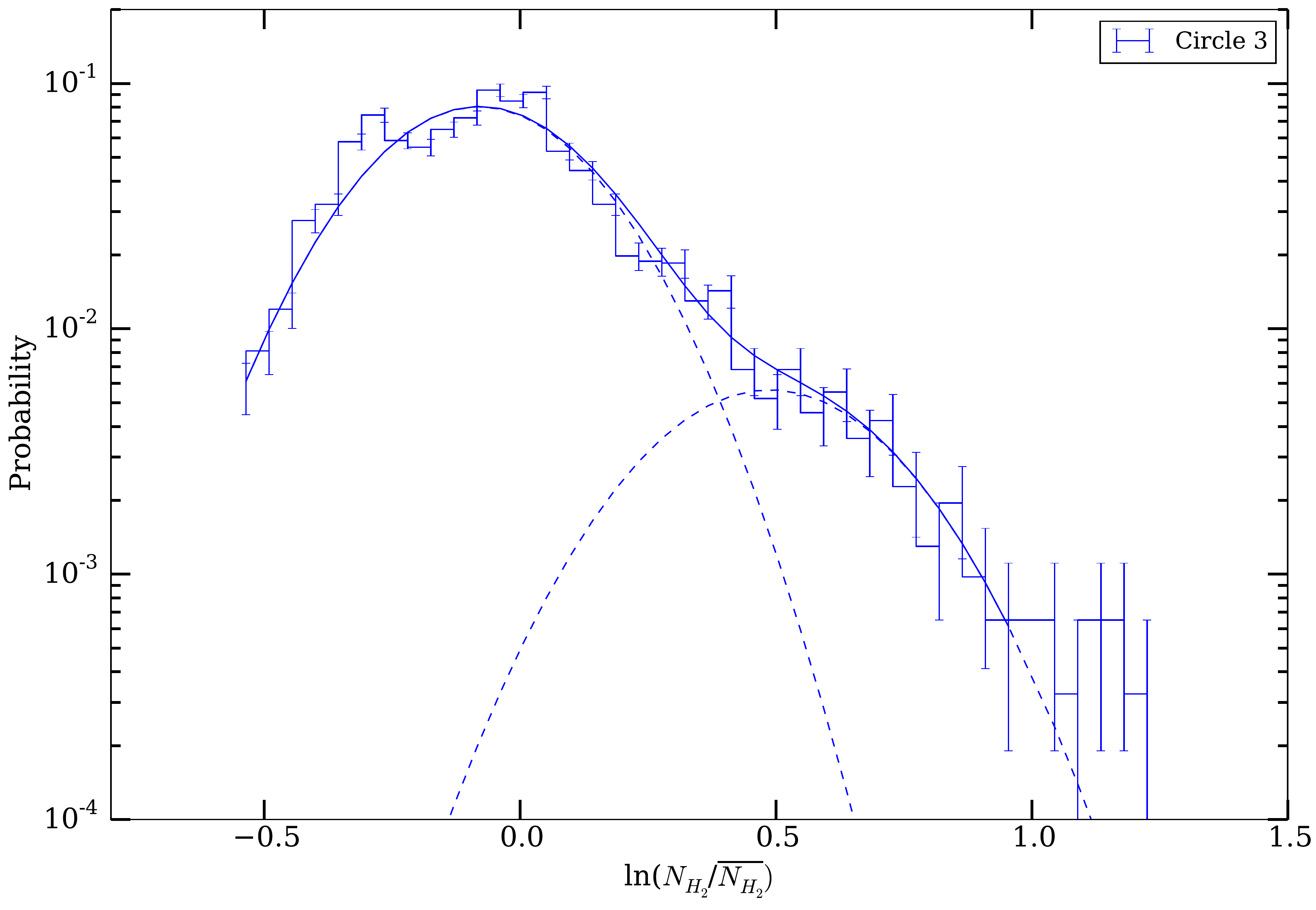}
    \caption{Column density PDFs over the target regions (see Figure~\ref{fig-NT}).
     $\overline{N_{\mathrm{H_{2}}}}$ represents mean the density.
     The best lognormal PDF fit (solid curve) is done by the sum of two
     lognormal distributions (dashed curves, see Eq. \ref{eq-pdf}). Error bars
     are determined via Poisson statistics.}
    \label{fig-pdf}
\end{center}
\end{figure}

Therefore, the exploration of the PDF form in the bubble N4 can help to improve our understanding of the
influences of the H\,{\sc ii}~region
on the surroundings. The column density PDF is investigated over three regions (i.e., Circles 1, 2 and 3,
see Figure~\ref{fig-NT}) covering the ionized gas and the major surroundings. The three regions are concentric with
an equal separation of 0.01 degree. Figure~\ref{fig-pdf} displays
the column density PDFs toward the three regions well fitted by the sum of two lognormal distributions.
The mean column density $\bar{N}=8.0\times10^{21}$ \cm2~is averaged over the region of Circle 3.
The fitting parameters
can be found in Table \ref{tbl-6}. As shown in Figure~\ref{fig-pdf}, all the column density PDFs of the three regions
 show the second lognormal forms. Given the association of the H\,{\sc ii}~region with the bubble N4,
these forms  might be caused by compression of ionized gas.
We note that the regions smaller than the Circle 1 do not fit two lognormal distributions well. This
may be due to the fact that in this region there are not enough sample points.
Furthermore, we observe that the integral of the second lognormal component ($p_{1}$) decreases as the radius of the
target region increases. As suggested by \citet{tre14}, this trend could be due to the fact that
the larger the region is, the less important the amplitude of the compressed lognormal becomes since
more unperturbed gas is added to the distribution, indicative of less significance of compression of the
ionized gas in larger regions.

\begin{deluxetable}{ccccccc}[tbh!]
\tabletypesize{\scriptsize} \tablecolumns{9} \tablewidth{0pt}
\tablecaption{Parameters of the PDF fitting \label{tbl-6}}
\tablehead{
\colhead{Region}&\colhead{$p_{0}$}&\colhead{$\mu_0$}& \colhead{$\sigma_0$}& \colhead{$p_{1}$}&\colhead{$\mu_1$}& \colhead{$\sigma_1$}}
\startdata

1& 0.035 & -0.091 & 0.200 & 0.006 & 0.430 & 0.238 \\
2& 0.035 & -0.078 & 0.181 & 0.005 & 0.383 & 0.271 \\
3& 0.040 & -0.081 & 0.200 & 0.003 & 0.488 & 0.221
\enddata
\end{deluxetable}

\subsection{Collect and Collapse Process}\label{s-cc}

    The association of collected molecular gas with PDR regions and the compression from ionized gas confirm the
    strong influence of the H\,{\sc ii}~region on the adjacent medium. In the context of the C \& C process,
    such an impact may stimulate the formation of a new generation of stars. To test the occurrence of this process,
    the dynamical time of the H\,{\sc ii}~region can be compared  with the fragmentation
    time of the surrounding molecular clouds \citep{deh03,deh05,zav06,zav07,zav10}.

    According to the well-known expansion law \citep{spi78,dys80}, if an H\,{\sc ii}~region evolves in an homogeneous molecular cloud,
    the dynamical age ($t_\mathrm{dyn}$) of the
    H\,{\sc ii}~region can be expressed as follows:
    \begin{eqnarray}\label{eq-9}
    \nonumber
    R_{\mathrm{St}} &=& (3N_{\mathrm{Ly}}/4\pi n_{\mathrm{H},0}^{2}\alpha_{\mathrm{B}})^{1/3}, \\
     t_{\mathrm{dyn}}&=&\frac{4}{7}\ \frac{R_{\mathrm{St}}}{C_{\mathrm{H{\,\sc{\mathrm{ii}}}}}}\ [(\frac{R_{\mathrm{IF}}}{R_{\mathrm{St}}})^{7/4}-1],
    \end{eqnarray}
    where $R_{\mathrm{St}}$ is the radius of the Str$\mathrm{\ddot{o}}$mgren sphere,
    $n_{0}$ is the initial particle density of ambient neutral gas,
    $\alpha_{B}$ = $2.6\times10^{-13}(10^{4} \mathrm{K}/T)^{0.7}$ cm$^{3}$ s$^{-1}$ \citep{kwa97}
    is the the coefficient of the radiative recombination, $C_{\mathrm{H{\,\sc{\mathrm{ii}}}}}$ is the isothermal
    sound speed of ionized gas assumed to be 10~km~s$^{-1}$, and $R_{\mathrm{IF}}$ is the radius of
    the IF. If we assume that $R_{\mathrm{IF}}$ is approximately equal to the radius of N4, the total mass of $\sim555$ \msun~of the neutral
    and ionized gas can yield an initial number density of $n_{\mathrm{H},0}= \sim1.2\times10^{4}$ \cmt.
    Given the estimated $N_{\mathrm{Ly}}$ (see Section \ref{s-HII}), $t_\mathrm{dyn}$ of N4 is therefore
    $\sim1.0$ Myr. This timescale, however, is uncertain since the actual evolution of the H\,{\sc ii}~regions is
    not in a strictly uniform medium. Bearing this in mind, the estimated dynamical age should be
    considered with this caveat.

    On the other hand, following \citet{whi94}, gravitational fragmentation of the shell of collected material can be expected
    when
    \begin{equation}\label{eqa11}
    t_{\mathrm{frag}}=1.56 \ a_{.2}{}^{7/11} N _{49}{}^{-1/11} n_3{}^{-5/11},
    \end{equation}
    where $a_{.2}$ is the sound speed ($a_{s}$) inside the shell in units of 0.2~\vel,
    $N_{49}$ is the ionizing photon flux ($N_{Ly}$) in units
    of $10^{49}$ ph s$^{-1}$, and $n_{3}$ is the initial gas atomic number density ($n_{\mathrm{H},0}$)
    in units of $10^{3}$ cm$^{-3}$. An estimate of $a_{s}=0.3$~\vel at a dust temperature of 22 K gives
    $t_{\mathrm{frag}}=\sim0.3$ Myr.

    In comparison, the fragmentation time is much shorter than the dynamical age.
    This indicates that the shell of collected gas has had enough time to fragment during the lifetime of N4, which is
    consistent with the signature of six dust fragments condensed out of the shell.
    Therefore, the C \& C process is presumably at work in the bubble N4.
    In addition, the spatial distribution of some YSO candidates
    shows an overdensity of the number of YSOs at the edge of the bubble (see Figure~\ref{fig-N4rgbIII} and \ref{fig-ysoStat}). In combination with such  an overdensity
    and the existence of ionization compression, the aforementioned timescales  show that
    a scenario of triggered star formation in the bubble N4 through the C \& C
    process is possible.

\section{Summary}\label{s5}

Taking advantage of observations of \emph{Herschel} and of four molecular lines (i.e., H$_{2}$O $6_{1,6}-5_{2,3}$,
CH$_{3}$OH $7_{0,7}-6_{1,6}$, HCO$^{+}$ (1-0), and o-H$_{2}$CO $2_{1,2}-1_{1,1}$), together with auxiliary archival data
involving  four public surveys (i.e., GLIMPSE, MIPSGAL, \emph{WISE}, 2MASS, and MAGPIS), we have investigated the interactions of
the bubble N4 with the adjacent medium  and explored the possibility of triggered
star formation. The main results are summarized below.
\begin{enumerate}
  \item The distributions of the dust temperature and column density toward N4 show an anti-correlation:
  the higher the column density, the colder the dust temperature. This is attributed to the penetration
  degree of the external heating from the associated H\,{\sc ii} region.
  \item A shell structure standing out of the column density map harbors six dense dust clumps.
     These clumps have a mean size of $\sim0.5$ pc, temperature of $\sim22$ K,
    column density of  $\sim1.7\times10^{22}$ \cm2, volume density of  $\sim4.4\times10^{3}$ cm$^{-3}$, and mass of
    $\sim3.2\times10^{2}$ \msun. Two out of the six may be massive enough to form high-mass stars, while the remaining
    could form low-mass stars.
     At the present sensitivity and angular resolution, the observations of the four molecular lines toward the six clumps
     did not reveal a clear evidence of
    ongoing star formation.
  \item The H\,{\sc ii} region associated with the bubble is likely to be excited by an O8.5V-O9V type star with a dynamical
  age of $\sim 1.0$ Myr. The velocity difference
  between the southeastern clumps and the northwestern ones as shown in the spectra of HCO$^{+}$ and o-H$_{2}$CO, suggests
  that the bubble is expanding.
  \item The shell of cold dust emission associated with PDRs, and the compressions of ionized gas characterized by the
  column density PDF of the target regions demonstrate that  the expanding  bubble has a strong influence  on the its surroundings.
  \item In the context of the C \& C mechanism, we find that the shell of collected matter has enough time to
  fragment during the lifetime of N4. From compression of ionized gas, the overdensity of the number of YSO candidates
  at the edge of the bubble, and the timescales involved,  we suggest
  that triggered star formation might have taken place in the bubble N4 but its definitive demonstration
  requires more detailed molecular lines observations.
\end{enumerate}

\begin{acknowledgements}
We thank the anonymous referee for the comments that much improved the quality of this paper.
This work is supported by the National Natural Science Foundation of China through grants of 11503035, 11573036, 11373009, 11433008, and 11403040; the International S\&T Cooperation Program of China through grand of  2010DFA02710; and the Beijing Natural Science Foundation through the grant of 1144015; and the
Chinese Academy of Sciences. G.D., S.P., and M.O. acknowledge support
from ANPCyT and CONICET (Argentina) grants.
   SPIRE has been developed by a consortium of institutes
led by Cardiff Univ. (UK) with Univ. Lethbridge (Canada); NAOC (China);
CEA, LAM (France); IFSI, Univ. Padua (Italy); IAC (Spain); Stockholm
Observatory (Sweden); Imperial College London, RAL, UCL-MSSL, UKATC,
Univ. Sussex (UK); Caltech, JPL, NHSC, Univ. Colorado (USA). This development
has been supported by national funding agencies: CSA (Canada);
NAOC (China); CEA, CNES, CNRS (France); ASI (Italy); MCINN (Spain);
SNSB (Sweden); STFC (UK); and NASA (USA). PACS has been developed
by a consortium of institutes led by MPE (Germany) with UVIE (Austria); KU
Leuven, CSL, IMEC (Belgium); CEA, LAM(France); MPIA (Germany); INAFIFSI/
OAA/OAP/OAT, LENS, SISSA (Italy); IAC (Spain). This development
has been supported by the funding agencies BMVIT (Austria), ESA-PRODEX
(Belgium), CEA/CNES (France), DLR (Germany), ASI/INAF (Italy), and
CICYT/MCYT (Spain). We are grateful to the KVN staff. The KVN is a facility operated by the Korea Astronomy and Space Science Institute. We have used the NASA/IPAC Infrared Science Archive
to obtain data products from the \emph{Spitzer}-GLIMPSE, \emph{Spitzer}-MIPSGAL,
\emph{WISE}, and 2MASS surveys.
\end{acknowledgements}

\appendix
\section{Identification of YSO candidates}\label{s-app}
\begin{deluxetable}{ll}[tbh!]
\tabletypesize{\scriptsize} \tablecolumns{2} \tablewidth{0pt}
\tablecaption{Point sources within 5$\arcmin$ from the bubble center \label{tbl-9}}
\tablehead{
\colhead{Source}   & \colhead{Number}
}
\startdata

Selected point sources (ALLWISE and MIPSGAL catalogs)                        &  191  \\
Good fittings of YSO SEDs ($\chi^{2}_{best}/N_{data}\leq7$)                     &  77     \\
PAH-feature emission                                                         &  17     \\
Final YSO candidadates                                                       &  60

\enddata
\end{deluxetable}

\begin{figure}
  \includegraphics[width=0.24\textwidth]{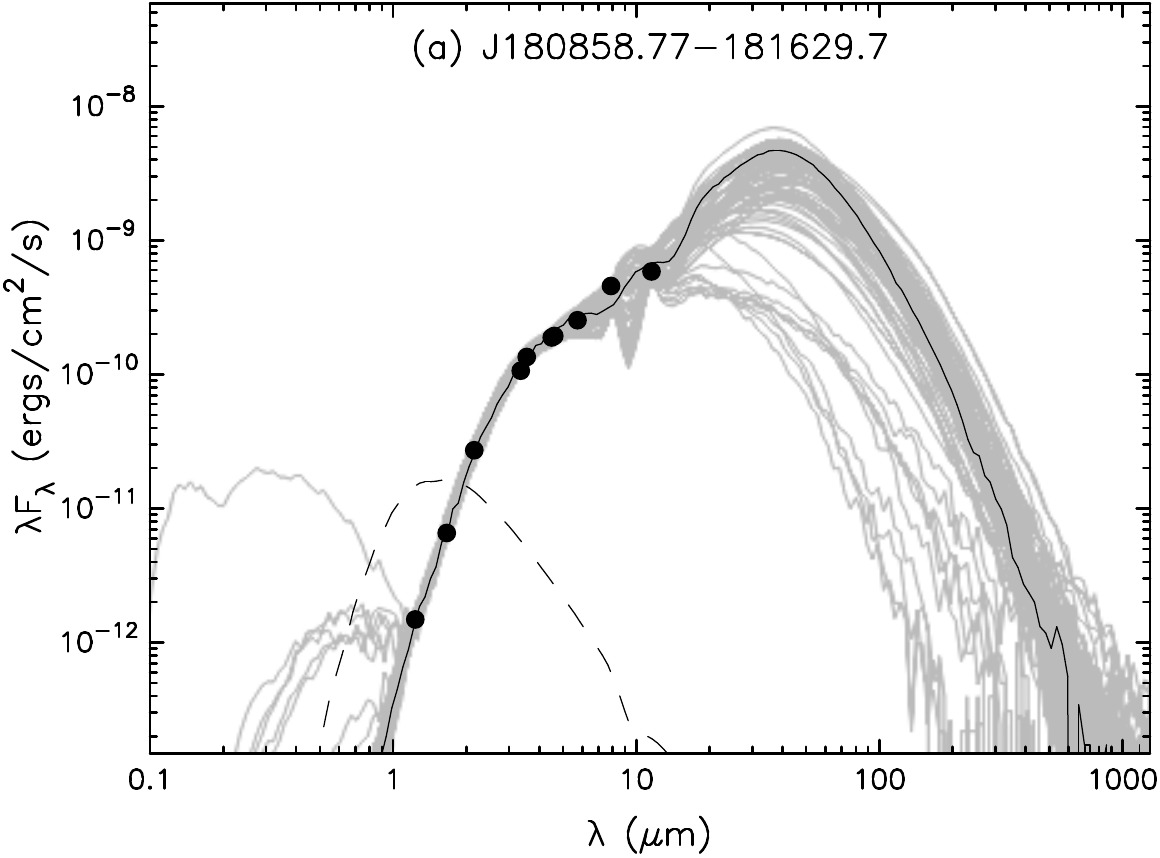}
  \includegraphics[width=0.24\textwidth]{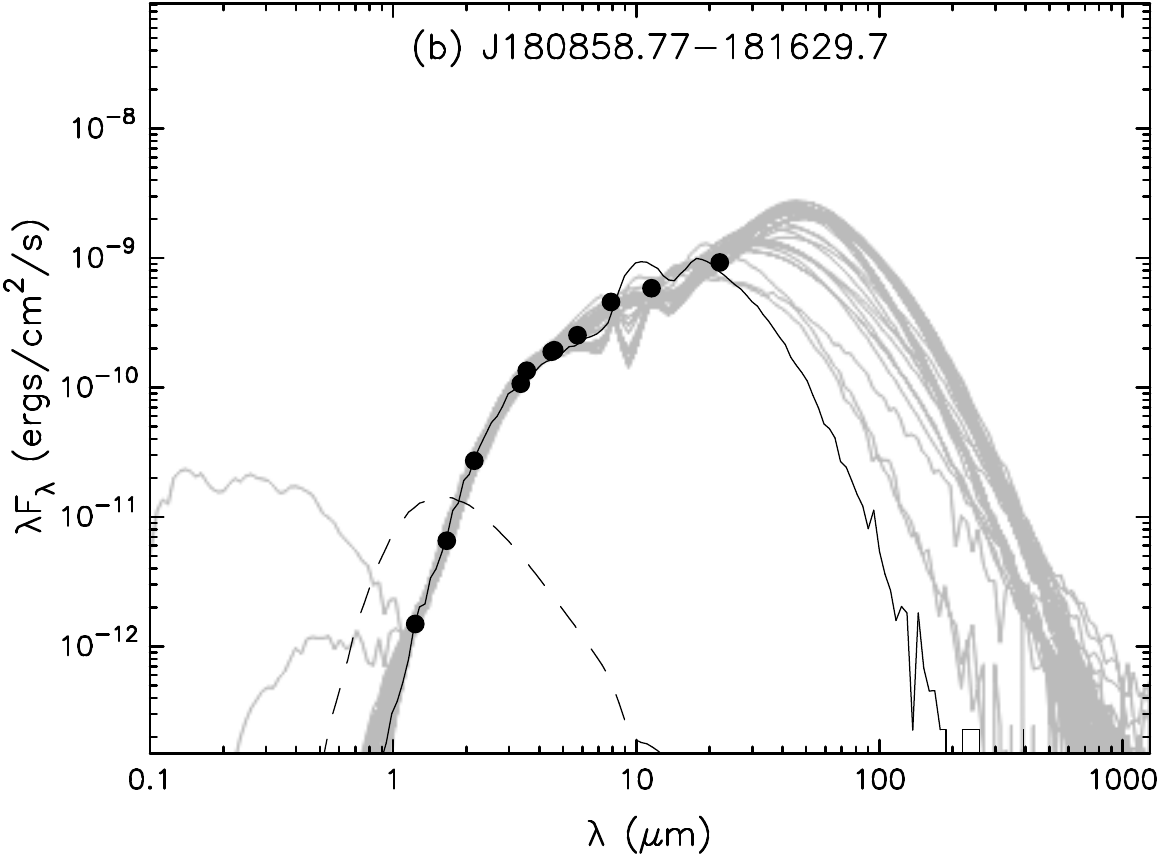}
  \includegraphics[width=0.24\textwidth]{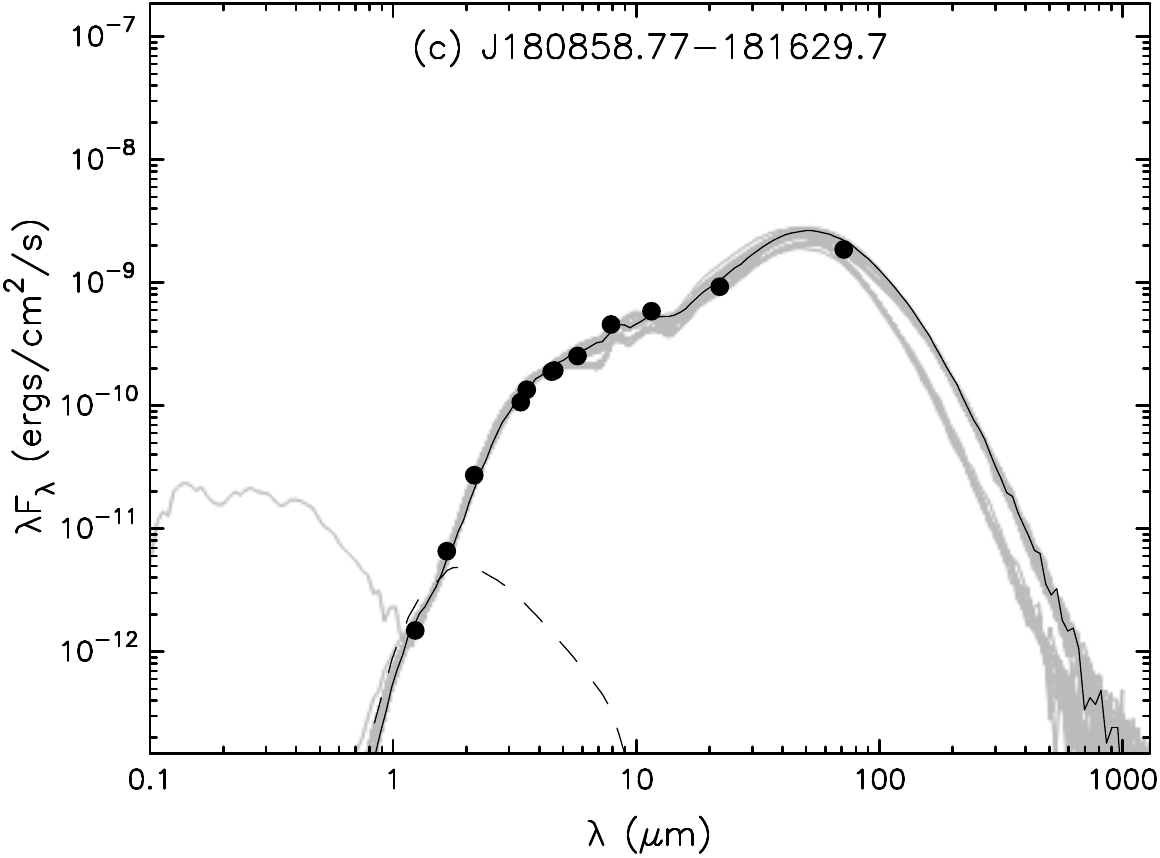}
  \includegraphics[width=0.24\textwidth]{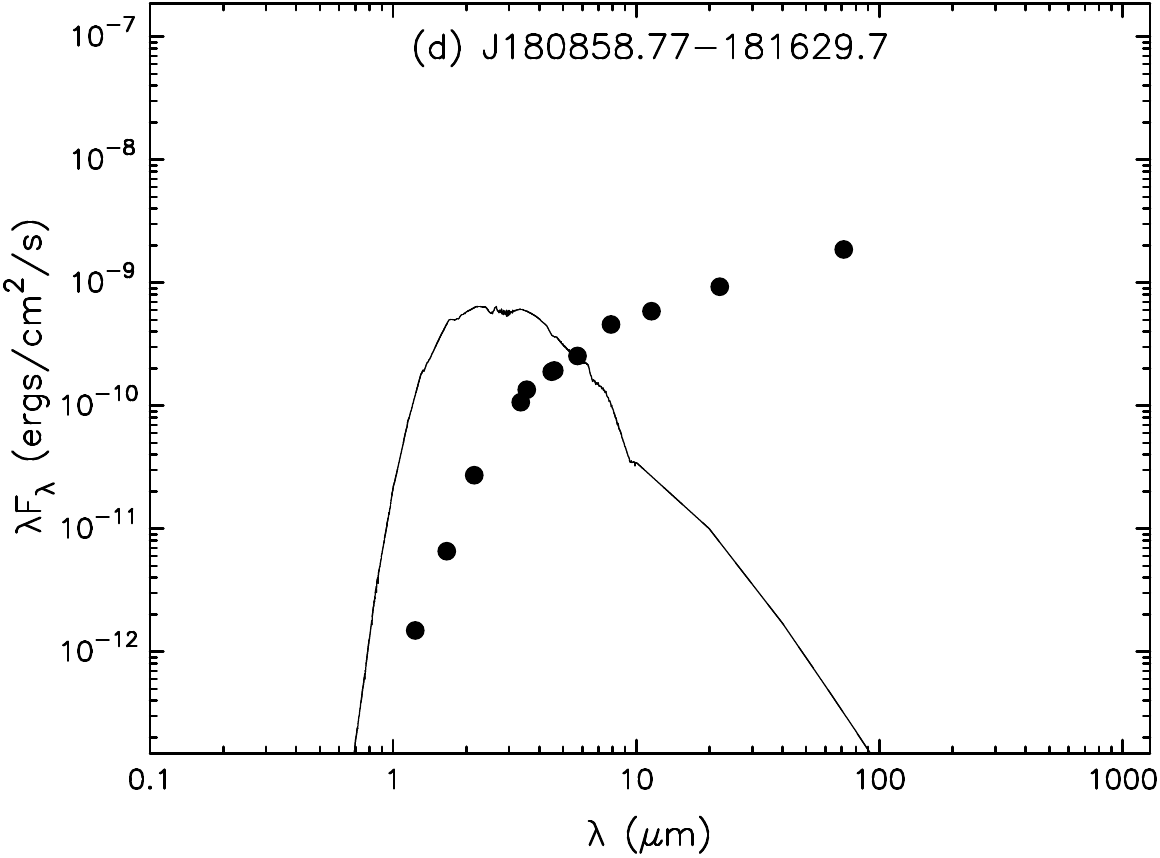}
\caption{ Example of the fitting of models to an object.
The filled circles and triangle symbolize the input fluxes and the upper limit, respectively.
The black line shows the best fit. The gray lines show a set of fits that satisfy
$\chi^{2}-\chi_{best}^{2}< 2 \times N_{data}$. The dashed
line means the stellar photosphere corresponding to the central source of the best
fitting model as it would look in the absence of circumstellar dust but with
interstellar extinction. The panels (a)-(c) show the resulting SED fittings without
the input fluxes at 24 and 70~\um, with the input flux at 24~\um, and with the input fluxes
at 24 and 70~\um, respectively. The panel (d) is to check whether the source cannot be fit
simply by a star with interstellar extinction.}
\label{fig-preyso}
\end{figure}

To our knowledge, if more data at longer wavelengths are
taken into account, the SED fitting will be better constrained. For example,
Figure~\ref{fig-preyso} shows the resulting SED fits with the same parameter inputs
except for different data inputs. Obviously, the SED fitting results with the 70~\um\ data
inputs (Figure~\ref{fig-preyso}~(c)) are best constrained, followed by the cases with
the 22~\um\ but without 70~\um\ data (Figure~\ref{fig-preyso}~(b)) inputs, whereas the cases
without the data inputs at both wavelengths (Figure~\ref{fig-preyso}~(a)) are most poorly constrained.
Therefore,  the \emph{ALLWISE} and MIPSGAL catalogs were taken into account. The former catalog
includes photometries at 22~\um and the latter one includes photometries at 24~\um. In addition,
these two catalogs have been cross matched with the 2MASS and/or GLIMPSE surveys, ensuring more
photometries to be used in the SED fitting.

From the two catalogs, 191 point sources were picked out.
First, we retrieved a sample of 164 sources detected with a ratio of signal-to-noise (S/N) of $>10$ in the
four \emph{WISE} bands and with an extension flag of $<2$ from the \emph{ALLWISE} Source catalog.
The former simply ensures sources with detections at more wavelengths and with good enough
data qualities and the latter guarantees the sources as a point source. Also, this sample has been
complemented by that retrieved from the MIPSGAL catalog. In this catalog, a sample of 32 point sources
was chosen with a photometry uncertainty of $<0.2$ in the MIPS 24~\um\ and the four IRAC
bands. These restrictions are equivalent to those mentioned previously.  Except for 5 sources overlapping
in these two samples, a total of 191 point sources were obtained.
In addition, we tried searching for photometries
at 70~\um\ for these selected sources. After investigating several published literatures, only one object, J180858.77-181629.7, has been found with 70~\um\ photometries \citep{lum13}. Moreover,
the 56 point sources from the CuTeX catalog (see Section~\ref{s-ob1}) were also inspected
by a cross matching with the chosen 191 sources within
a search radii of $2\arcsec$ in the software Topcat.\footnote{\url{http://www.star.bris.ac.uk/~mbt/topcat/}}
In a combination of visual inspections,
only five photometries at 70~\um\ from the CuTeX catalog were considered in the following
SED fittings. In total, 6 out of 191 point sources have 70~\um\ photometries.

In the fittings to the selected 191 sources, the distance
was constrained in the range of 2.3-4.1 kpc, and the extinction was in the range of 0 to
16 Mag, which was estimated from the column density map by a relation of $A_{\mathrm{v}} = 5.34 \times 10^{-22}
N(\mathrm{H_2})$ \citep{deh09}. After the SED fitting to the sources, the YSO candidates were determined
by the following criteria. That is, the candidates should be those
(1) with $\chi^{2}_{best}/N_{\mathrm{data}}\leq7$
and (2) fitted well only by the YSO models, not by a model of a stellar photosphere with a foreground
extinction. The former criterion has been determined by visually inspecting
dozens of profiles of resulting SED fittings. We note that the value of 7 is a
little greater than those adopted in the literature \citep[e.g.,][]{pov09,pov13,kan09}.
This could be due to the fact that more data points ($N_{\mathrm{data}}$) have been taken into account in
this work.
Figure~\ref{fig-preyso} exemplifies a source satisfied with the above criteria.
The candidate J180858.77-181629.7 has a good fit of the YSO models (Figure~\ref{fig-preyso}~(c))
 but a poor fit of a stellar photosphere (Figure~\ref{fig-preyso}~(d)),
indicating that this object is not a field star. These criteria result in 77 YSO candidates including
74 sources from the \emph{ALLWISE} catalog and the remaining 3 from the MIPSGAL catalog.

\begin{figure}
  \includegraphics[height=4cm,width=8.75cm]{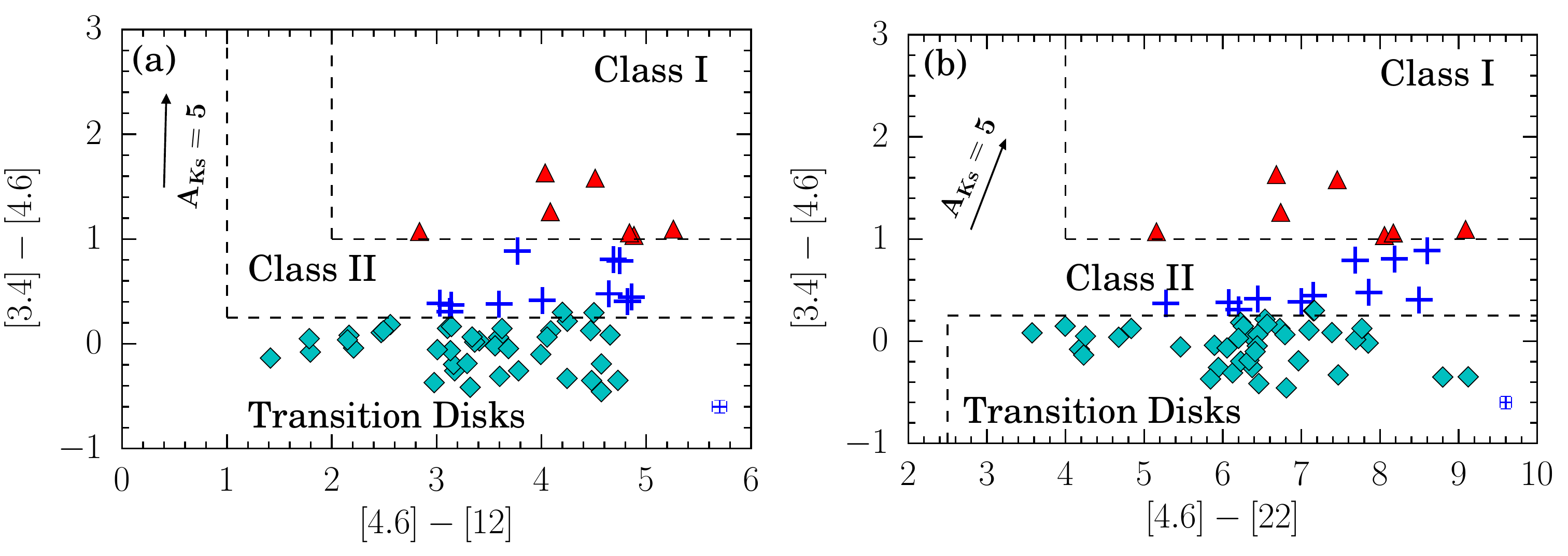}
  \includegraphics[height=3.8cm,width=8.75cm]{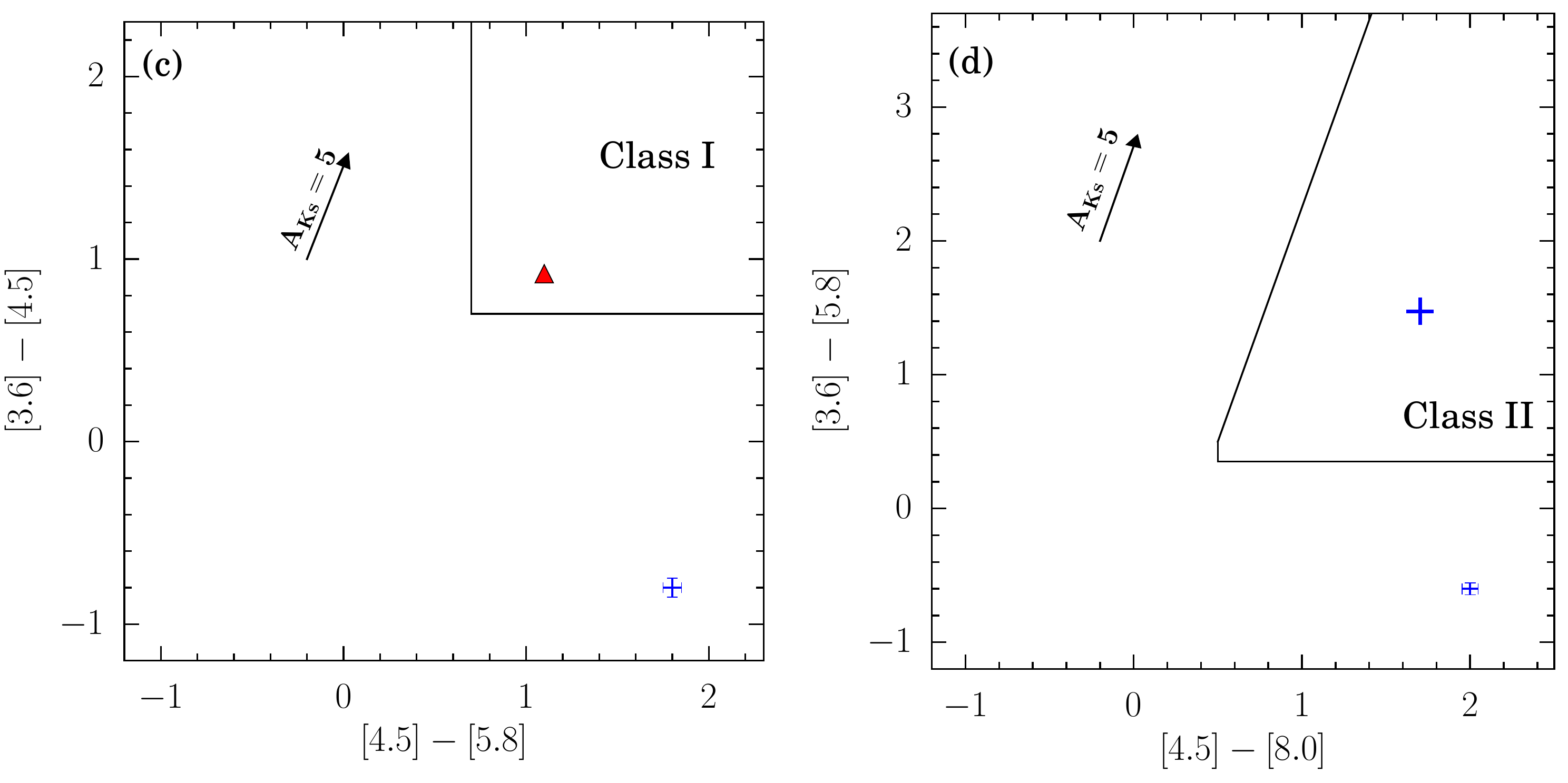}
\caption{Color-color diagrams for (a) \emph{WISE} bands 1, 2, and 3;
(b) WISE bands 1, 2, and 4; (c) IRAC bands 1, 2, and 3; (d) IRAC bands 1, 2, 3, and 4.
Red triangles symbolize Class I YSOs, blue pluses Class II YSOs, and cyan diamond
transition disk YSOs. The arrows show an extinction vector of $A_{\mathrm{K}}$ = 5 mag
following the extinction laws of \citet{koe14,fla07}.
The combined measurement errors are shown on the bottom right. }
\label{fig-cc}
\end{figure}

The color selection schemes were used to categorize the 77 YSO candidates.
To begin,  the classification method of \citet{koe12} was adopted for the 74 candidates.
According to the criteria in various color spaces, which can kick out possible contaminants
such as star-forming galaxies, broad-line AGNs, and knots of shock emission, 16 PAH-contaminated
emission sources were removed. The remaining uncontaminated sources are considered to be Class I
YSOs if their colors follow (1) $[3.4]-[4.6]>1.0$ and (2) $[4.6]-[12]>2.0$;
Class II YSOs if their colors follow (1) $[3.4]-[4.6]-\sigma_{12}>0.25$ and (2)
$[4.6]-[12]-\sigma_{23}>1.0$, where $ \sigma_{12}, \sigma_{23}$ are the combined
 measurement errors of two corresponding  \emph{WISE} bands (i.e., [3.4] and [4.6], [4.6] and [12],
 respectively); and TD YSOs if their colors follow (1) $[4.6]-[22]>2.5$ and (2) $[3.4]<14$.
In the end, 7 Class~I, 11 Class~II, and 40 TD YSOs are obtained  by the above constraints,
which are also shown in Figure~\ref{fig-cc}~(a)-(b). For the three candidates from the MIPSGAL catalog,
 the color selection method of \citet{gut09} was considered. Similarly, one PAH-feature emission
 was ruled out. The remaining 2 sources are classified into one Class I YSO since its colors follow
 (1) $[4.5]-[5.8]>0.7$, and (2) $[3.6]-[4.5]>0.7$; and one Class II YSO since its colors follow
 (1) $[4.5]-[8.0]-\sigma_{24}>0.5$, (2)  $[3.6]-[5.8]-\sigma_{13}>0.35$,
 (3) $[3.6]-[5.8]+\sigma_{13}\leq \frac{0.14}{0.04} \times (([4.5]-[8.0]-\sigma_{24})-0.5)+0.5$,
 and (4) $[3.6]-[4.5]-\sigma_{12}>0.35$, where $\sigma_{24}, \sigma_{13}, \sigma_{12}$ are the combined
 measurement errors of two corresponding  IRAC bands (i.e., [4.5] and [8.0], [3.6] and [5.8], [3.6] and [4.5],
 respectively). The above constraints are displayed in Figure~\ref{fig-cc}(c)-(d). In total, built on
 the 77 SED-identified YSO candidates, the color schemes confirm 60 out of them as potential YSOs:
 8 Class I, 12 Class II and 40 TD objects. The results of the classification can be found in
 Table~\ref{tbl-5}. Table \ref{tbl-9} summarizes the number of sources in each step for identifying
 YSO candidates. Their photometries are tabulated in Table~\ref{tbl-10} and Table~\ref{tbl-11}.
 The plots of the SED fitting to the 60 YSO candidates are available in Figures~10 and 11.

The resulting 60 potential YSOs are incomplete toward N4 but robust. The strict criteria on
 the aforementioned source selections  result in the incompleteness. For instance, if the
 S/Ns of photometries at the selected wavelengths were set to 5, there would be more sources to
 be in question, resulting in more YSO candidates. However, the lower S/Ns are directly related
 to the photometry qualities, reducing the robustness of the resulting YSO candidates.
 For the sake of the robustness, the strict criteria were be carried out. Additionally,
 the combination of the SED fitting method with the color selection schemes makes the
 60 YSO candidates reliable enough.

\begin{deluxetable*}{llllllll}[tbh!]
\tabletypesize{\scriptsize} \tablecolumns{8} \tablewidth{0pt}
\tablecaption{Photometries of YSO candidates from the archived ALLWISE Source catalog\label{tbl-10}}
\tablehead{\colhead{Name} & \colhead{$W1$}   &  \colhead{$W2$}  & \colhead{$W3$}   & \colhead{$W4$}
           &\colhead{$J$}        & \colhead{$H$}    &\colhead{$K$}     \\
           \colhead{ }           & \colhead{(mag)}  &  \colhead{(mag)} & \colhead{(mag)}  & \colhead{(mag)}
           & \colhead{(mag)}     & \colhead{(mag)}  &  \colhead{(mag)}  }

\startdata

J180831.89-181639.1 & 10.697 $\pm$     0.042 & 10.311 $\pm$     0.031 &  7.278 $\pm$     0.083 &   3.314 $\pm$     0.055 &    16.589 $\pm$      -- &    13.393  $\pm$    0.022 &    11.821 $\pm$  0.027 \\
J180832.49-181616.6 & 10.541 $\pm$      0.03 & 10.434 $\pm$     0.032 &  7.963 $\pm$     0.045 &   3.337 $\pm$     0.028 &      11.7 $\pm$   0.026 &    11.194  $\pm$    0.024 &    10.823 $\pm$  0.024 \\
J180834.88-181751.0 & 10.074 $\pm$     0.031 &  9.927 $\pm$     0.031 &  6.823 $\pm$     0.048 &   5.931 $\pm$     0.067 &    13.385 $\pm$   0.029 &     11.75  $\pm$    0.037 &    11.113 $\pm$  0.034 \\
J180835.68-181814.8 &  9.551 $\pm$     0.025 &   9.47 $\pm$     0.025 &  7.305 $\pm$     0.049 &   5.895 $\pm$     0.108 &    13.831 $\pm$   0.049 &    12.006  $\pm$       -- &    10.925 $\pm$     -- \\
J180839.36-181813.8 & 11.207 $\pm$     0.056 & 11.146 $\pm$      0.07 &  7.554 $\pm$     0.101 &   4.809 $\pm$     0.087 &    15.853 $\pm$   0.065 &    13.108  $\pm$     0.03 &    12.014 $\pm$  0.021 \\
J180839.71-181841.5 & 11.005 $\pm$      0.05 & 10.881 $\pm$     0.061 &  6.792 $\pm$     0.046 &   4.154 $\pm$     0.036 &    16.762 $\pm$      -- &     13.47  $\pm$    0.037 &    11.907 $\pm$  0.026 \\
J180840.11-181902.3 & 11.032 $\pm$     0.042 & 10.849 $\pm$     0.057 &   8.29 $\pm$      0.09 &   4.622 $\pm$     0.043 &     15.16 $\pm$      -- &    13.997  $\pm$    0.091 &    12.005 $\pm$     -- \\
J180841.62-181600.0 & 10.152 $\pm$     0.031 &  10.41 $\pm$     0.029 &  7.238 $\pm$     0.042 &   4.466 $\pm$     0.097 &    11.908 $\pm$   0.023 &    10.873  $\pm$    0.021 &    10.554 $\pm$  0.019 \\
J180842.08-181457.5 & 10.534 $\pm$     0.037 & 10.949 $\pm$     0.043 &  7.628 $\pm$     0.049 &   4.491 $\pm$      0.04 &    12.144 $\pm$   0.023 &    11.197  $\pm$    0.022 &     10.89 $\pm$  0.026 \\
J180842.19-181329.1 & 10.587 $\pm$     0.034 & 10.666 $\pm$     0.036 &   8.87 $\pm$     0.062 &   6.489 $\pm$     0.102 &    14.399 $\pm$   0.043 &    12.301  $\pm$    0.042 &    11.352 $\pm$  0.031 \\
J180842.69-181930.1 & 10.072 $\pm$     0.026 & 10.023 $\pm$      0.03 &  8.238 $\pm$     0.101 &   5.769 $\pm$     0.076 &    15.183 $\pm$   0.058 &    12.526  $\pm$     0.07 &    11.322 $\pm$  0.045 \\
J180843.58-181441.8 &  9.652 $\pm$     0.035 &  9.506 $\pm$      0.03 &  5.882 $\pm$     0.021 &   3.244 $\pm$      0.03 &    15.211 $\pm$   0.053 &    12.255  $\pm$    0.037 &    10.903 $\pm$  0.024 \\
J180844.11-181456.4 & 11.345 $\pm$     0.045 & 11.696 $\pm$     0.051 &  7.217 $\pm$     0.028 &   2.897 $\pm$     0.024 &    16.351 $\pm$      -- &    14.165  $\pm$    0.055 &    12.771 $\pm$  0.055 \\
J180844.15-181401.9 & 10.359 $\pm$      0.03 & 10.617 $\pm$     0.032 &  6.835 $\pm$     0.042 &   4.243 $\pm$     0.041 &     16.05 $\pm$   0.094 &    13.476  $\pm$    0.035 &    12.264 $\pm$  0.031 \\
J180844.97-181457.3 & 11.433 $\pm$     0.053 & 11.782 $\pm$     0.048 &  7.052 $\pm$     0.036 &    2.66 $\pm$      0.03 &    16.942 $\pm$      -- &    14.029  $\pm$    0.035 &    12.598 $\pm$  0.033 \\
J180845.97-181612.4$^{(a)}$ & 10.931 $\pm$     0.083 & 10.484 $\pm$     0.055 &  5.624 $\pm$     0.053 &    3.33 $\pm$     0.047 &    16.876 $\pm$      -- &    13.625  $\pm$    0.035 &    12.049 $\pm$  0.024 \\
J180846.14-181844.0 & 10.287 $\pm$     0.055 & 10.479 $\pm$      0.05 &  5.907 $\pm$     0.018 &   3.516 $\pm$      0.03 &    15.098 $\pm$   0.035 &    12.466  $\pm$    0.021 &    11.317 $\pm$  0.023 \\
J180847.47-181416.2 & 10.423 $\pm$     0.036 & 10.442 $\pm$     0.031 &  6.884 $\pm$     0.031 &   2.591 $\pm$     0.033 &     16.78 $\pm$      -- &    13.575  $\pm$    0.055 &     12.05 $\pm$  0.045 \\
J180847.65-181450.0 & 10.064 $\pm$     0.053 &  9.849 $\pm$      0.06 &  5.602 $\pm$     0.054 &   3.312 $\pm$     0.084 &    15.599 $\pm$   0.061 &    12.603  $\pm$    0.046 &    11.213 $\pm$  0.039 \\
J180847.67-181441.7 & 10.371 $\pm$     0.067 & 10.074 $\pm$      0.07 &  5.573 $\pm$     0.063 &    2.92 $\pm$      0.06 &    16.419 $\pm$      -- &    13.655  $\pm$    0.034 &    11.849 $\pm$  0.026 \\
J180848.33-181357.2 &    9.8 $\pm$     0.042 &  9.635 $\pm$     0.045 &  6.494 $\pm$     0.047 &   3.071 $\pm$     0.042 &    16.221 $\pm$   0.085 &    12.853  $\pm$    0.079 &    11.306 $\pm$  0.058 \\
J180849.08-181441.8 & 11.845 $\pm$      0.09 & 10.265 $\pm$     0.056 &  5.752 $\pm$     0.036 &   2.808 $\pm$      0.05 &        --  &        --   &             -- \\
J180849.78-181340.5 &  8.674 $\pm$     0.027 &  8.714 $\pm$     0.028 &  6.506 $\pm$     0.024 &    2.82 $\pm$     0.034 &    10.041 $\pm$   0.026 &     9.383  $\pm$    0.032 &     9.162 $\pm$   0.03 \\
J180850.19-181816.2 &    9.1 $\pm$     0.028 &   8.73 $\pm$     0.026 &  5.591 $\pm$     0.061 &   3.453 $\pm$     0.071 &    14.469 $\pm$   0.037 &    11.537  $\pm$    0.024 &    10.137 $\pm$  0.024 \\
J180850.83-181836.6 &  9.318 $\pm$      0.03 &   9.01 $\pm$      0.03 &  5.884 $\pm$     0.036 &   2.808 $\pm$     0.034 &    15.385 $\pm$      -- &    12.583  $\pm$    0.022 &    10.647 $\pm$  0.027 \\
J180851.55-181831.9 &  8.991 $\pm$     0.026 &  8.963 $\pm$     0.026 &  5.555 $\pm$     0.024 &   2.763 $\pm$     0.034 &     9.631 $\pm$   0.024 &      9.25  $\pm$    0.021 &     9.161 $\pm$  0.021 \\
J180852.42-181744.2 &  9.108 $\pm$      0.05 &  8.729 $\pm$     0.028 &  5.133 $\pm$     0.024 &    2.65 $\pm$     0.079 &    15.479 $\pm$      -- &    12.151  $\pm$    0.032 &    10.418 $\pm$  0.027 \\
J180852.59-181356.6 & 10.289 $\pm$     0.049 &  9.885 $\pm$      0.04 &  5.064 $\pm$     0.054 &   1.387 $\pm$     0.019 &    15.335 $\pm$    0.06 &    12.364  $\pm$       -- &    11.101 $\pm$     -- \\
J180852.97-181336.2 &  9.418 $\pm$     0.027 &  9.402 $\pm$      0.03 &  6.039 $\pm$     0.042 &   1.711 $\pm$     0.016 &     14.05 $\pm$   0.033 &    11.437  $\pm$    0.037 &    10.218 $\pm$  0.029 \\
J180853.04-181742.2$^{(b)}$ & 10.116 $\pm$     0.054 &  8.856 $\pm$      0.03 &  4.771 $\pm$     0.025 &   2.119 $\pm$     0.044 &    13.289 $\pm$   0.036 &    12.471  $\pm$    0.049 &    11.942 $\pm$  0.051 \\
J180853.35-181130.9 &  11.43 $\pm$      0.04 & 11.741 $\pm$     0.035 &  8.139 $\pm$     0.085 &   5.616 $\pm$     0.071 &    16.122 $\pm$   0.079 &    13.566  $\pm$    0.037 &    12.434 $\pm$  0.037 \\
J180853.65-181208.4 & 11.036 $\pm$     0.045 & 11.233 $\pm$     0.046 &  8.071 $\pm$     0.051 &   5.005 $\pm$     0.089 &    15.465 $\pm$   0.064 &    13.051  $\pm$     0.03 &    12.031 $\pm$  0.026 \\
J180853.94-181422.4$^{(c)}$ & 10.641 $\pm$     0.052 &  9.607 $\pm$     0.036 &  4.724 $\pm$     0.047 &    1.55 $\pm$     0.037 &        --  &        --   &        --  \\
J180854.47-181406.5 & 10.503 $\pm$     0.046 &  9.409 $\pm$      0.04 &   4.15 $\pm$     0.022 &    0.32 $\pm$     0.009 &        --  &        --   &         \\
J180854.69-181230.5 & 10.053 $\pm$      0.04 &  9.929 $\pm$     0.035 &  7.437 $\pm$     0.096 &   5.093 $\pm$     0.071 &    13.979 $\pm$   0.024 &    11.683  $\pm$    0.021 &    10.695 $\pm$  0.021 \\
J180854.87-181409.5 & 10.185 $\pm$     0.041 &  9.126 $\pm$     0.035 &  4.286 $\pm$      0.03 &   0.956 $\pm$      0.03 &        --  &        --   &        --  \\
J180855.80-181136.3 & 12.069 $\pm$     0.064 & 12.526 $\pm$     0.074 &  7.954 $\pm$     0.072 &   5.717 $\pm$     0.093 &    15.822 $\pm$   0.087 &    13.738  $\pm$    0.074 &    12.892 $\pm$  0.045 \\
J180856.26-181505.7 & 10.283 $\pm$     0.058 &  9.398 $\pm$     0.043 &  5.627 $\pm$     0.075 &   0.797 $\pm$     0.062 &     15.32 $\pm$      -- &    13.878  $\pm$    0.062 &    12.337 $\pm$  0.035 \\
J180856.64-181420.4 &  9.631 $\pm$     0.041 &  9.154 $\pm$     0.047 &   4.51 $\pm$     0.021 &   1.295 $\pm$     0.022 &    16.241 $\pm$   0.084 &    12.831  $\pm$    0.028 &    11.059 $\pm$  0.024 \\
J180857.00-181354.7 & 11.014 $\pm$     0.066 & 10.209 $\pm$     0.047 &  5.523 $\pm$     0.019 &   2.025 $\pm$     0.022 &        --  &        --   &         \\
J180857.48-181853.3 & 10.859 $\pm$     0.057 & 10.794 $\pm$     0.066 &  7.455 $\pm$     0.029 &   4.348 $\pm$     0.038 &    15.821 $\pm$   0.075 &     13.12  $\pm$    0.068 &    11.663 $\pm$  0.039 \\
J180858.07-181506.0 &   8.04 $\pm$     0.026 &  8.086 $\pm$     0.029 &    4.4 $\pm$     0.016 &   1.647 $\pm$     0.032 &     9.067 $\pm$   0.034 &     8.456  $\pm$    0.053 &      8.27 $\pm$   0.02 \\
J180858.12-181807.2 &  10.55 $\pm$     0.044 & 10.425 $\pm$     0.049 &  5.956 $\pm$     0.033 &   2.655 $\pm$     0.033 &    14.648 $\pm$      -- &    12.495  $\pm$       -- &    12.088 $\pm$  0.072 \\
J180858.75-181806.5 & 10.519 $\pm$     0.042 & 10.435 $\pm$     0.054 &   5.78 $\pm$     0.017 &   3.046 $\pm$     0.037 &     15.81 $\pm$   0.065 &    12.795  $\pm$    0.022 &    11.453 $\pm$  0.021 \\
J180858.77-181629.7$^{(e)}$ &  8.532 $\pm$     0.024 &  6.901 $\pm$     0.022 &  2.865 $\pm$     0.018 &   0.219 $\pm$     0.026 &    16.035 $\pm$      -- &    13.621  $\pm$    0.045 &    11.328 $\pm$  0.026 \\
J180859.32-181316.1 &  11.16 $\pm$     0.047 & 10.745 $\pm$      0.04 &  6.736 $\pm$     0.091 &   4.302 $\pm$     0.074 &    15.747 $\pm$   0.078 &    13.754  $\pm$    0.033 &    12.725 $\pm$  0.033 \\
J180859.44-181332.8 &  8.857 $\pm$      0.03 &  7.786 $\pm$     0.023 &  4.949 $\pm$     0.021 &   2.629 $\pm$     0.027 &    14.689 $\pm$   0.039 &    12.413  $\pm$    0.047 &    10.703 $\pm$  0.027 \\
J180859.47-181855.8 & 11.603 $\pm$     0.047 & 11.933 $\pm$     0.051 &  7.688 $\pm$     0.044 &   4.464 $\pm$     0.034 &     14.72 $\pm$   0.049 &    13.208  $\pm$    0.045 &    12.366 $\pm$  0.035 \\
J180900.26-181347.5 &  9.892 $\pm$     0.037 &  9.958 $\pm$     0.038 &  6.826 $\pm$     0.055 &   3.903 $\pm$     0.037 &    11.435 $\pm$   0.023 &    10.455  $\pm$    0.022 &    10.073 $\pm$  0.019 \\
J180900.28-181921.8 & 11.454 $\pm$     0.039 & 11.642 $\pm$     0.045 &  8.351 $\pm$     0.081 &   5.308 $\pm$     0.089 &    15.867 $\pm$    0.08 &    13.298  $\pm$    0.035 &    12.136 $\pm$  0.027 \\
J180900.48-181903.2 & 11.585 $\pm$     0.053 & 11.687 $\pm$     0.073 &  7.693 $\pm$     0.068 &   5.275 $\pm$     0.094 &    16.271 $\pm$   0.104 &    14.157  $\pm$    0.085 &    13.071 $\pm$  0.055 \\
J180902.92-181316.2 &  9.691 $\pm$     0.028 &  9.827 $\pm$     0.034 &  8.412 $\pm$     0.045 &   5.596 $\pm$     0.078 &    11.372 $\pm$   0.022 &    10.364  $\pm$    0.022 &     9.922 $\pm$  0.019 \\
J180904.56-181638.3 &  9.987 $\pm$     0.032 &  9.923 $\pm$     0.034 &  5.869 $\pm$     0.015 &   3.132 $\pm$     0.028 &    14.576 $\pm$   0.032 &    12.207  $\pm$    0.023 &    11.139 $\pm$  0.021 \\
J180904.95-181806.2 & 10.578 $\pm$     0.035 &  10.54 $\pm$     0.041 &   8.39 $\pm$     0.094 &   5.864 $\pm$     0.095 &     15.07 $\pm$   0.083 &    12.745  $\pm$    0.065 &    11.608 $\pm$   0.05 \\
J180905.60-181621.4$^{(d)}$ & 11.084 $\pm$     0.066 & 10.292 $\pm$     0.039 &  5.544 $\pm$     0.015 &   2.607 $\pm$     0.024 &    12.639 $\pm$   0.023 &    11.881  $\pm$    0.022 &    11.549 $\pm$  0.019 \\
J180906.19-181639.1 & 10.814 $\pm$     0.033 & 10.514 $\pm$     0.039 &  6.314 $\pm$     0.017 &   3.348 $\pm$     0.032 &    15.044 $\pm$   0.045 &    12.744  $\pm$    0.038 &     11.67 $\pm$  0.029 \\
J180906.60-181847.9 & 11.425 $\pm$     0.057 & 11.481 $\pm$     0.077 &  8.474 $\pm$     0.059 &   6.018 $\pm$     0.065 &    14.951 $\pm$   0.052 &    12.994  $\pm$    0.033 &    12.082 $\pm$   0.03 \\
J180908.60-181706.1 & 11.316 $\pm$      0.05 & 11.686 $\pm$     0.073 &  8.709 $\pm$     0.055 &   5.842 $\pm$     0.054 &    15.388 $\pm$   0.104 &    13.396  $\pm$    0.145 &    12.225 $\pm$  0.068

\enddata
\tablenotetext{Notes}{The symbol `--' indicates no available photometries. The 70~\um\ photometries
 from the CuTeX catalog are (a): $19.2538\pm0.331$ Jy; (b): $36.1023\pm0.296$ Jy; (c): $19.1234\pm0.242$ Jy;
 (d): $7.9266\pm0.176$ Jy. The 70~\um\ photometry from \citet{lum13} is (e): $44.7\pm4.5$ Jy.}
\end{deluxetable*}

\begin{deluxetable*}{lllllllll}[tbh!]
\tabletypesize{\scriptsize} \tablecolumns{9} \tablewidth{0pt}
\tablecaption{Photometries of YSO candidates from the archived MIPSGAL catalog\label{tbl-11}}
\tablehead{\colhead{Name} & \colhead{$[24]$}   &  \colhead{$J$}  & \colhead{$H$}   & \colhead{$K$}
           &\colhead{I1}        & \colhead{I2}    &\colhead{I3}   & \colhead{I4}  \\
           \colhead{ }           & \colhead{(mag)}  &  \colhead{(mag)}  & \colhead{(mag)}  & \colhead{(mag)}
           & \colhead{(mag)}     & \colhead{(mag)}  &  \colhead{(mag)}  & \colhead{(mag)}  }

\startdata

MG011.8545+00.7327$^{(a)}$ &   2.04 $\pm$ 0.02 & 14.739 $\pm$  0.037 & 13.029 $\pm$ 0.046 & 11.361 $\pm$ 0.034 &   8.786 $\pm$  0.031 &   7.866 $\pm$  0.039 &   6.767 $\pm$ 0.033 &   5.163 $\pm$  0.024 \\
MG011.9455+00.7481 &   2.78 $\pm$ 0.02 & 14.689 $\pm$  0.039 & 12.413 $\pm$ 0.047 & 10.703 $\pm$ 0.027 &   8.567 $\pm$  0.031 &   7.882 $\pm$  0.041 &   7.093 $\pm$ 0.032 &    6.18 $\pm$  0.025

\enddata
\tablenotetext{(a)}{ The 70~\um\ photometry is $1.9156\pm0.111$ Jy from the CuTeX catalog.}
\end{deluxetable*}

\begin{figure*}
  \centering
\includegraphics[width=0.24\textwidth]{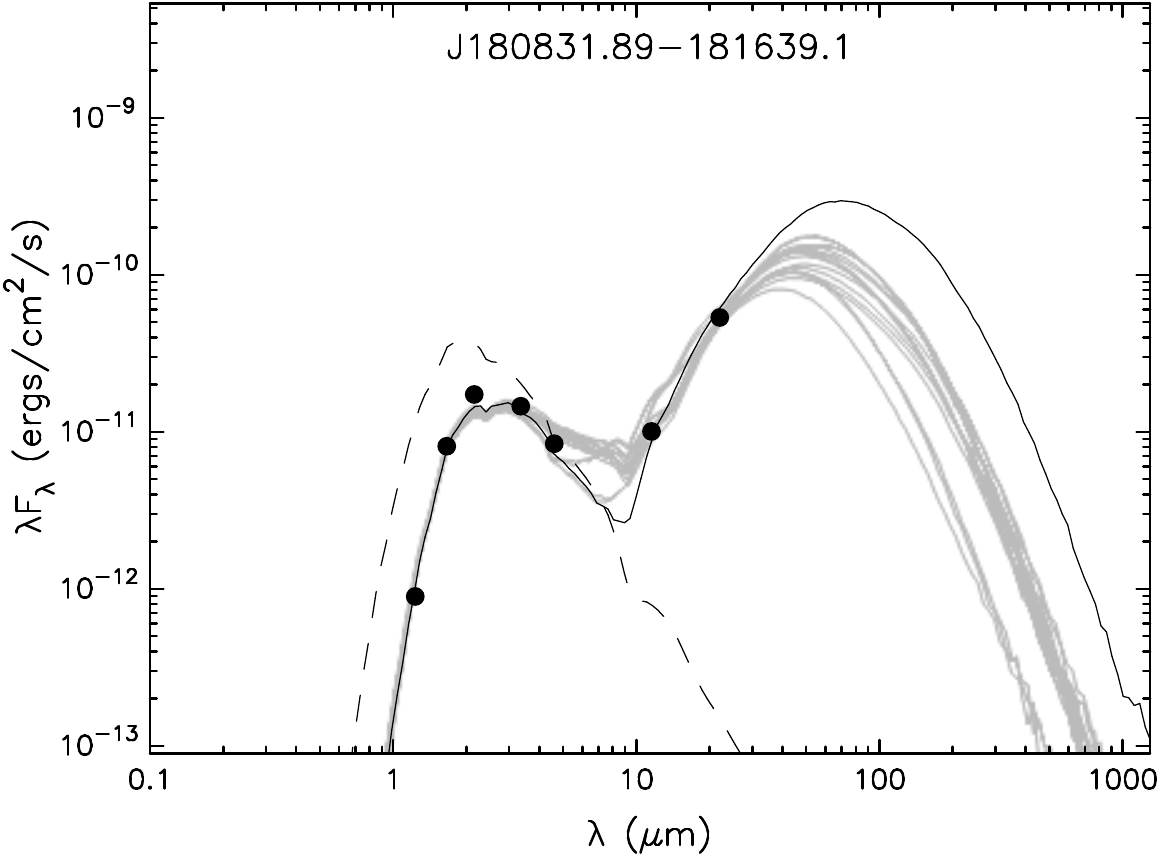}
\includegraphics[width=0.24\textwidth]{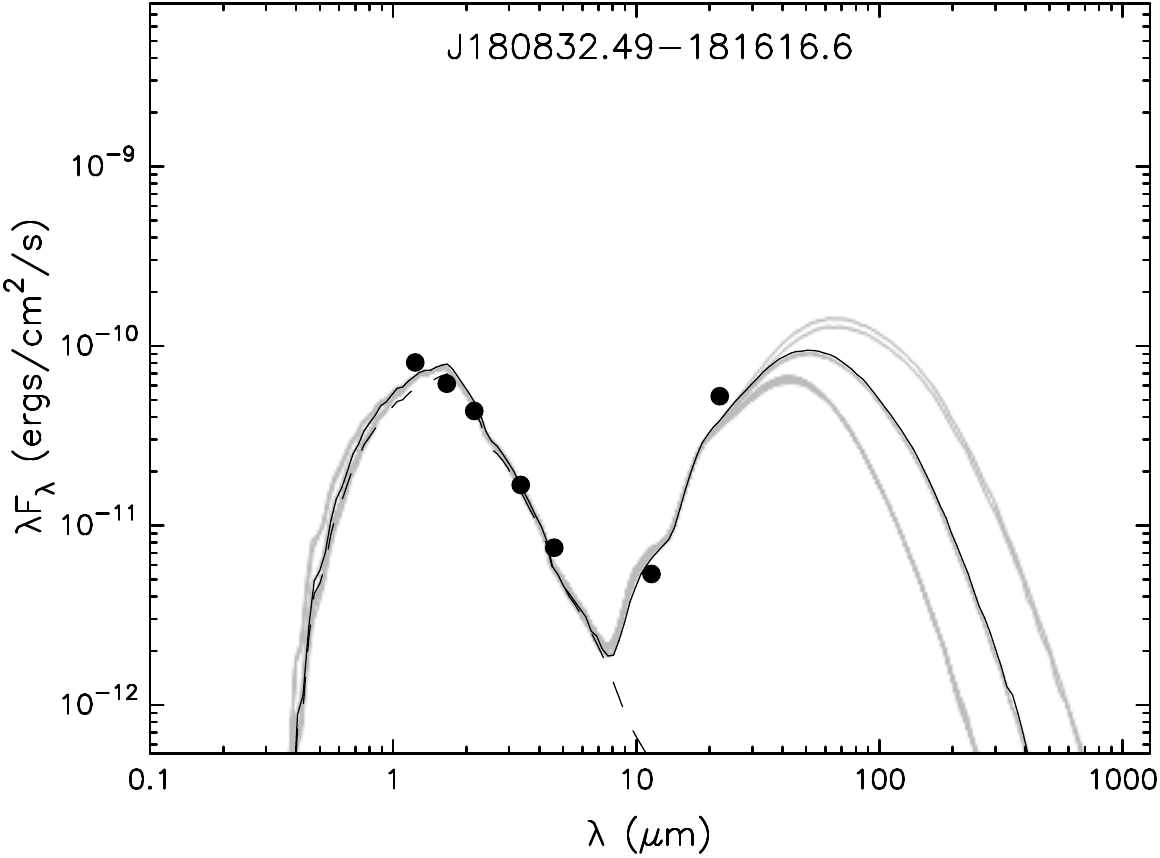}
\includegraphics[width=0.24\textwidth]{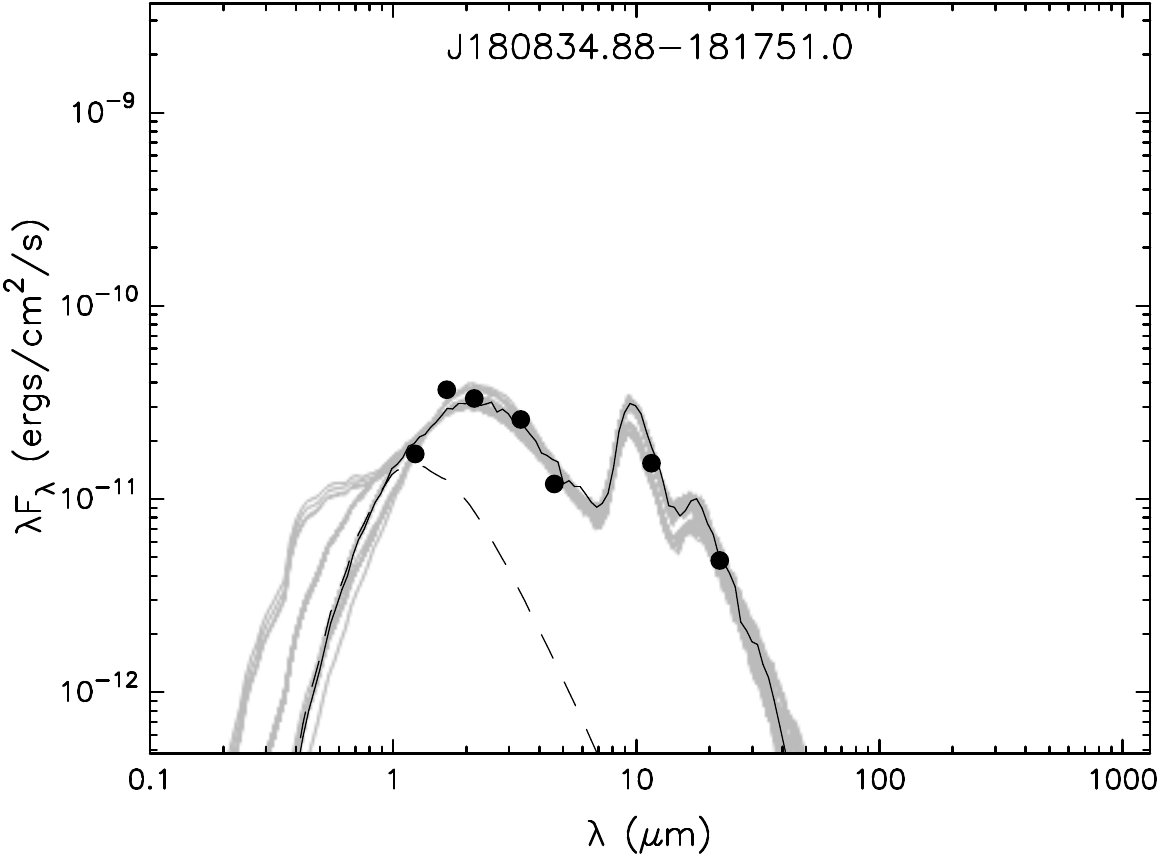}
\includegraphics[width=0.24\textwidth]{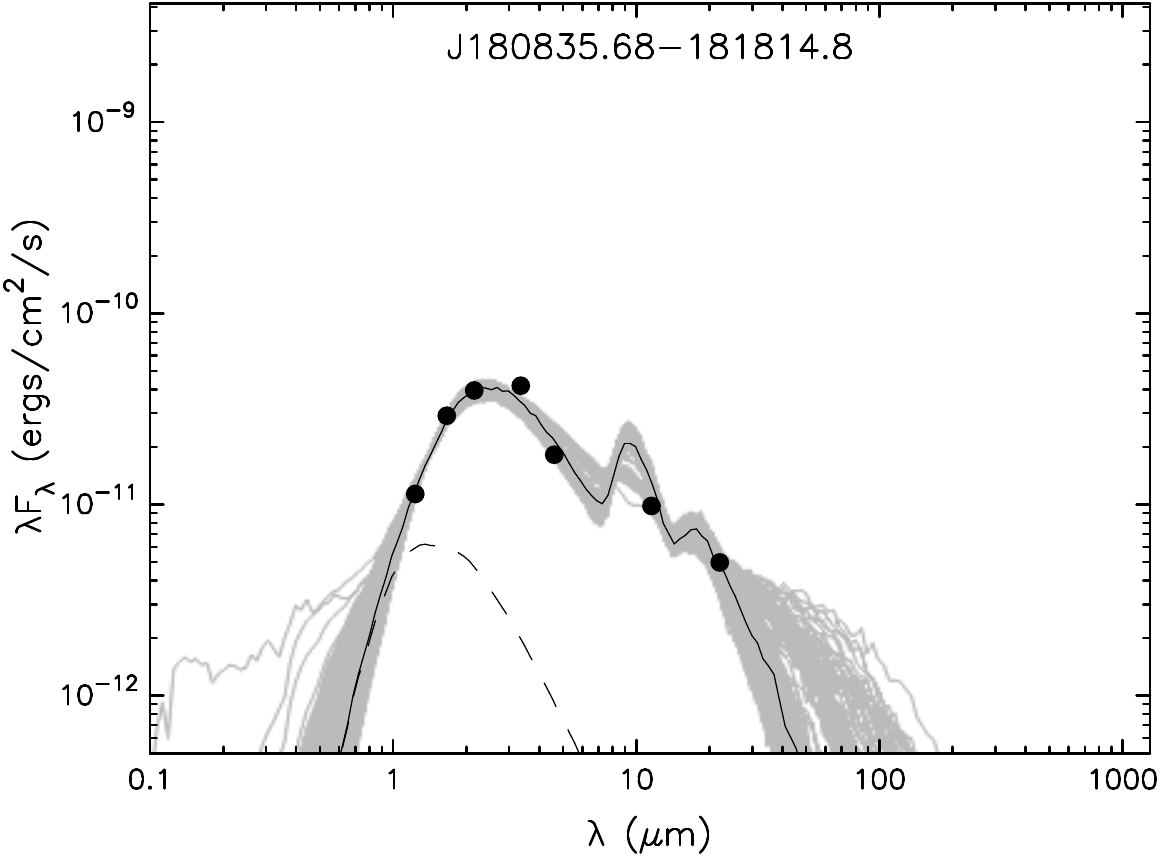}
\includegraphics[width=0.24\textwidth]{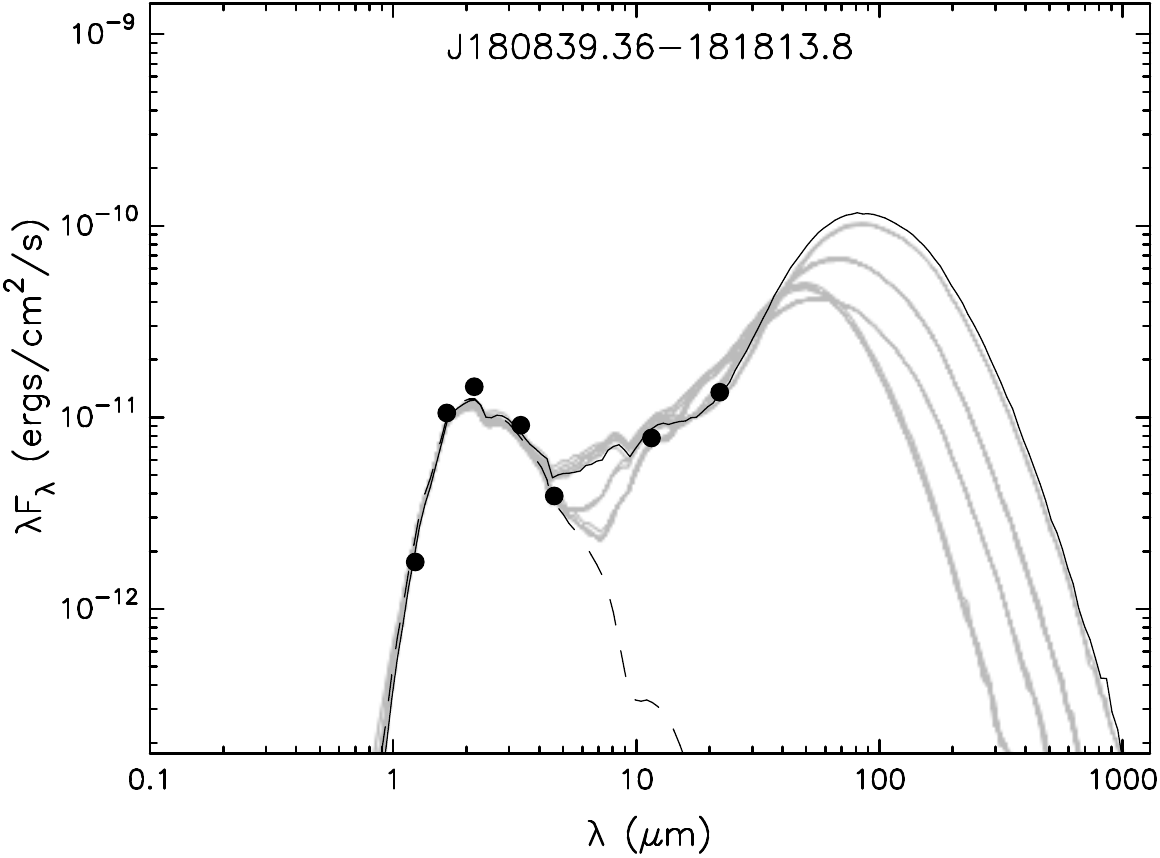}
\includegraphics[width=0.24\textwidth]{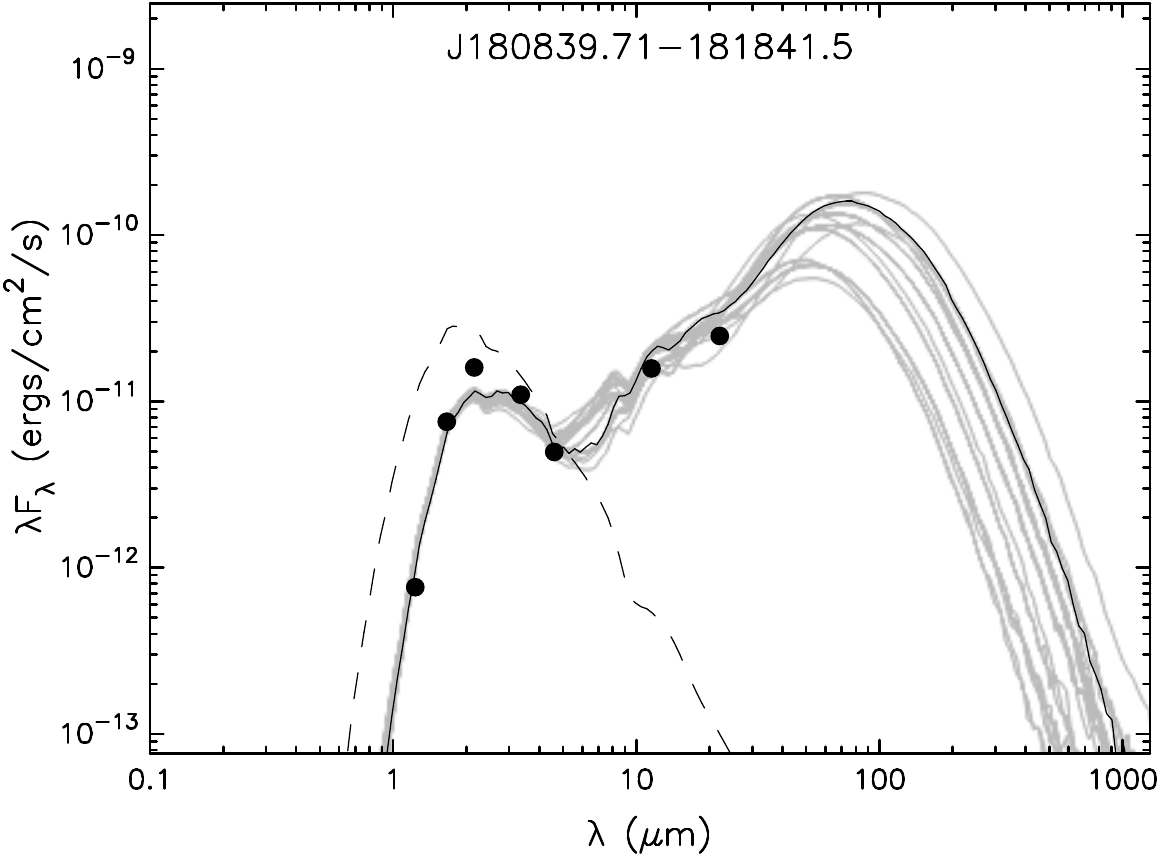}
\includegraphics[width=0.24\textwidth]{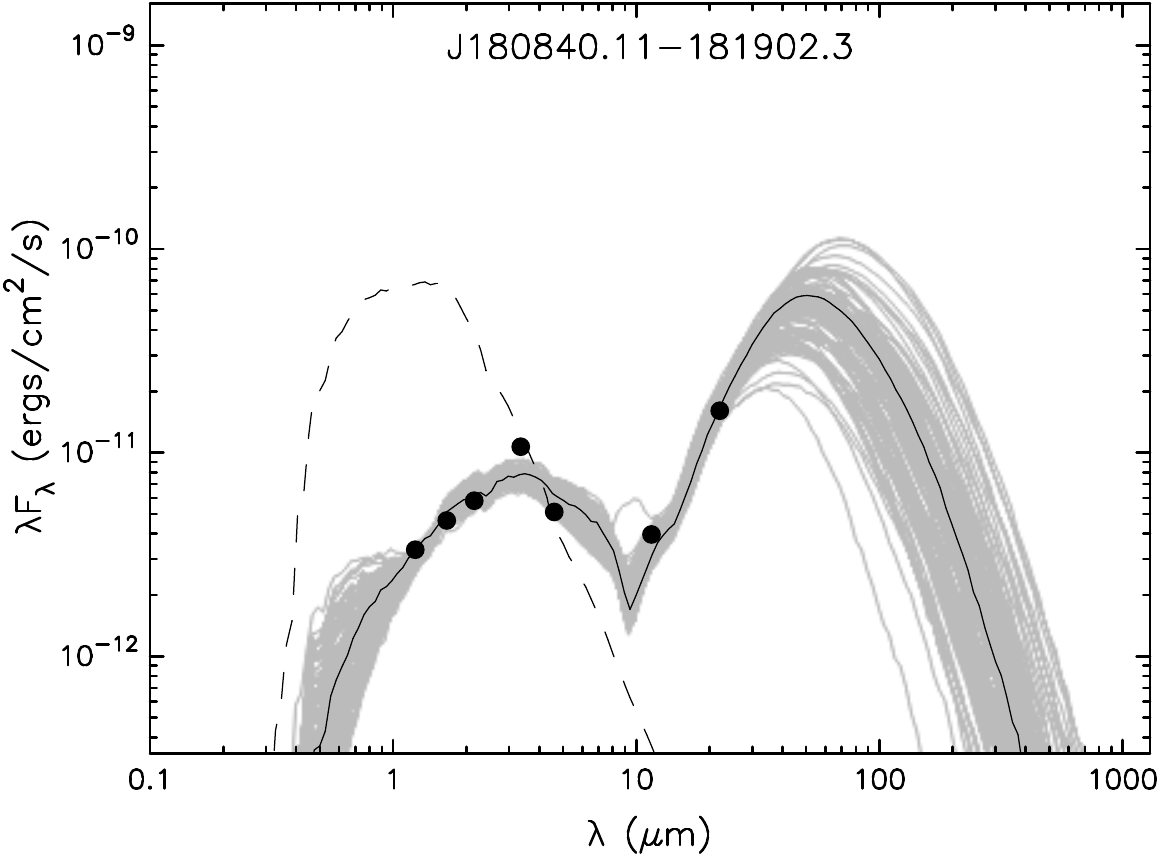}
\includegraphics[width=0.24\textwidth]{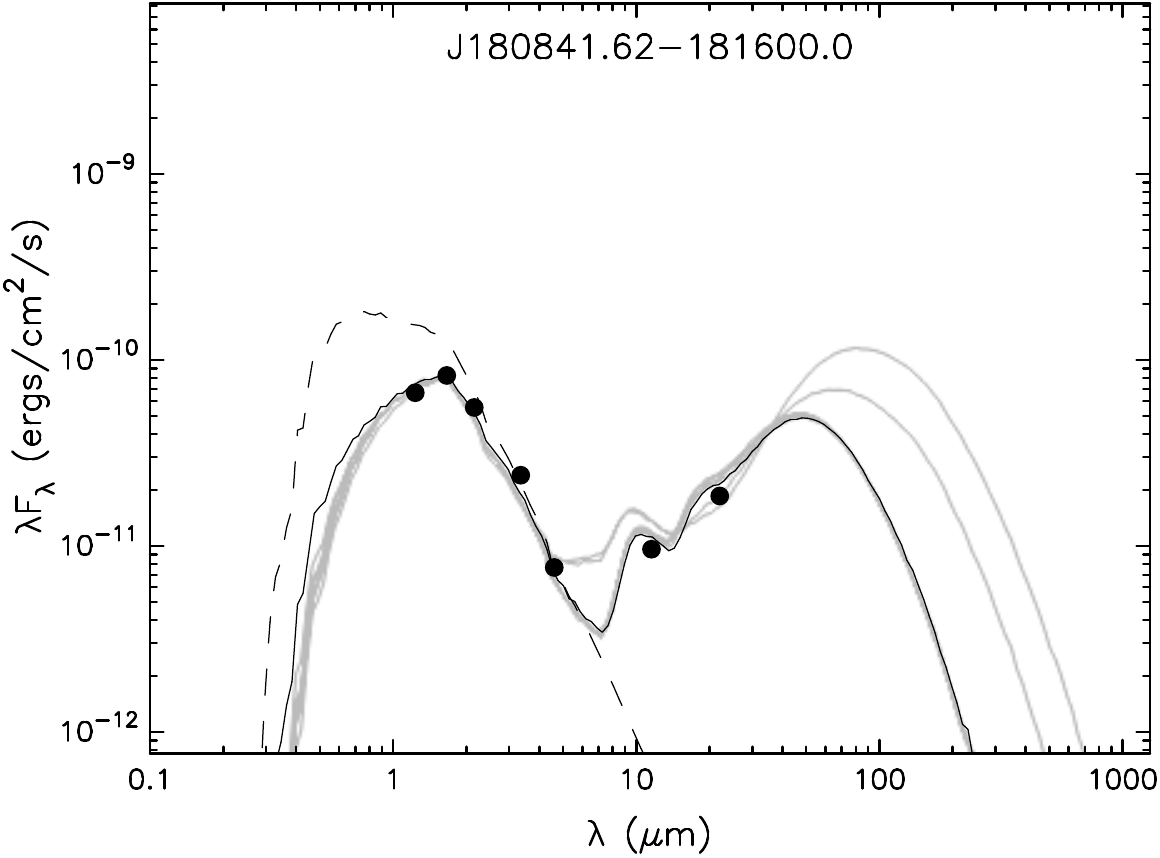}
\includegraphics[width=0.24\textwidth]{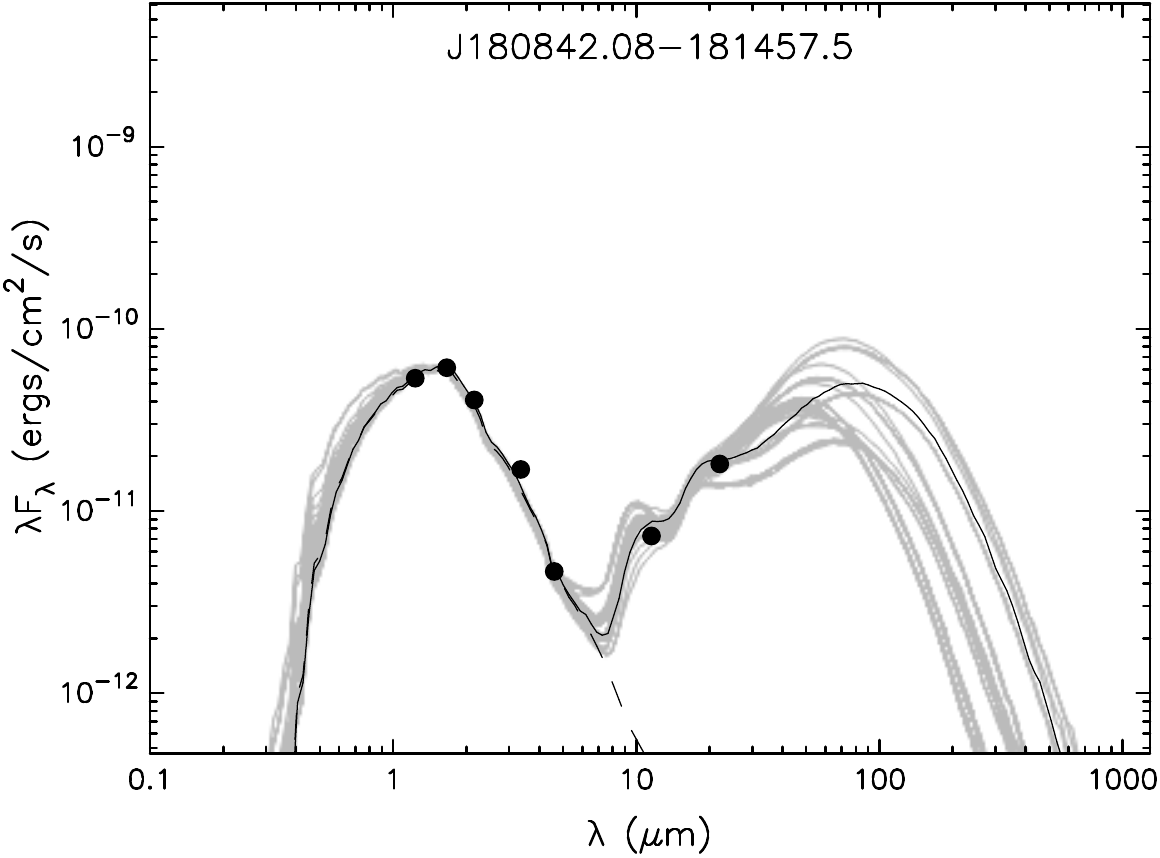}
\includegraphics[width=0.24\textwidth]{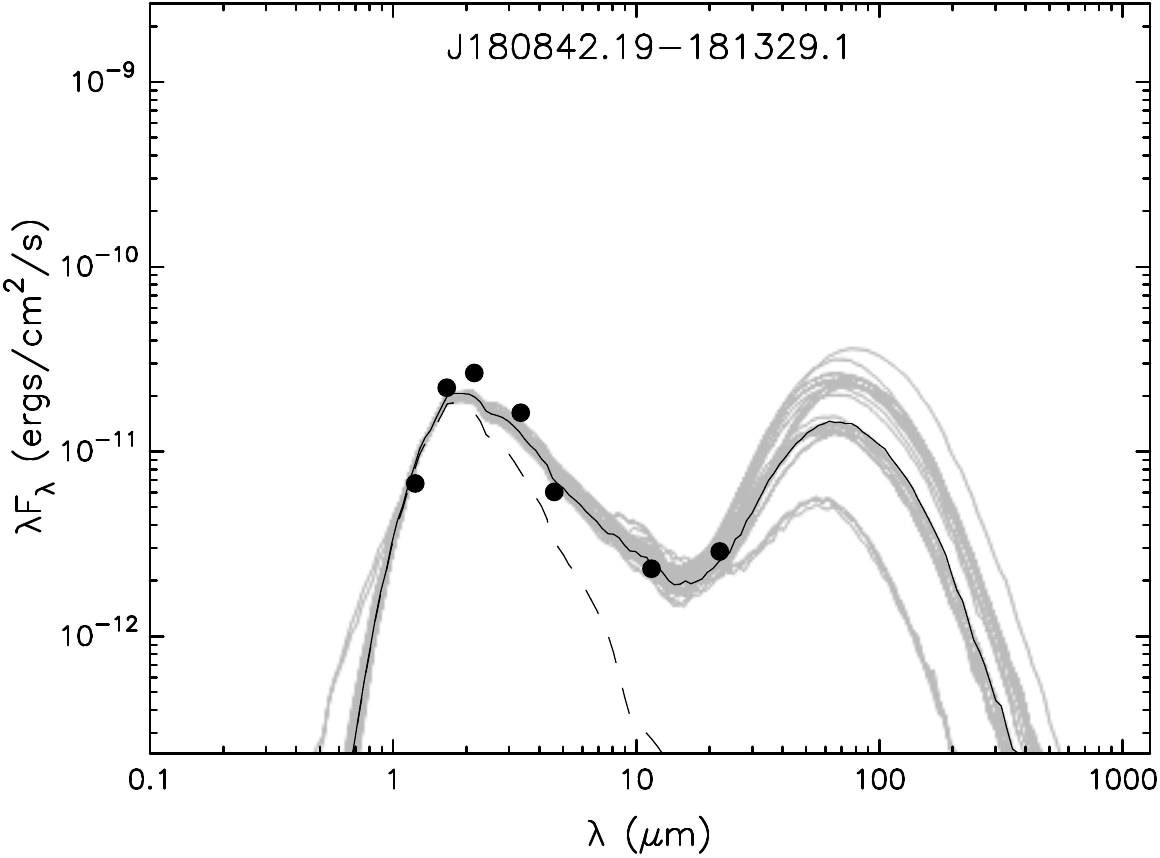}
\includegraphics[width=0.24\textwidth]{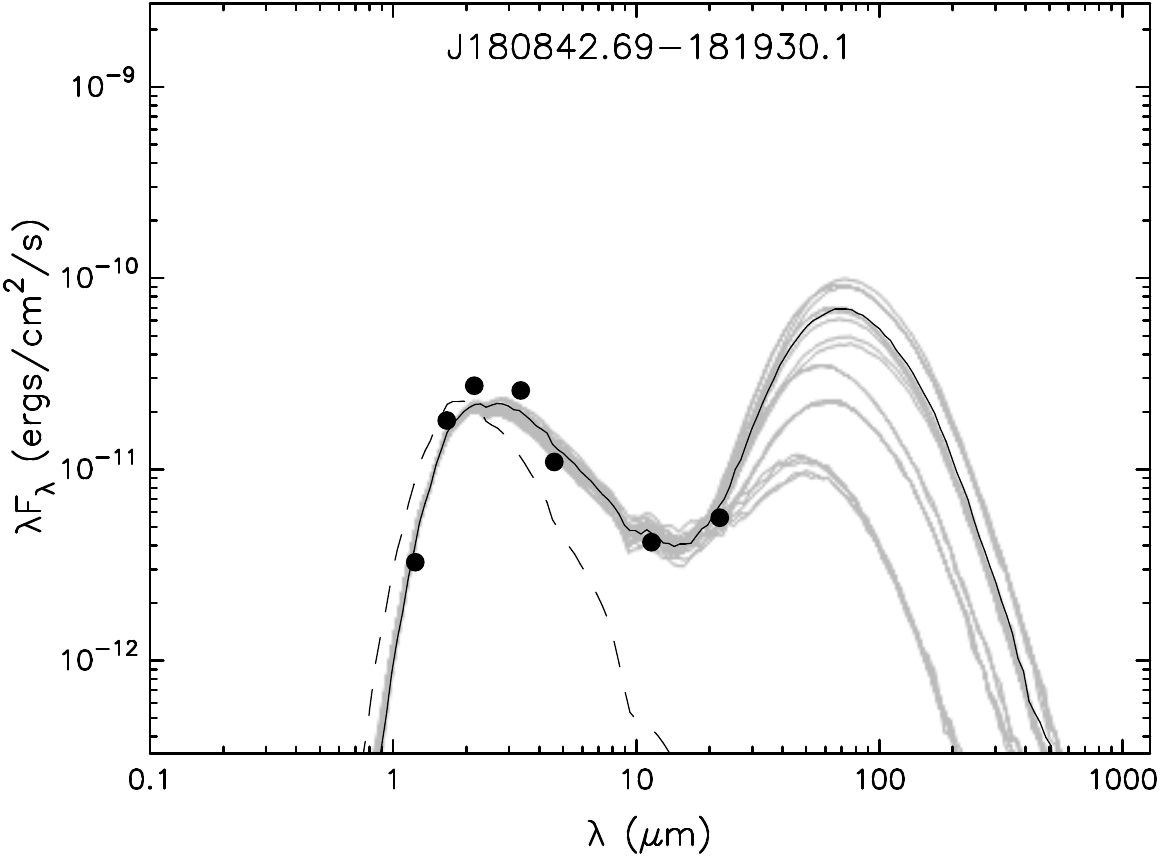}
\includegraphics[width=0.24\textwidth]{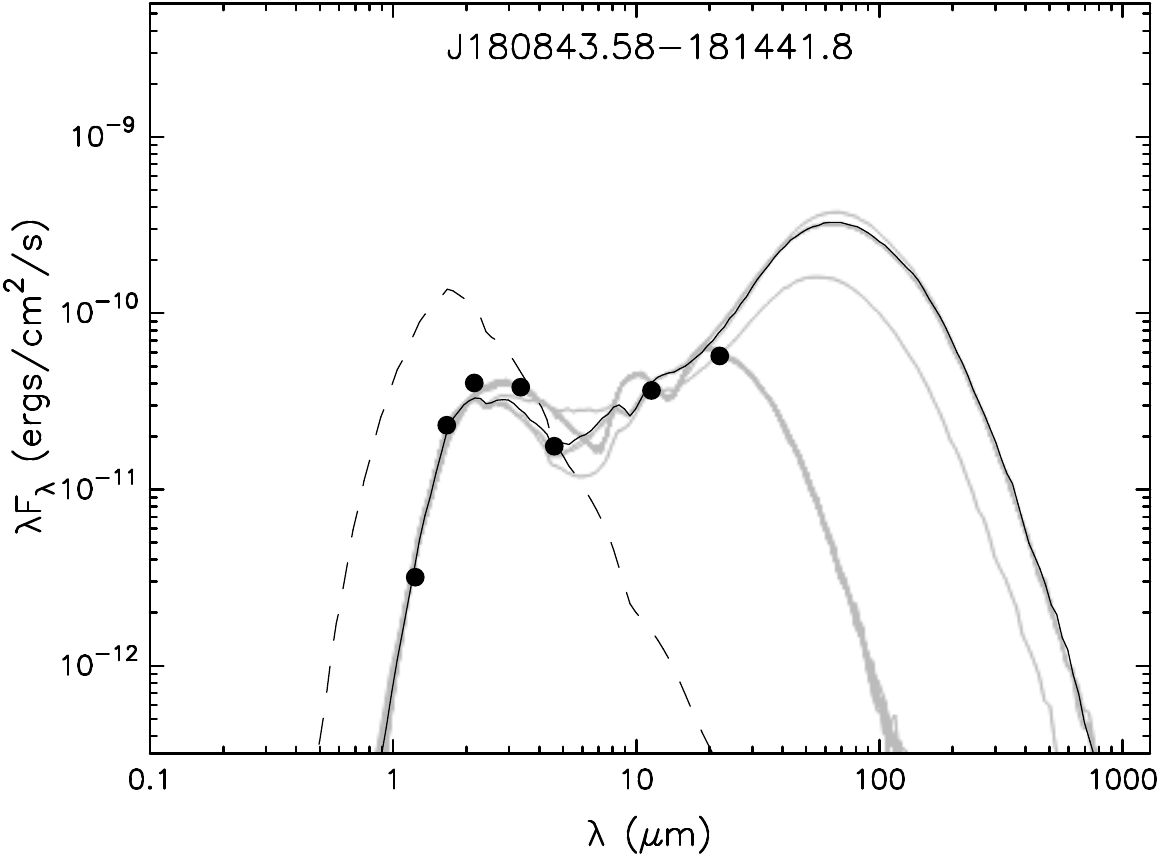}
\includegraphics[width=0.24\textwidth]{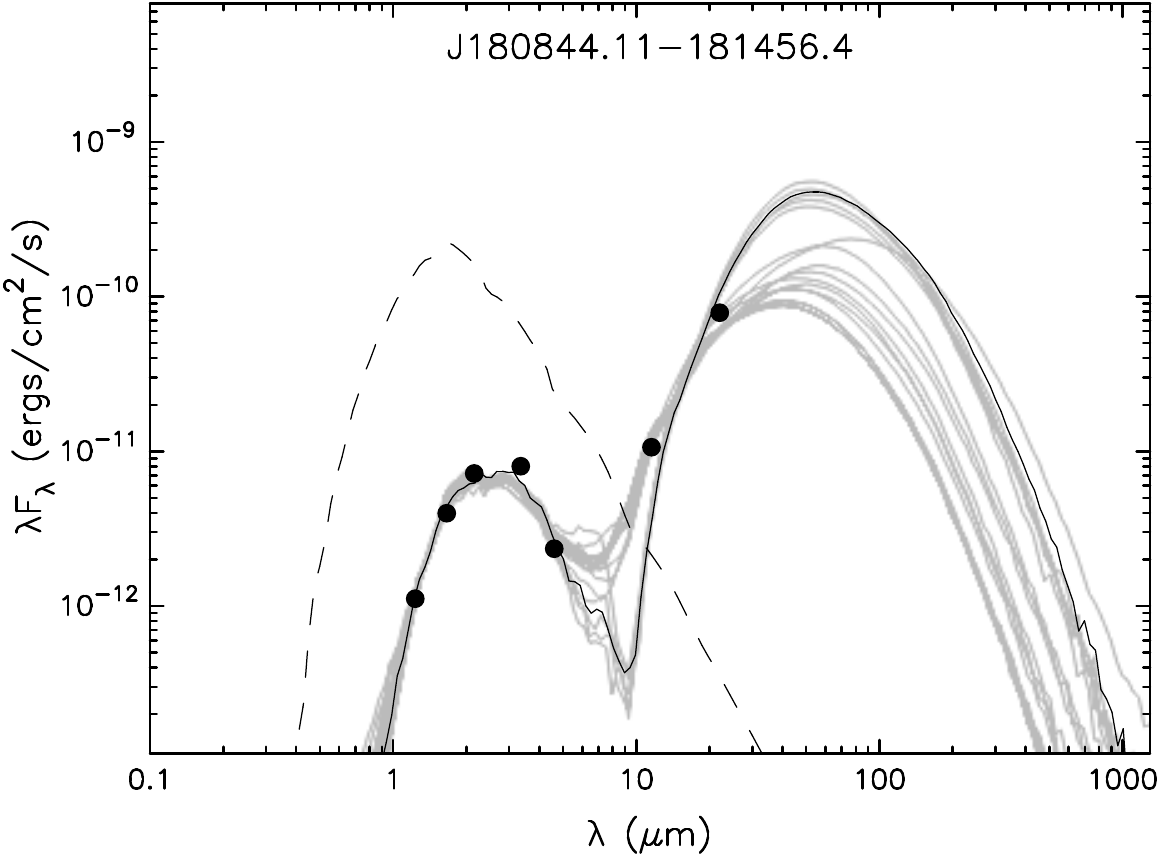}
\includegraphics[width=0.24\textwidth]{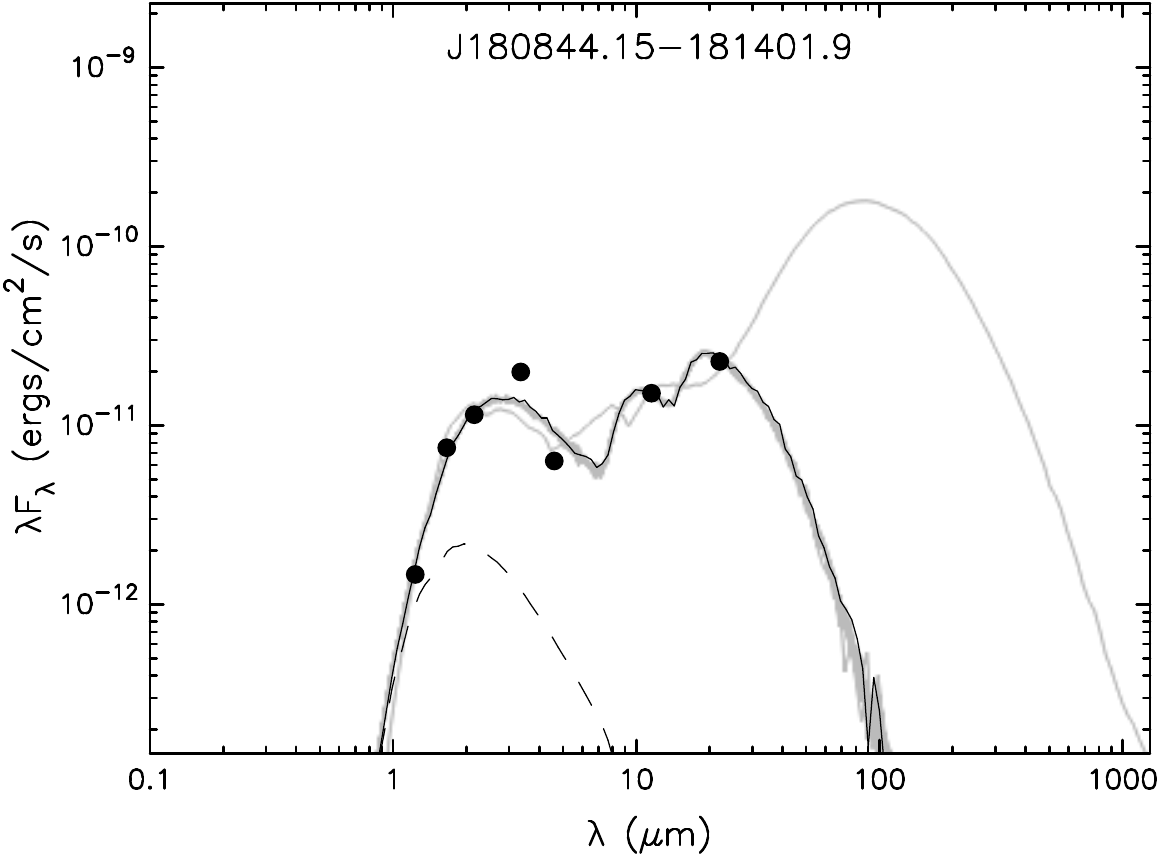}
\includegraphics[width=0.24\textwidth]{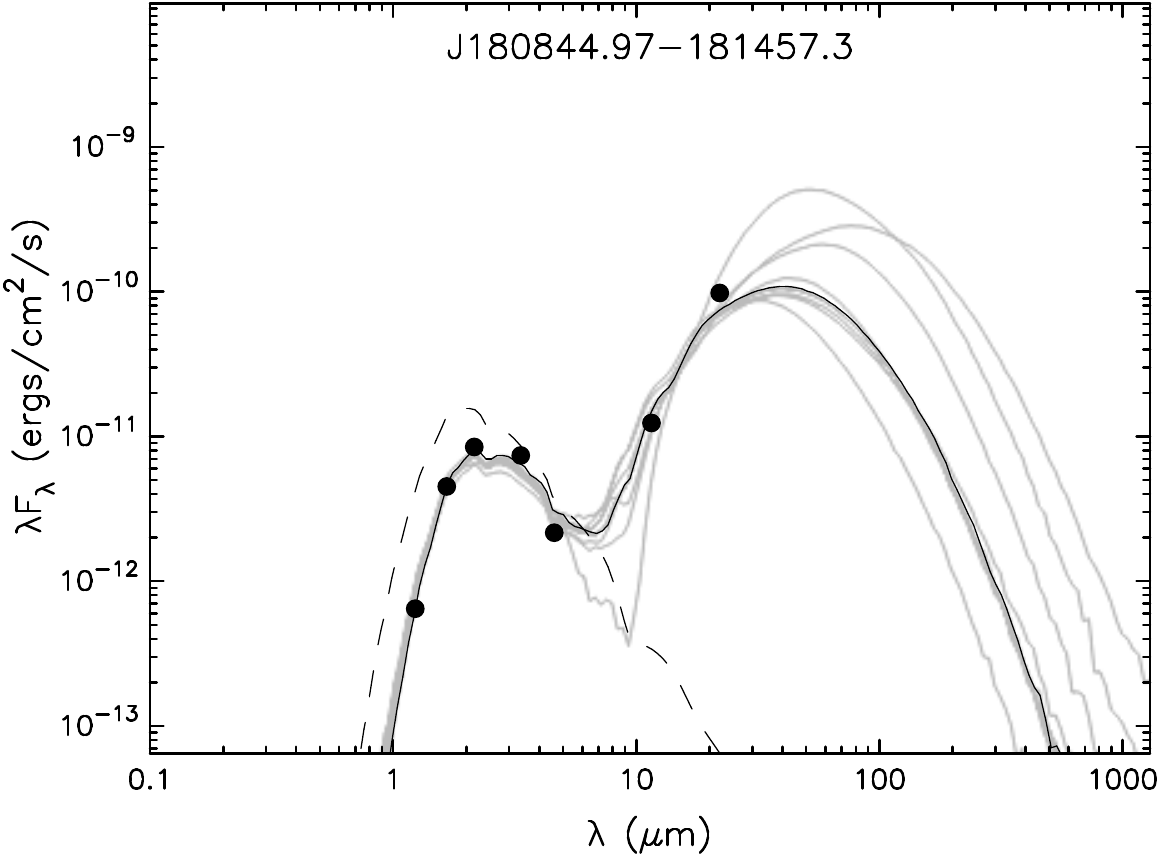}
\includegraphics[width=0.24\textwidth]{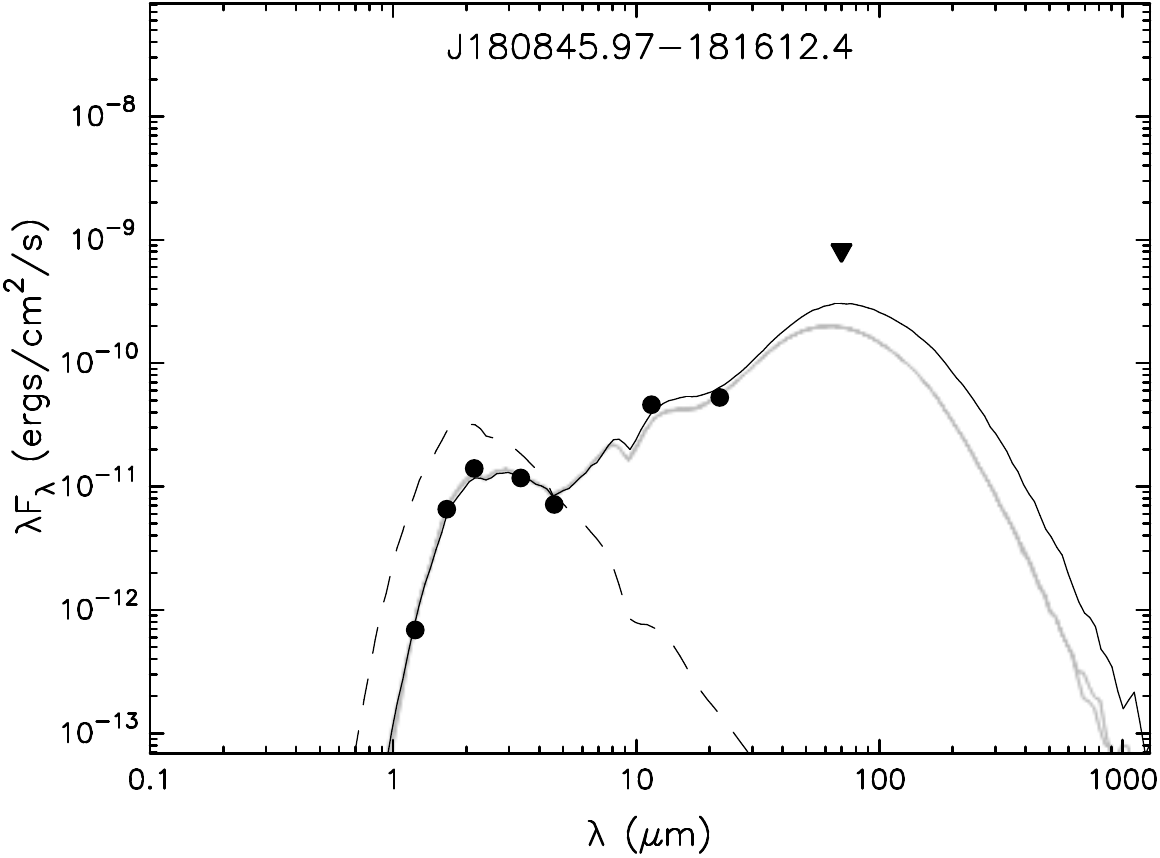}
\includegraphics[width=0.24\textwidth]{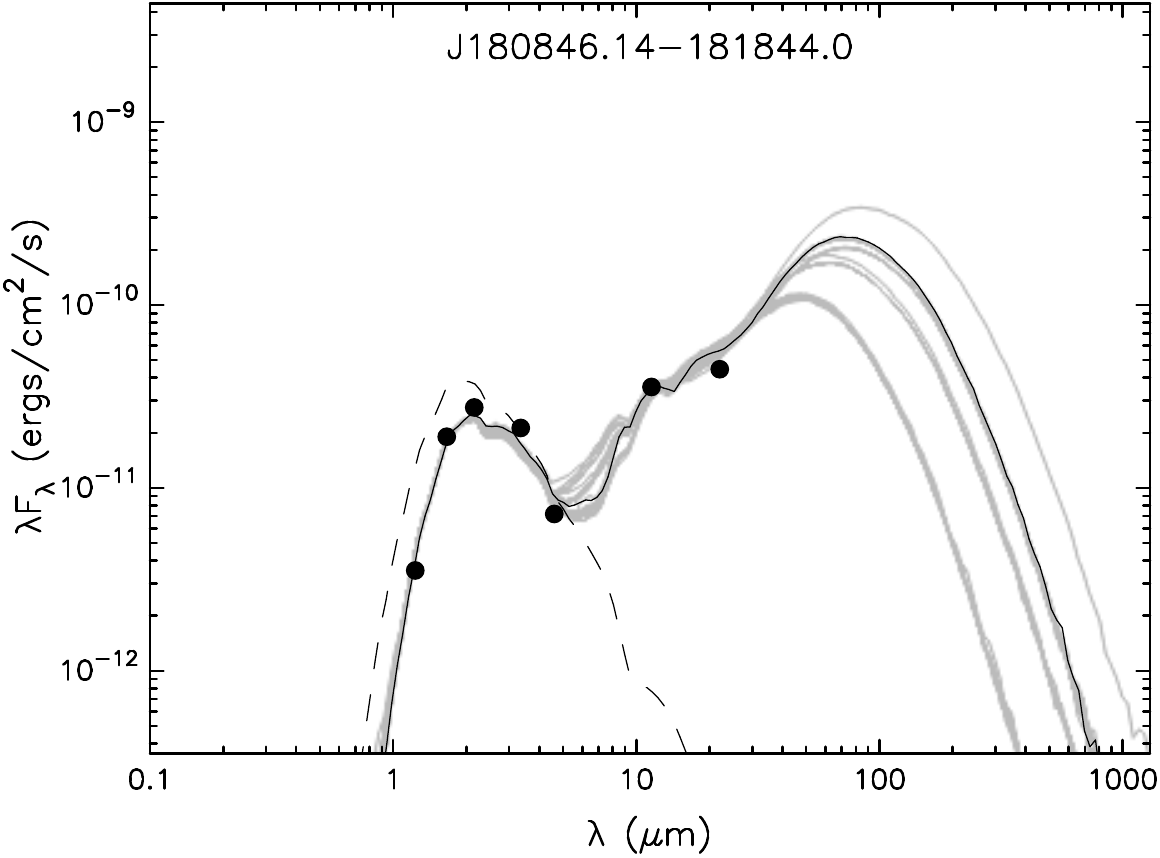}
\includegraphics[width=0.24\textwidth]{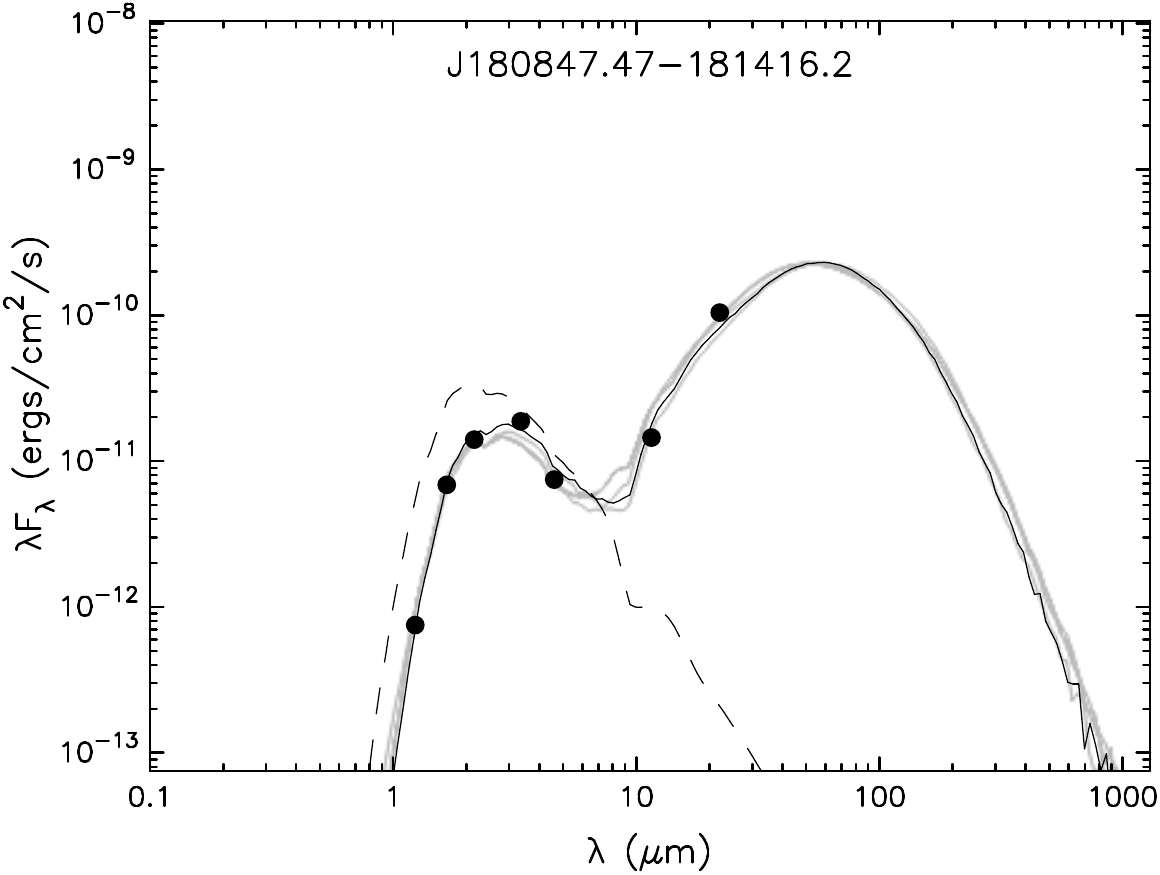}
\includegraphics[width=0.24\textwidth]{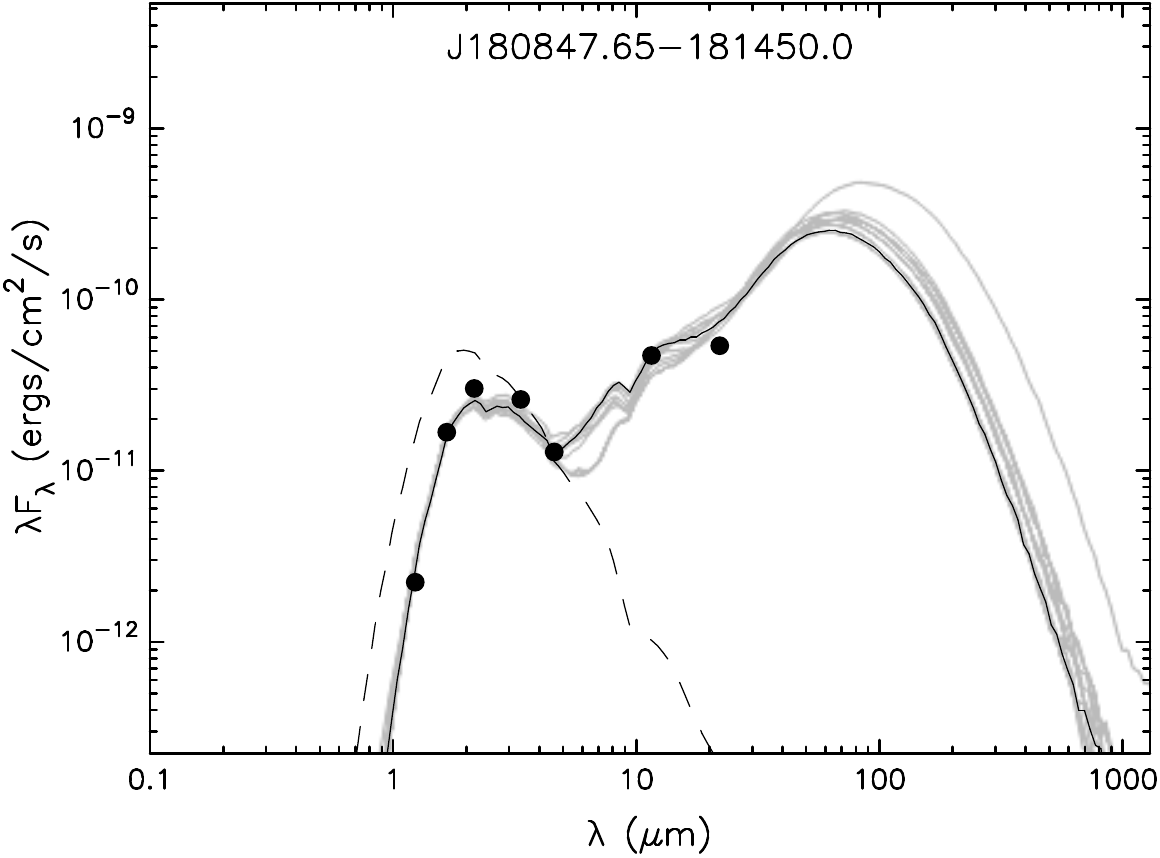}
\includegraphics[width=0.24\textwidth]{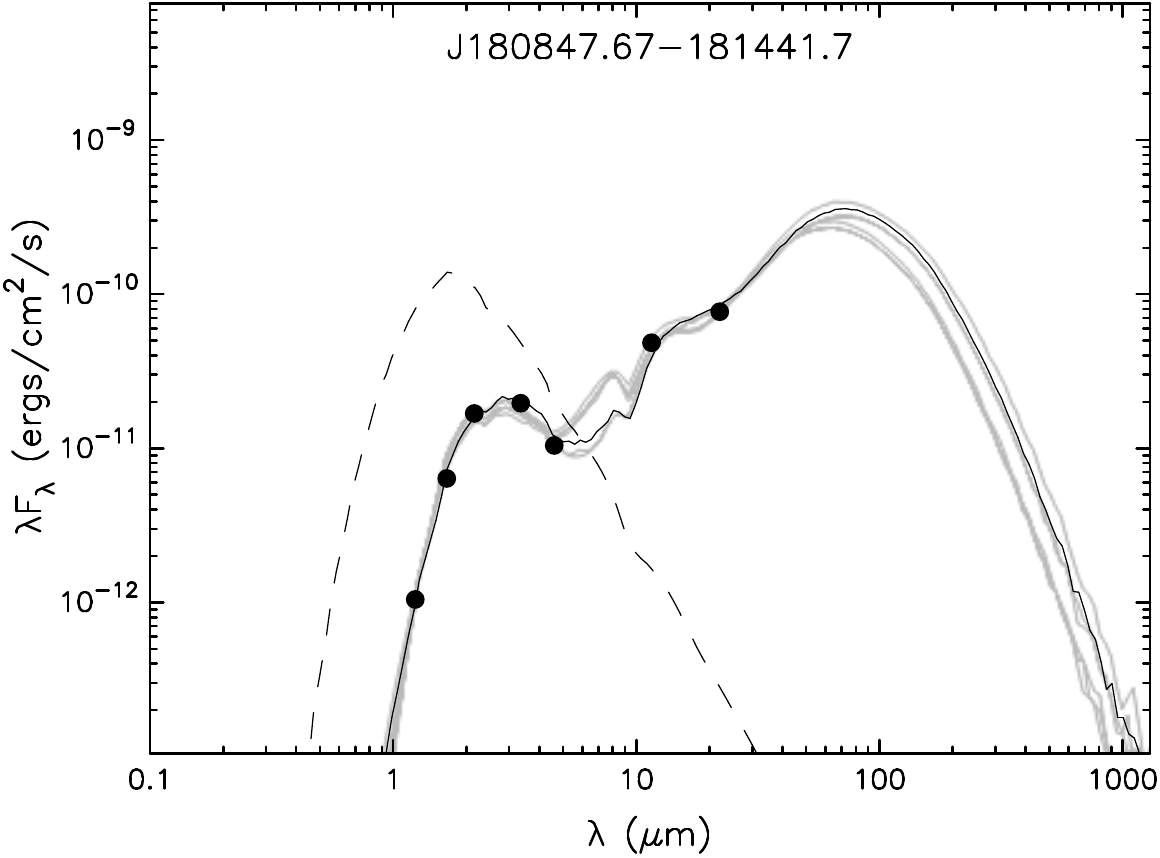}
\includegraphics[width=0.24\textwidth]{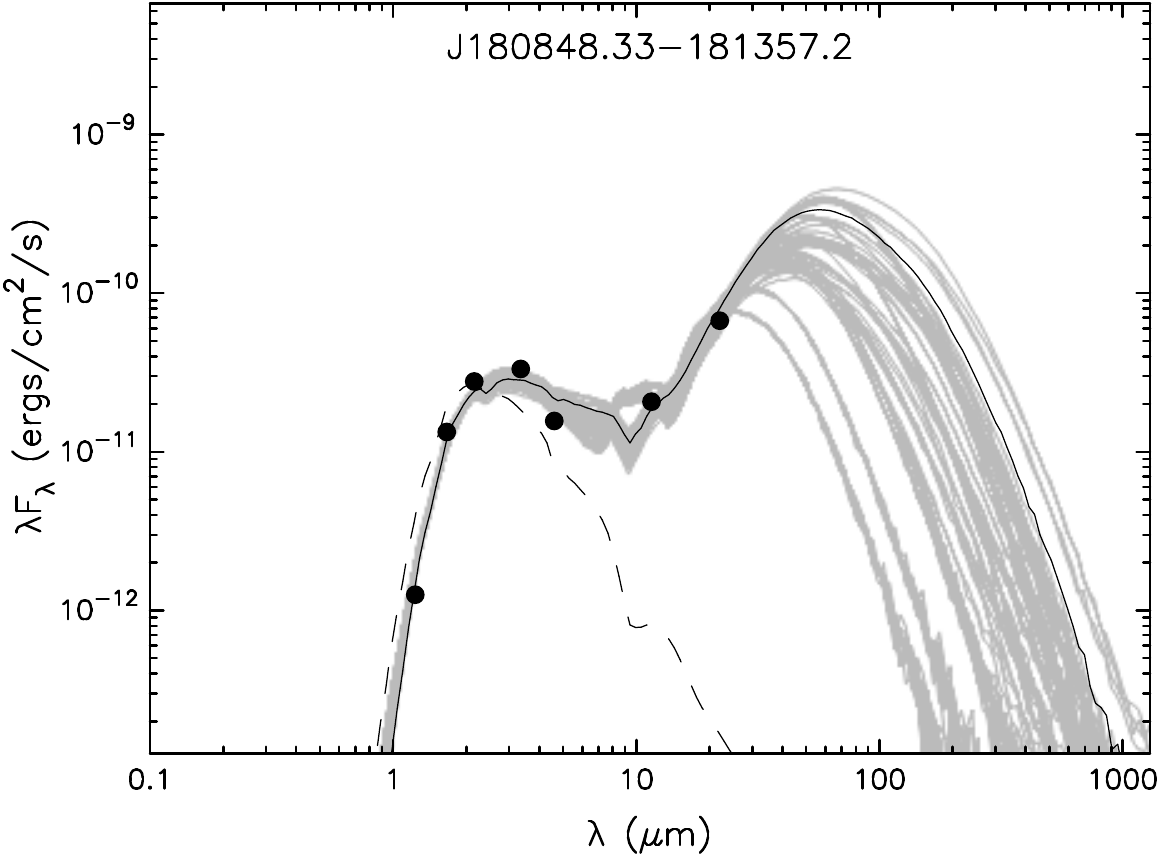}
\includegraphics[width=0.24\textwidth]{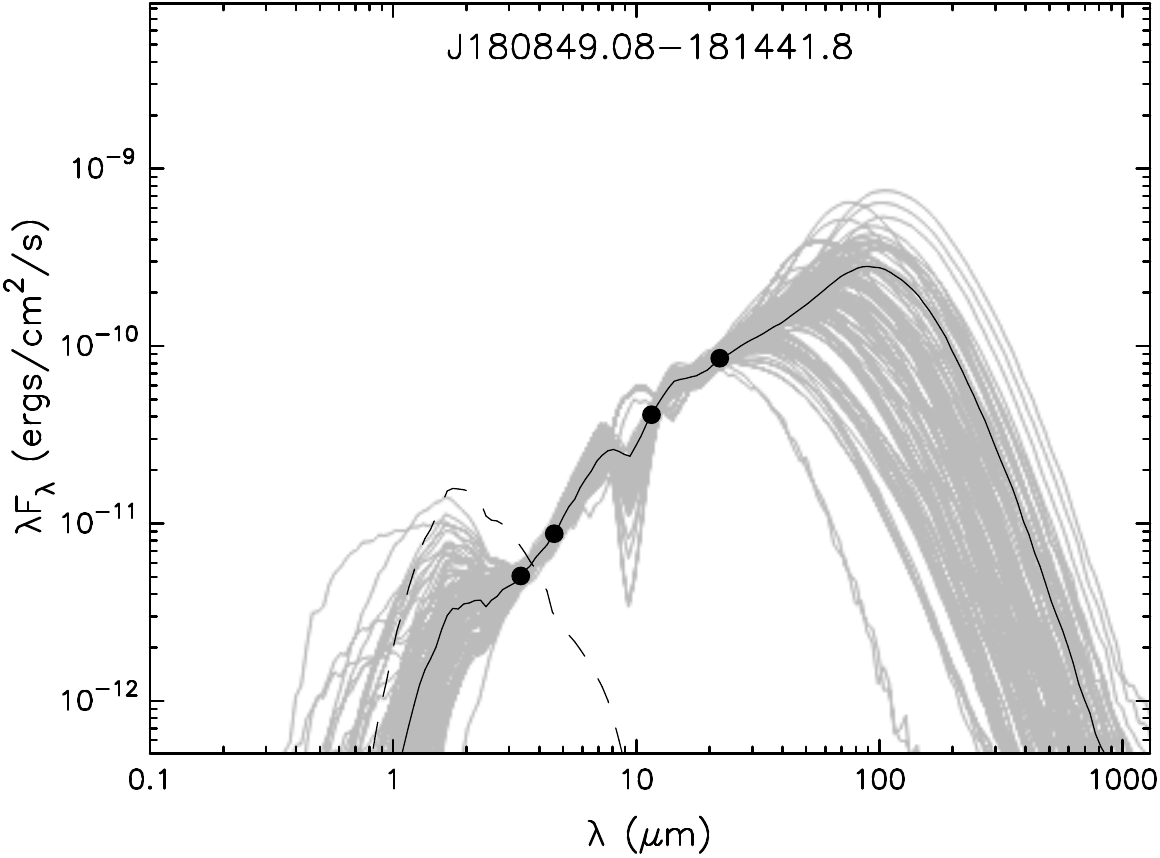}
\includegraphics[width=0.24\textwidth]{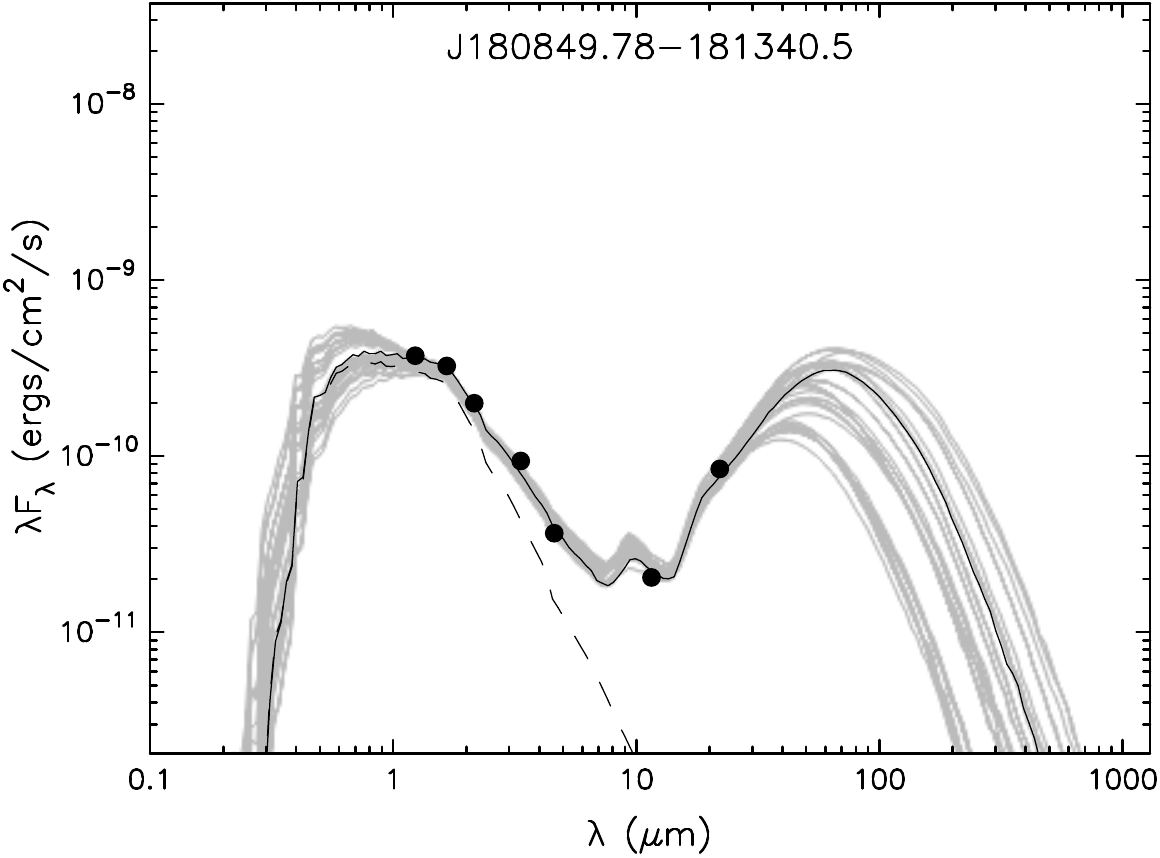}
\includegraphics[width=0.24\textwidth]{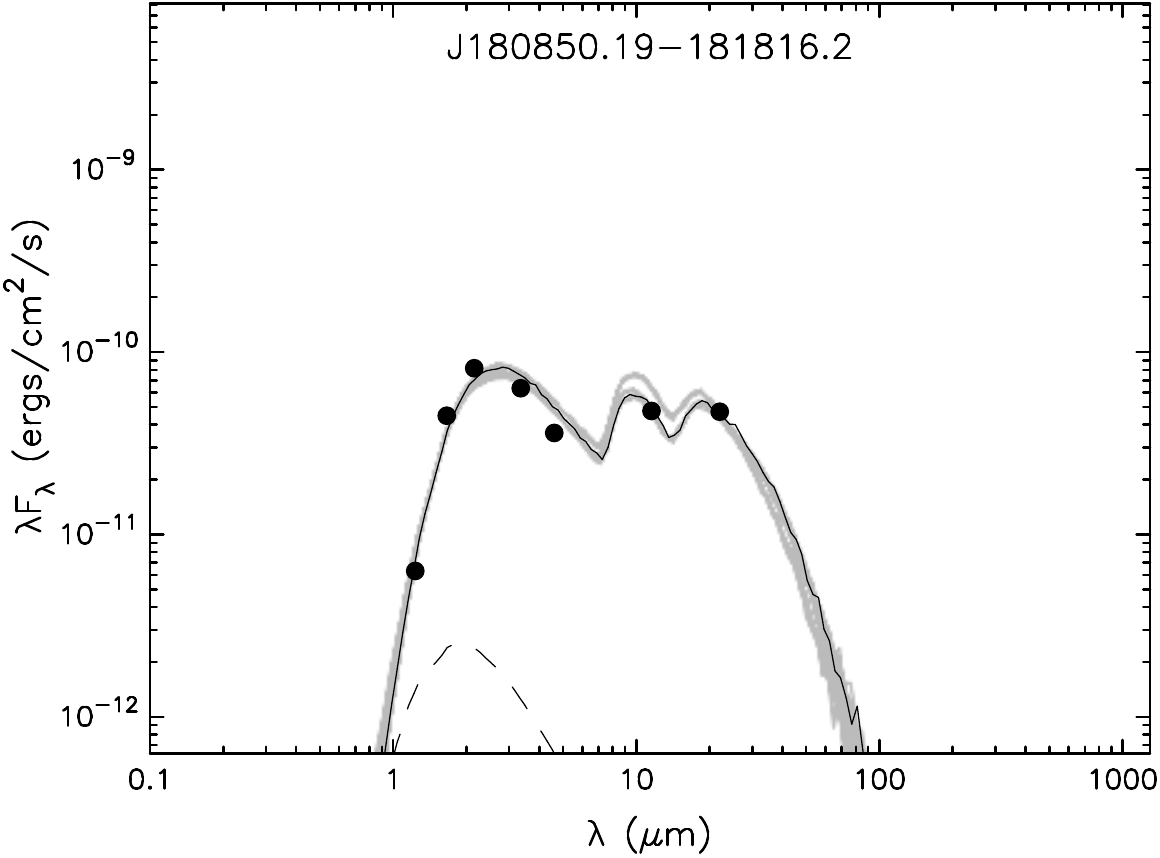}
\includegraphics[width=0.24\textwidth]{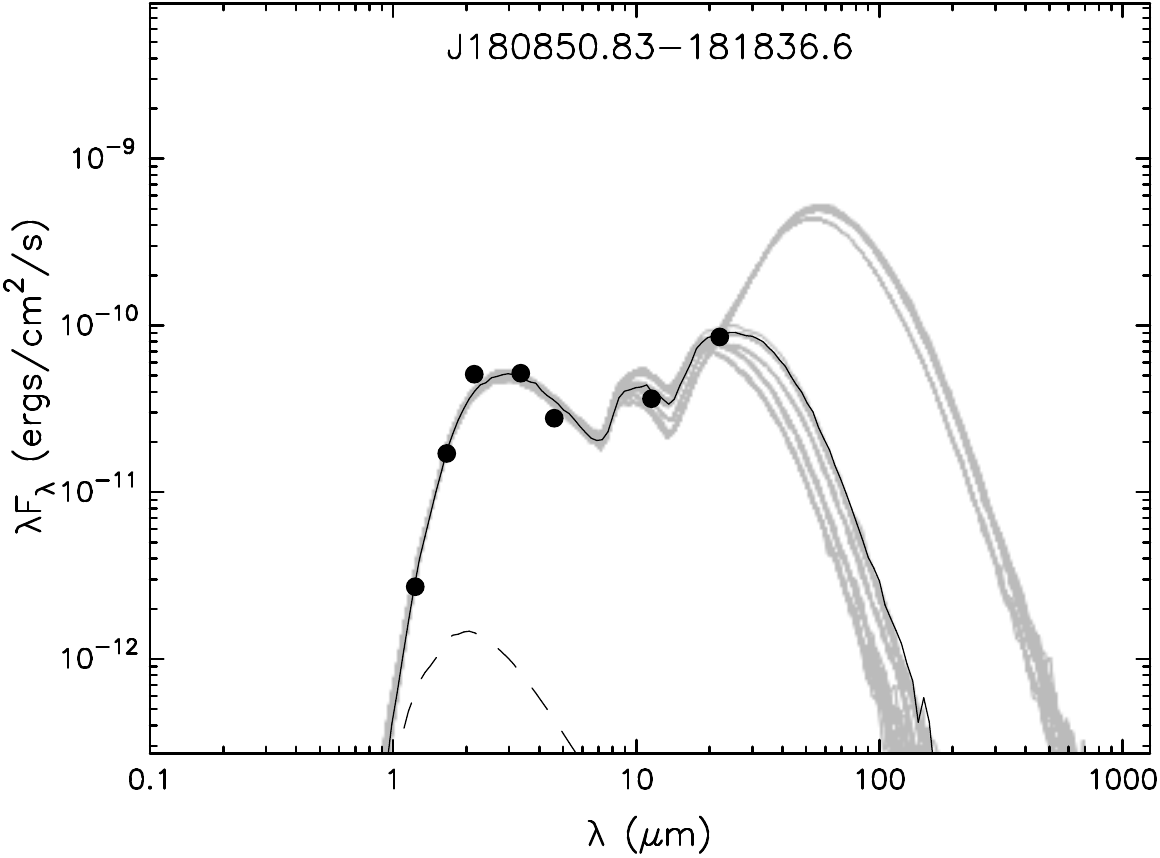}
\includegraphics[width=0.24\textwidth]{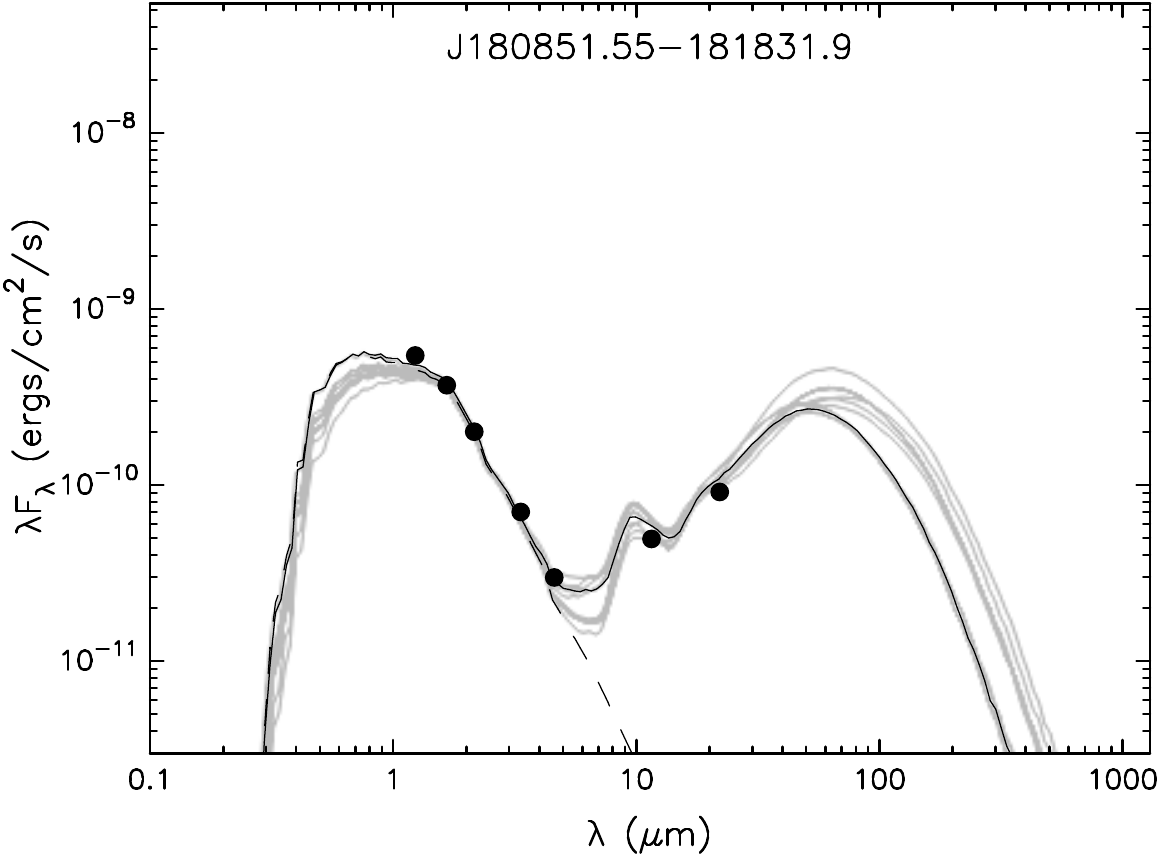}
\includegraphics[width=0.24\textwidth]{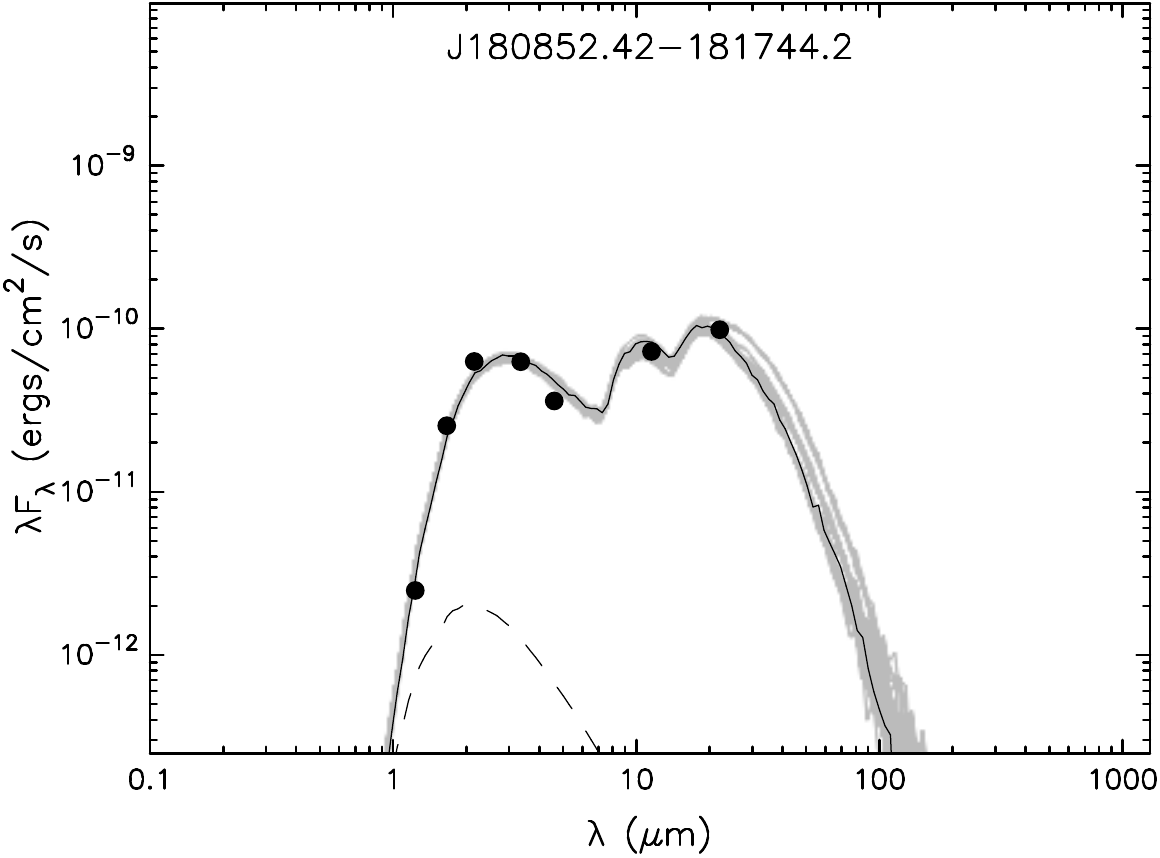}
\includegraphics[width=0.24\textwidth]{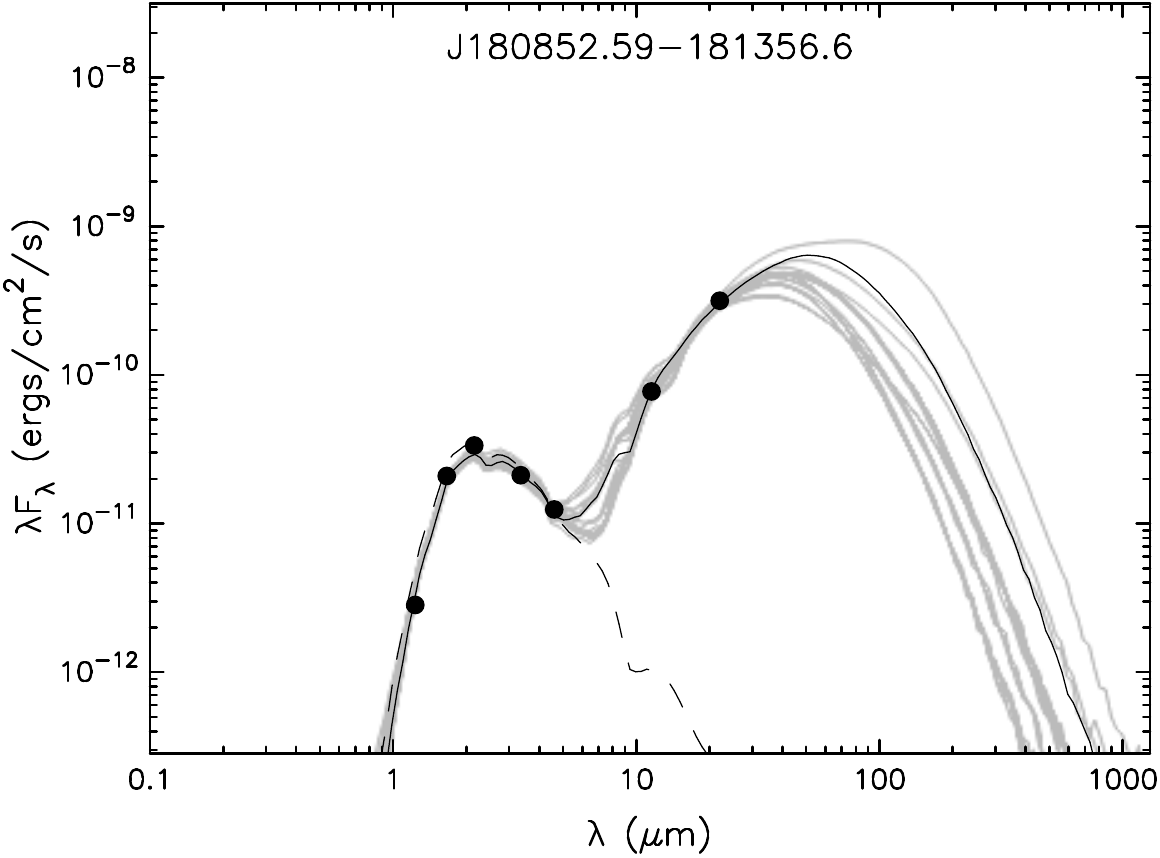}
\includegraphics[width=0.24\textwidth]{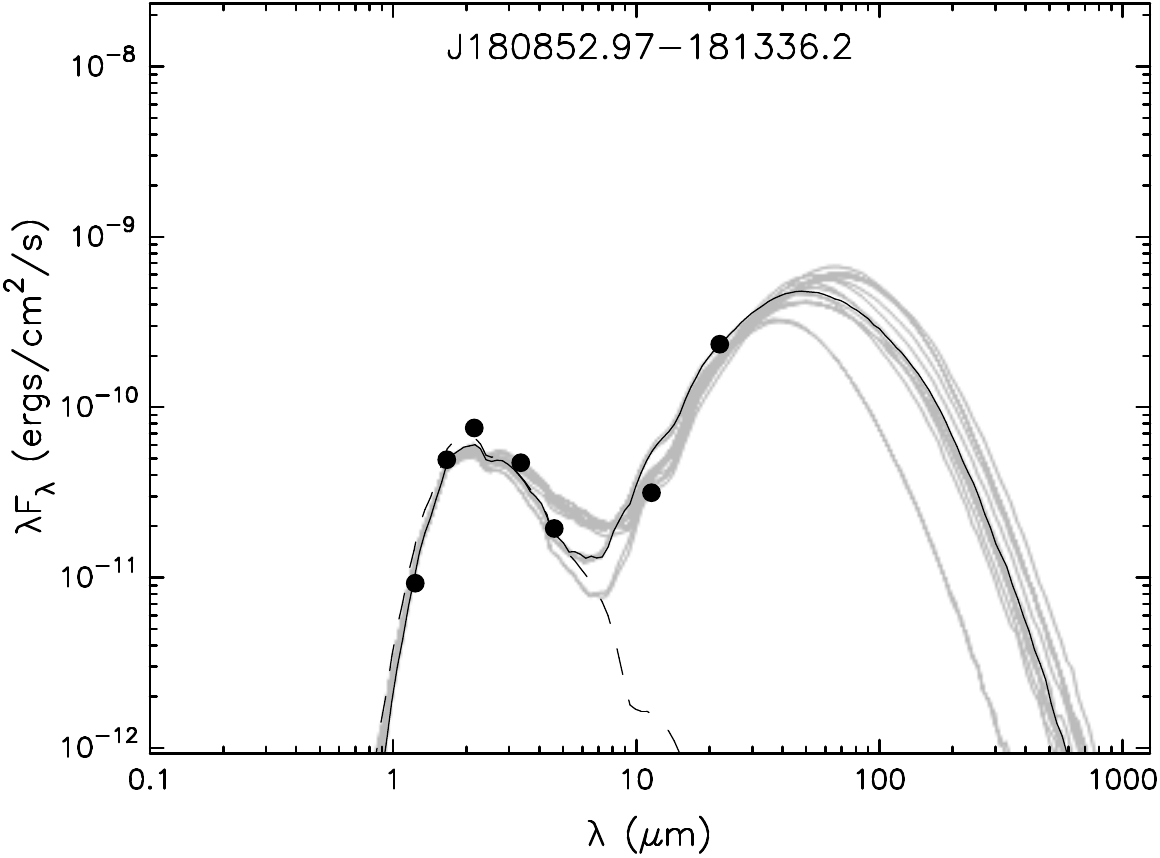}
\includegraphics[width=0.24\textwidth]{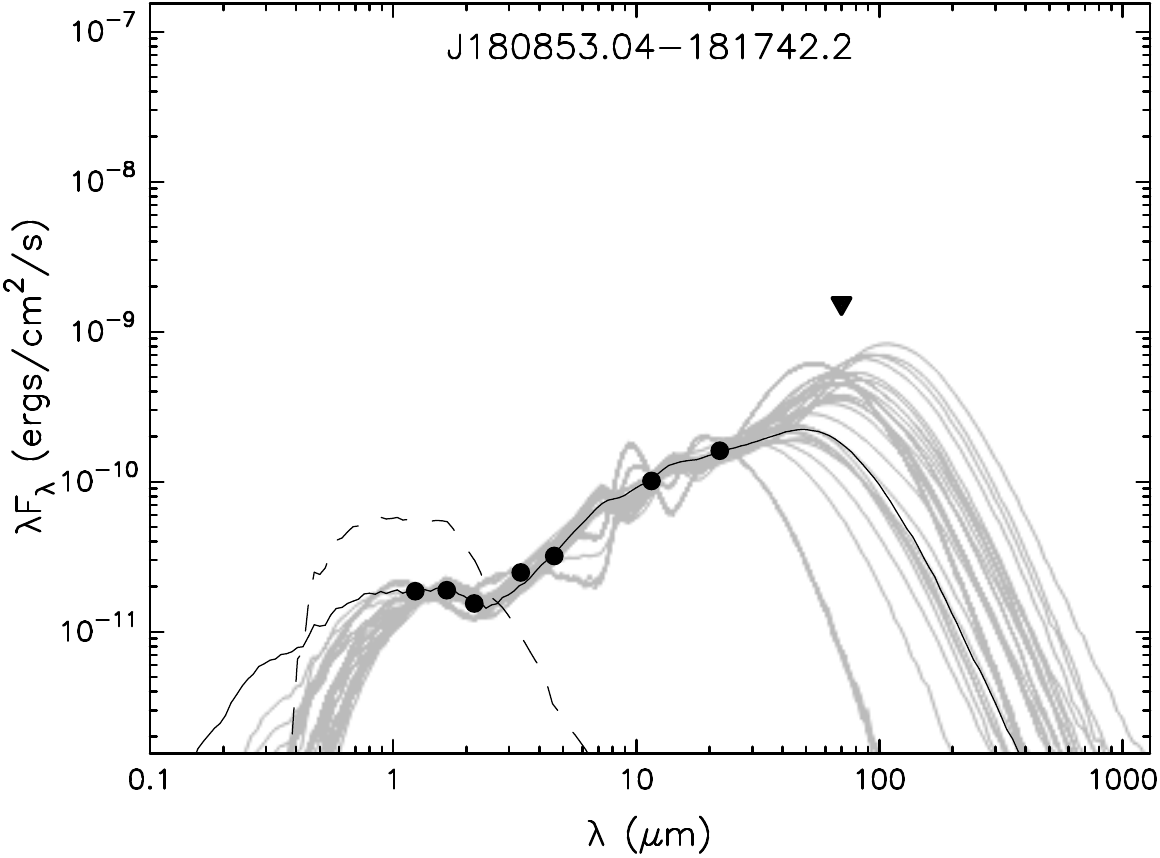}
\includegraphics[width=0.24\textwidth]{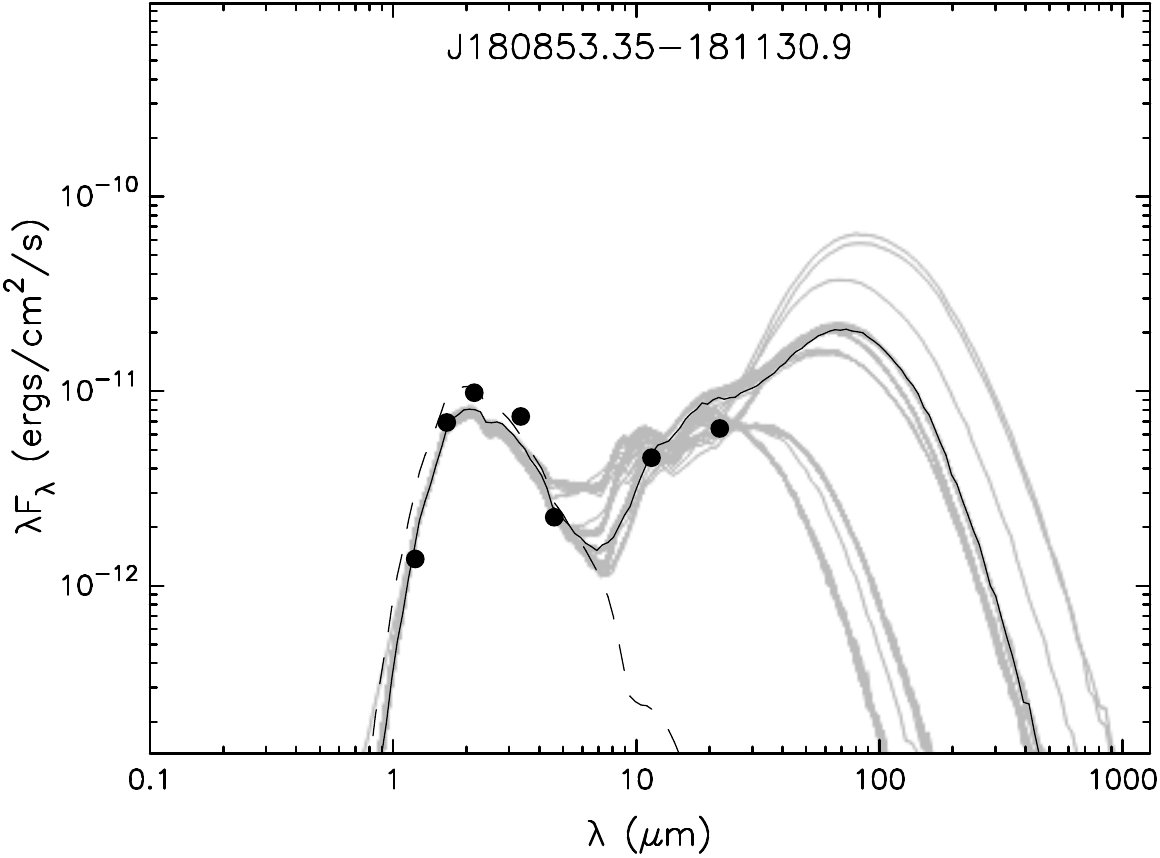}
\includegraphics[width=0.24\textwidth]{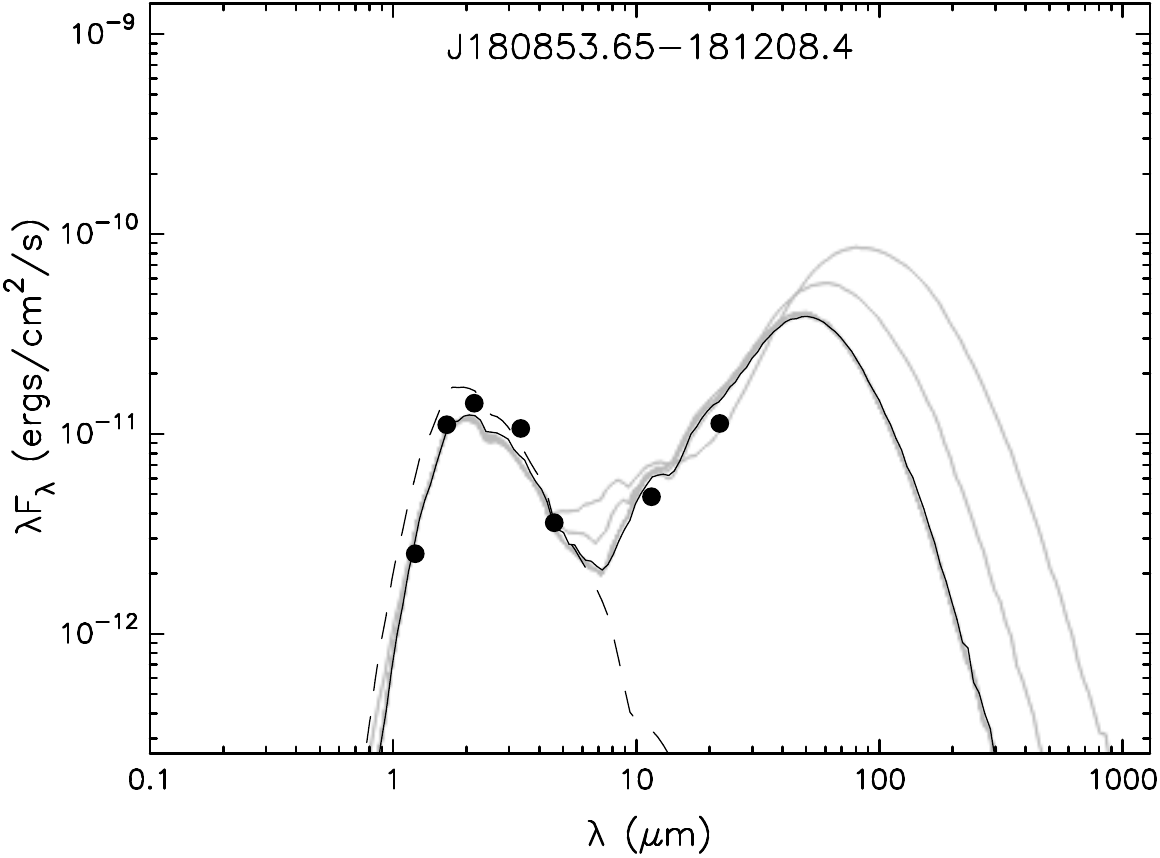}

\caption{ SED fitting of candidate YSOs.
The filled circles and triangle symbolize the input fluxes and the upper limit, respectively.
The black line shows the best fit. The grey lines show a set of fits that satisfy
$\chi^{2}-\chi_{best}^{2}< 2\times N_{\mathrm{data}}$. The dashed line means the stellar
photosphere corresponding to the central source of the best fitting model, as it
would look in the absence of circumstellar dust but with interstellar extinction.}
\label{sed-1}
\end{figure*}

\begin{figure*}
  \centering
\includegraphics[width=0.24\textwidth]{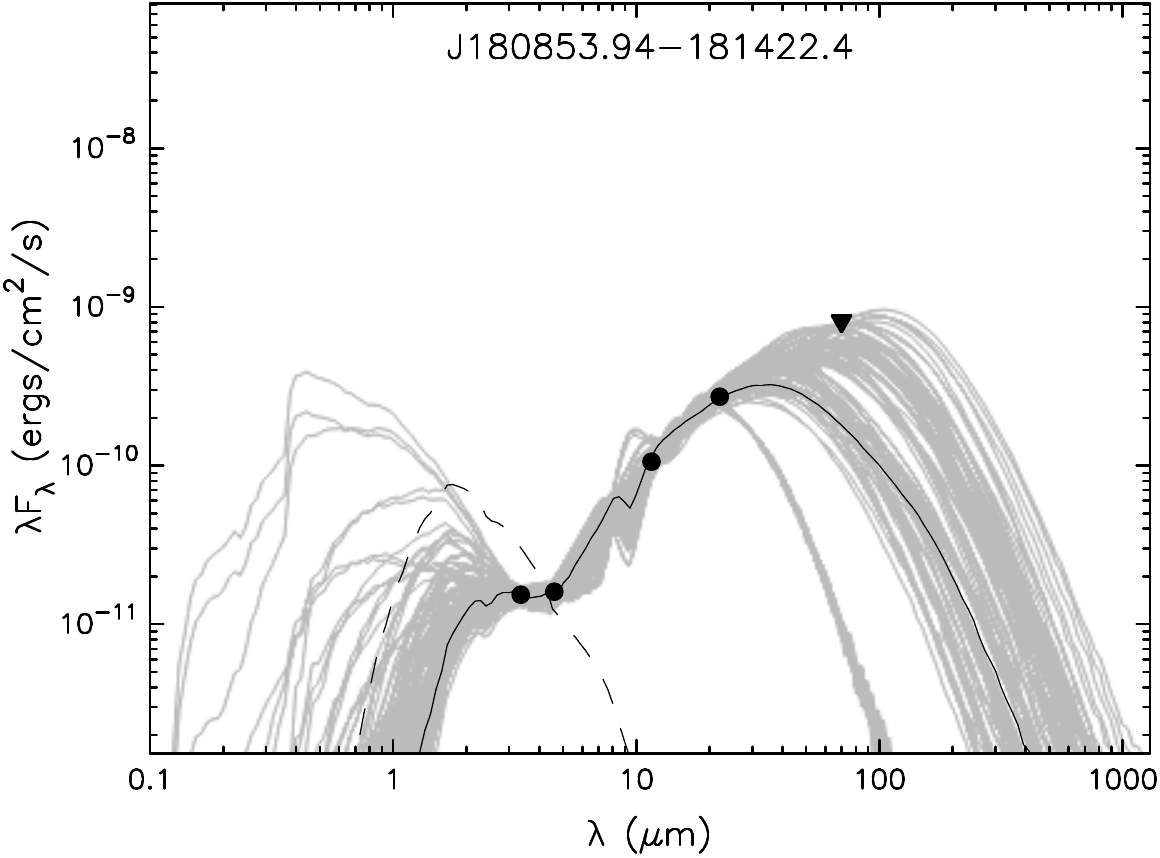}
\includegraphics[width=0.24\textwidth]{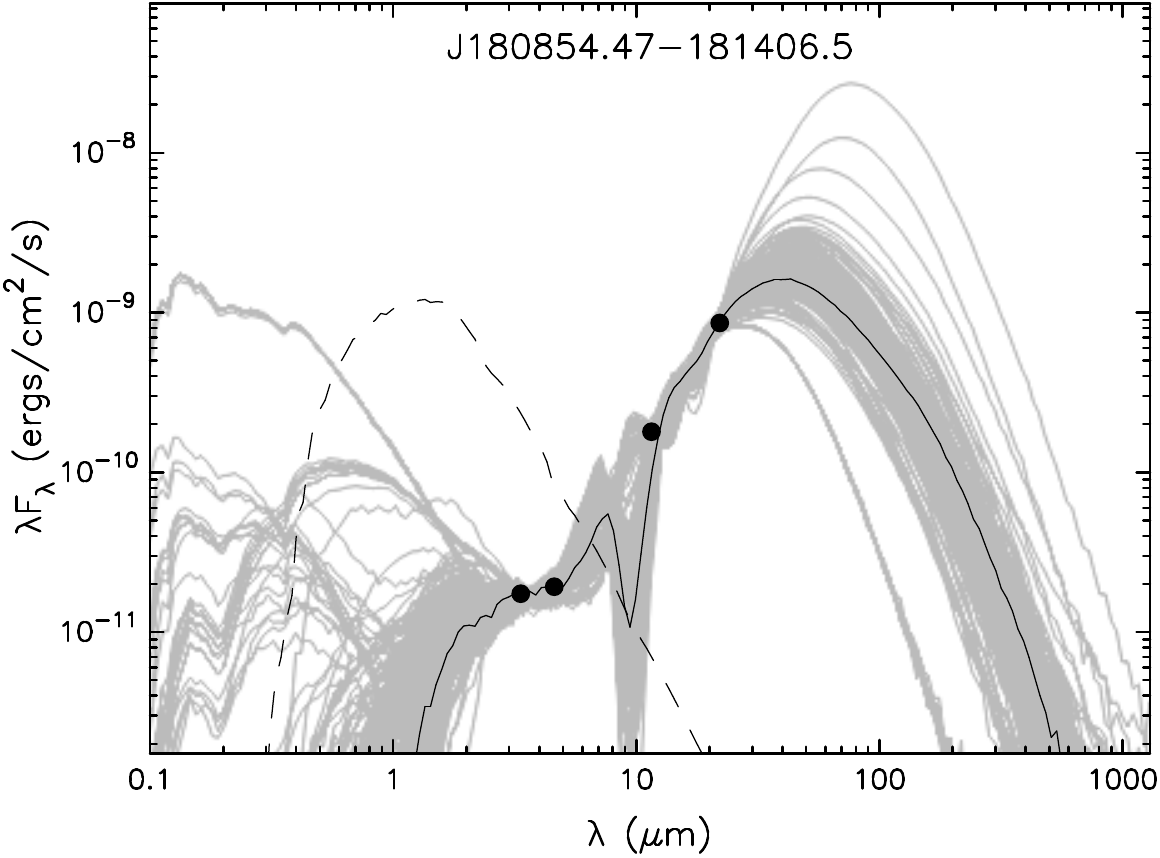}
\includegraphics[width=0.24\textwidth]{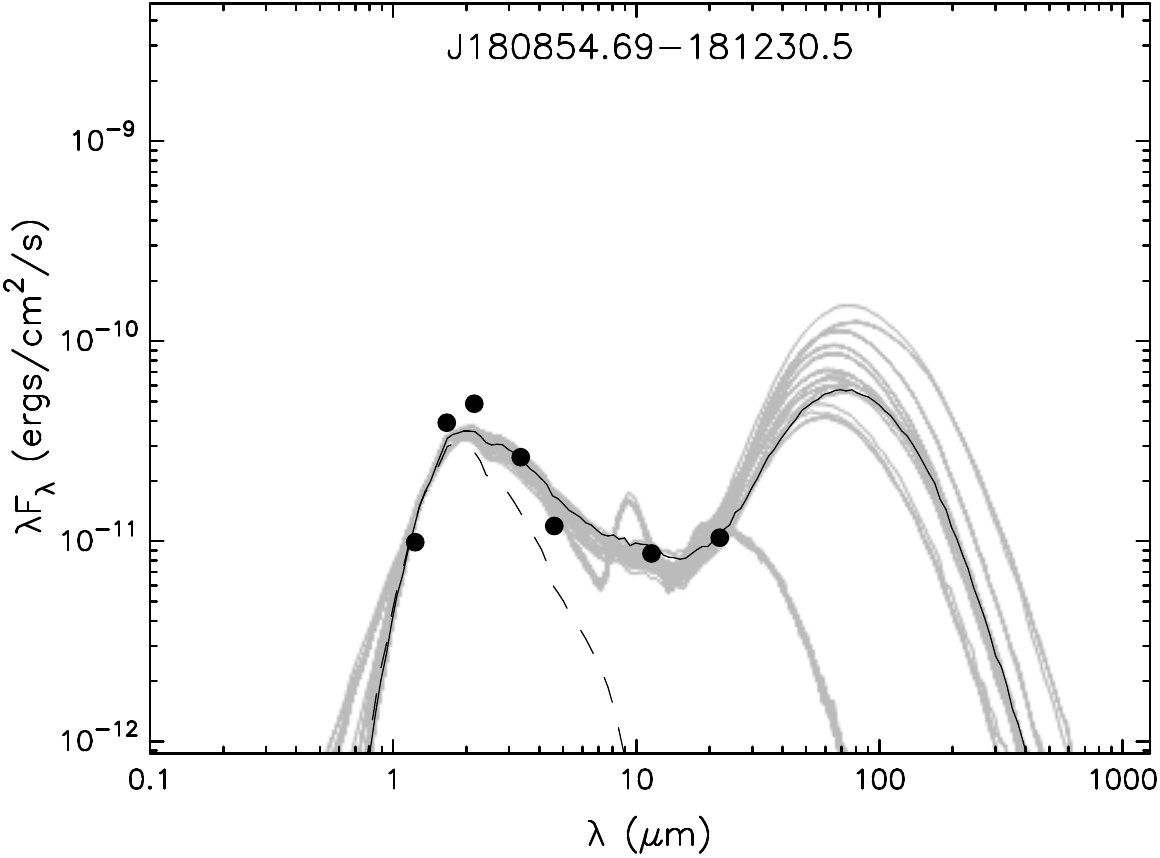}
\includegraphics[width=0.24\textwidth]{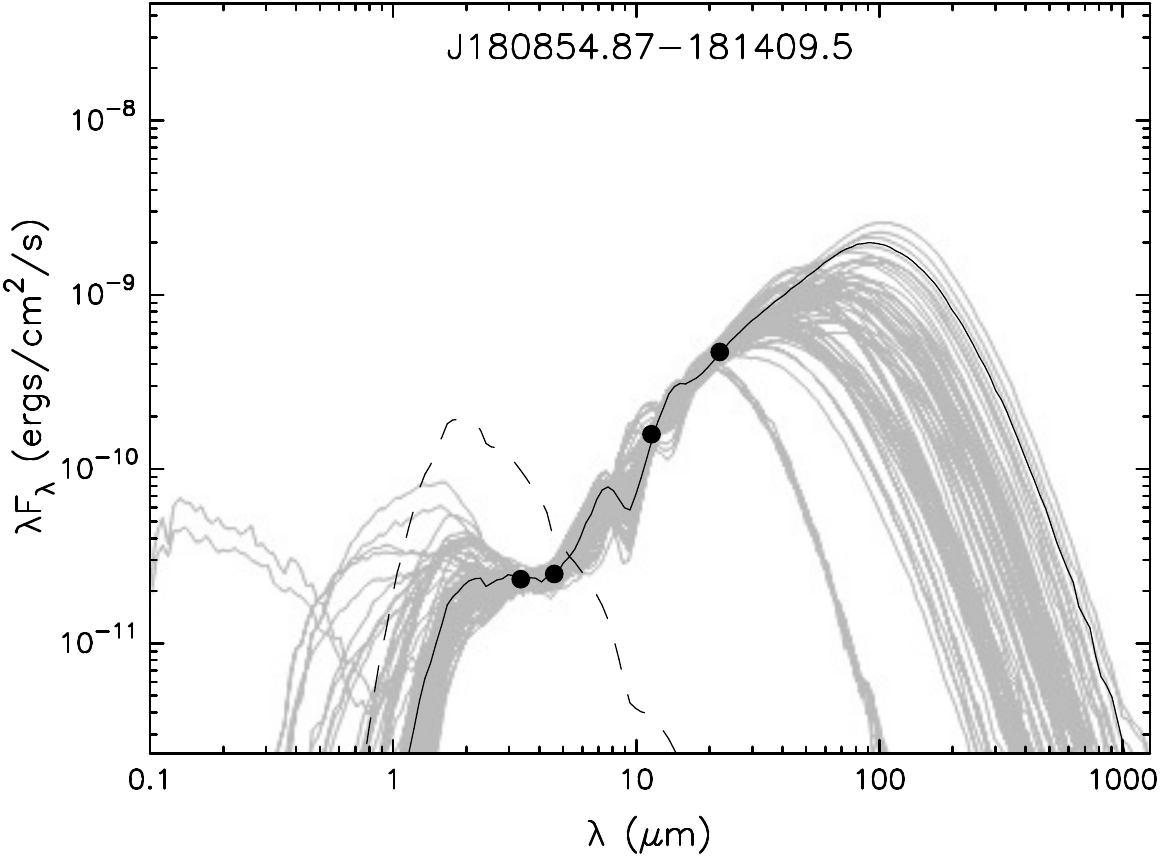}
\includegraphics[width=0.24\textwidth]{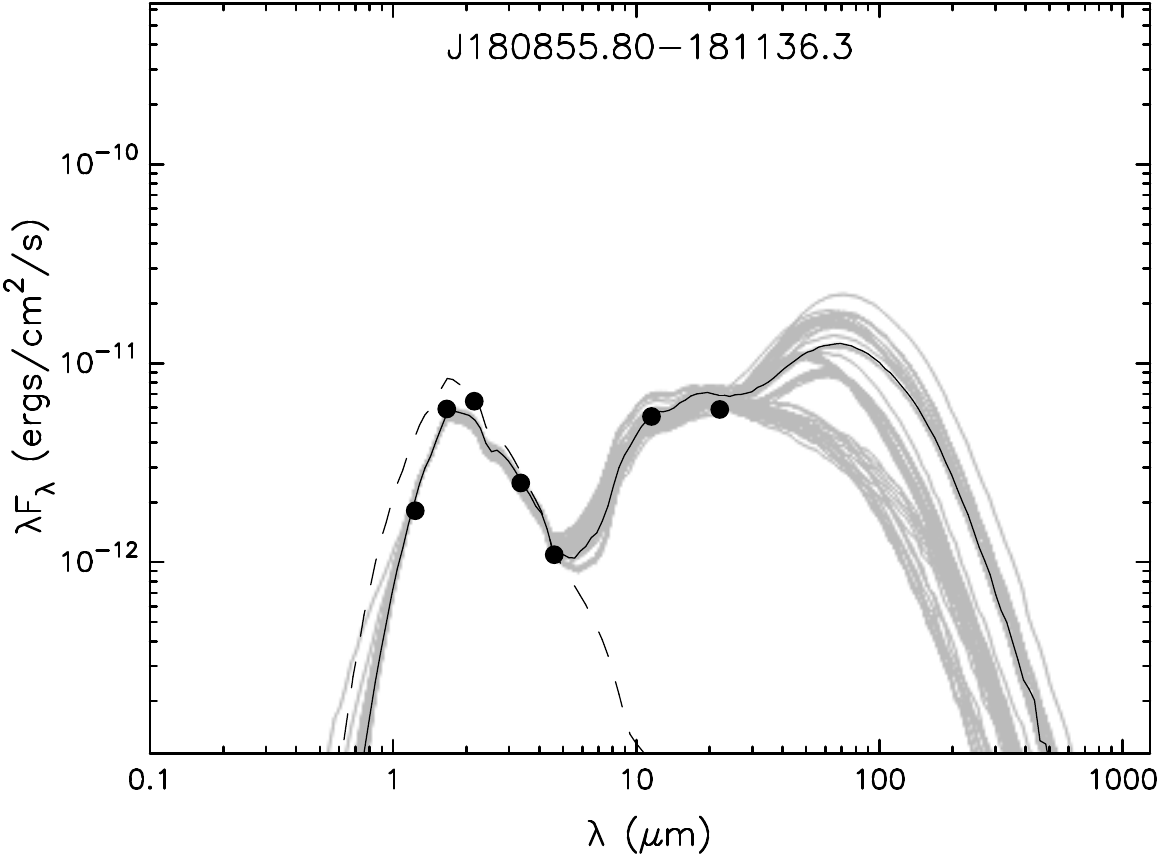}
\includegraphics[width=0.24\textwidth]{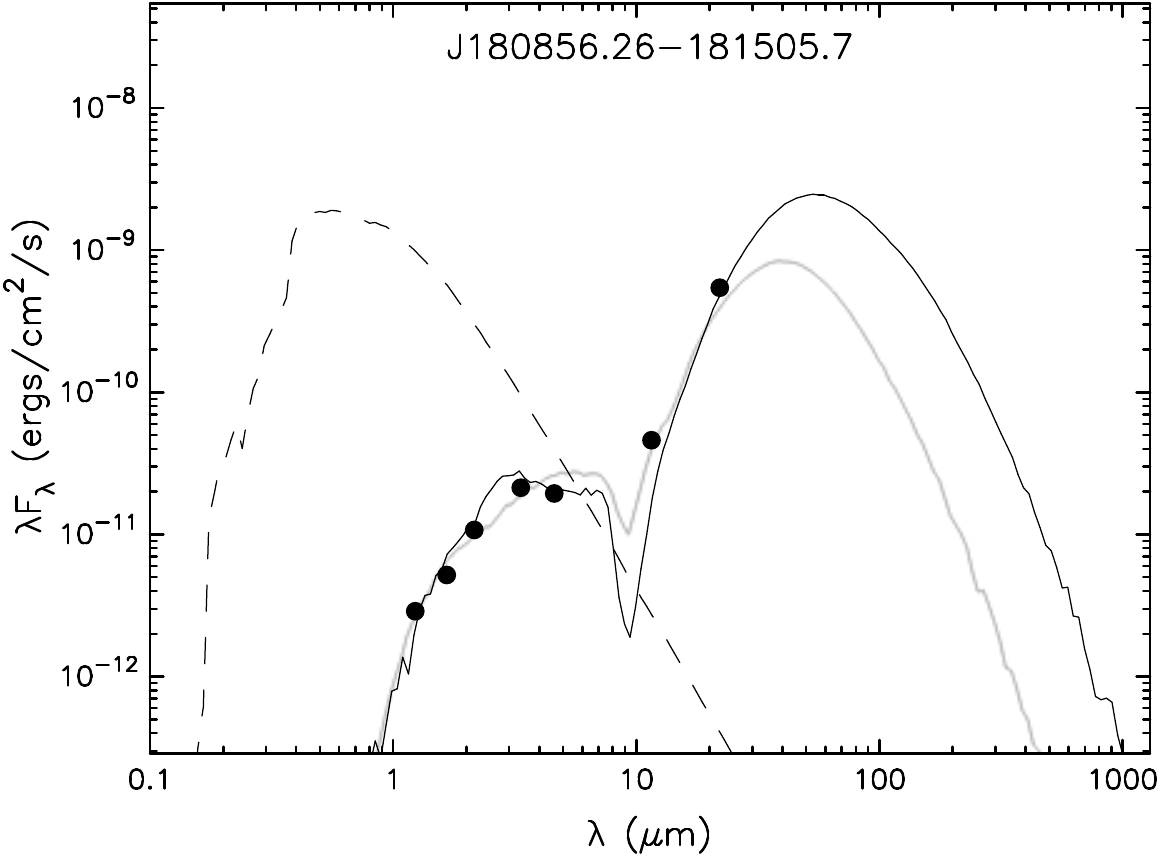}
\includegraphics[width=0.24\textwidth]{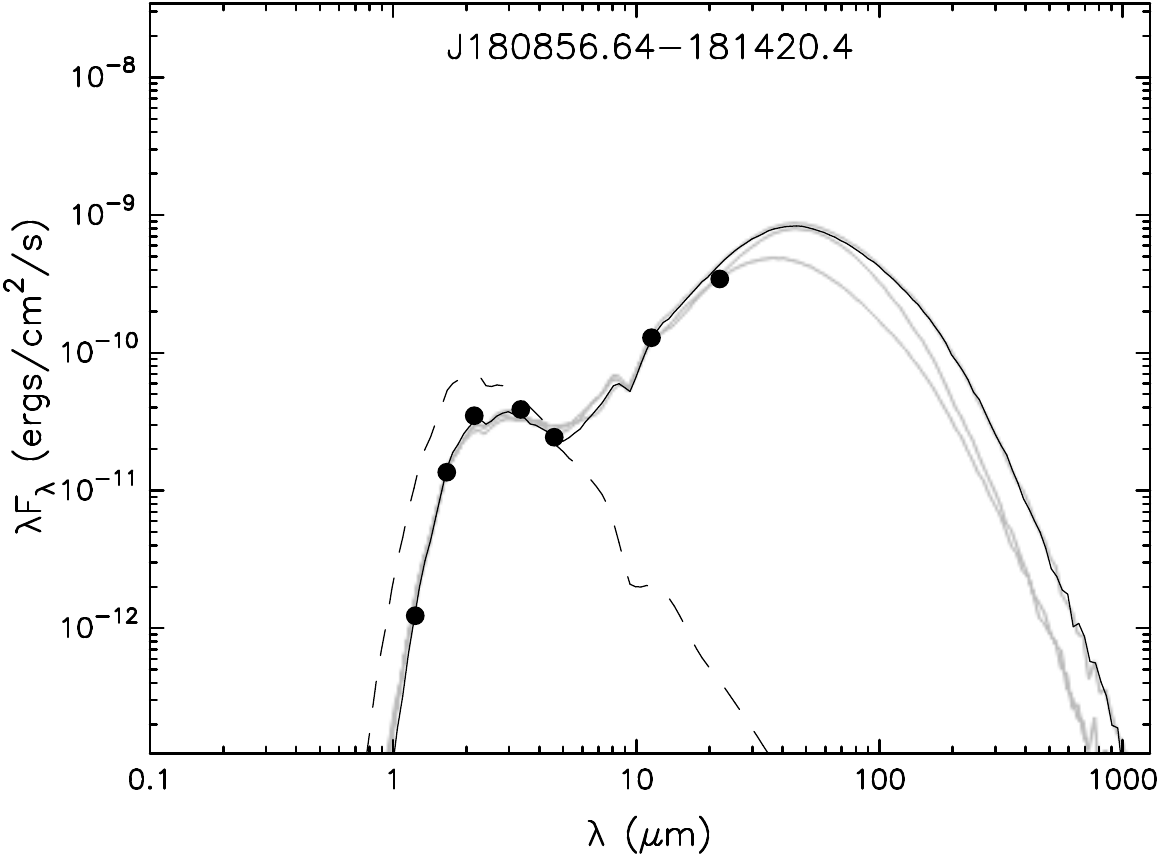}
\includegraphics[width=0.24\textwidth]{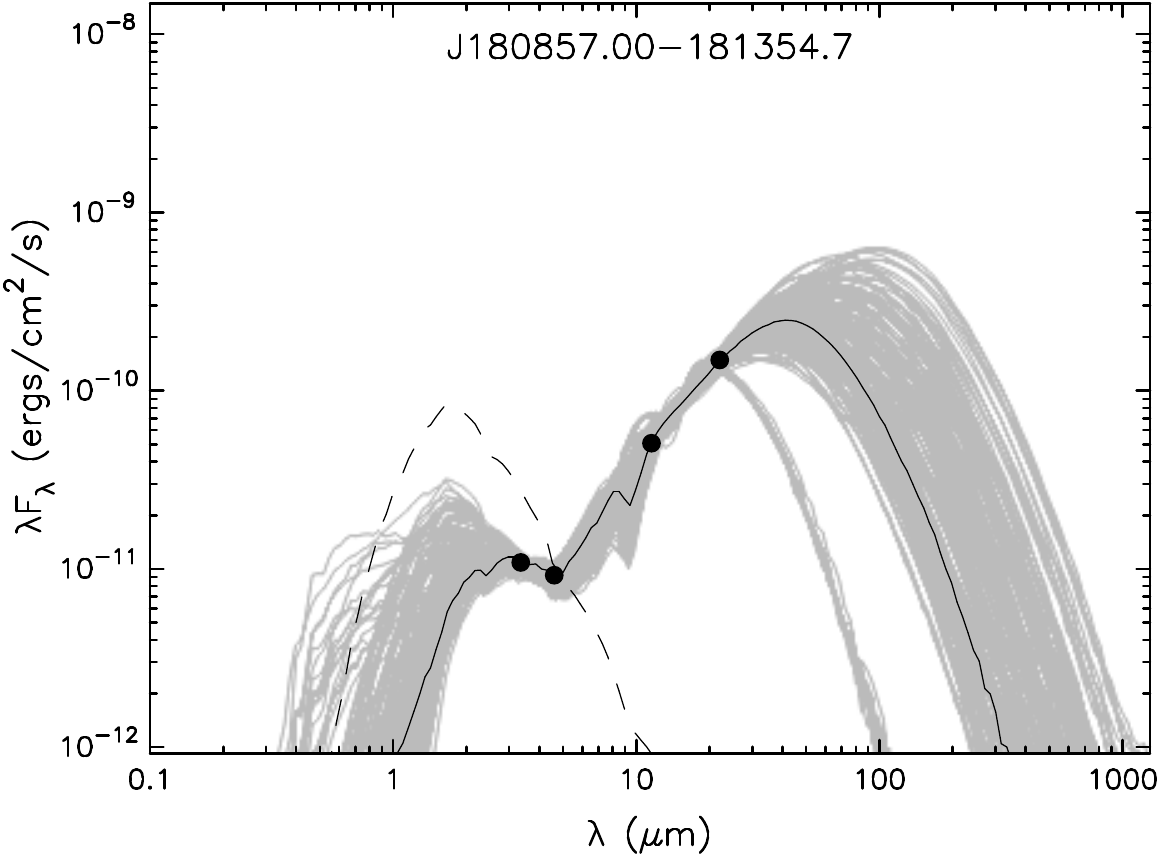}
\includegraphics[width=0.24\textwidth]{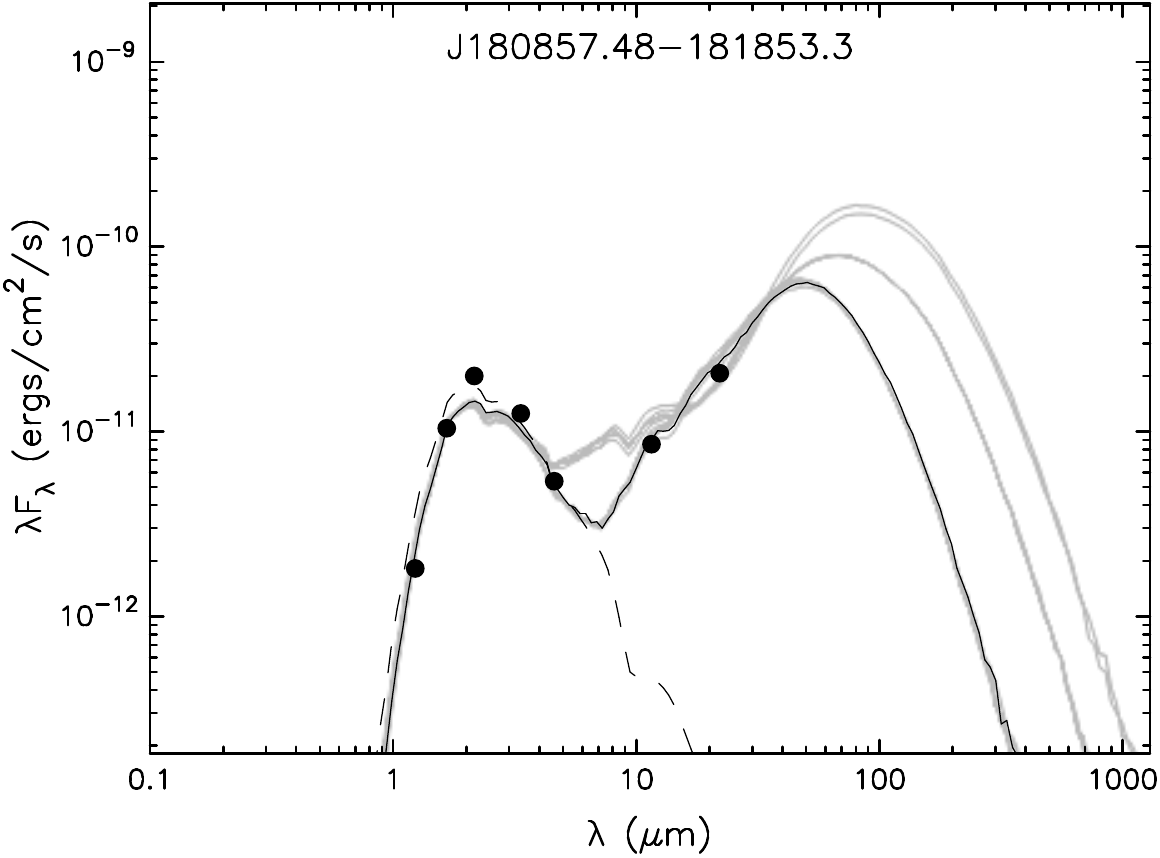}
\includegraphics[width=0.24\textwidth]{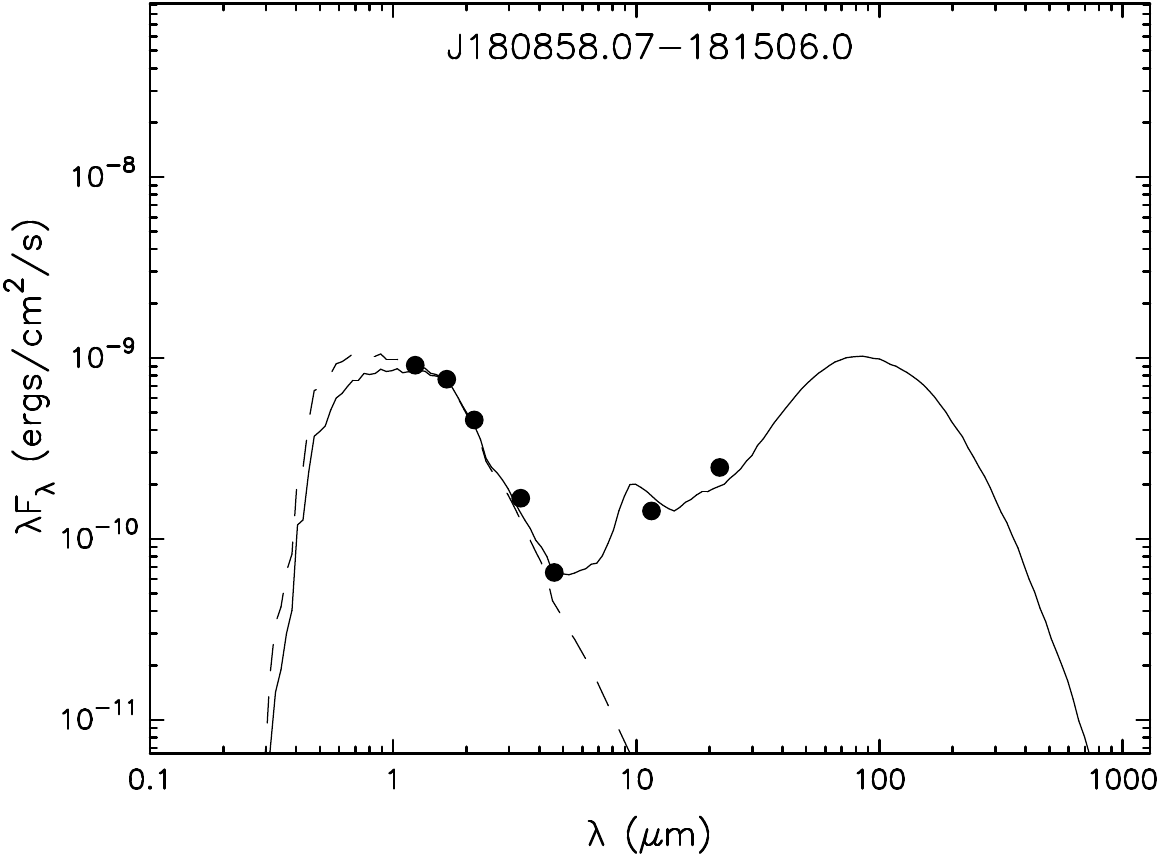}
\includegraphics[width=0.24\textwidth]{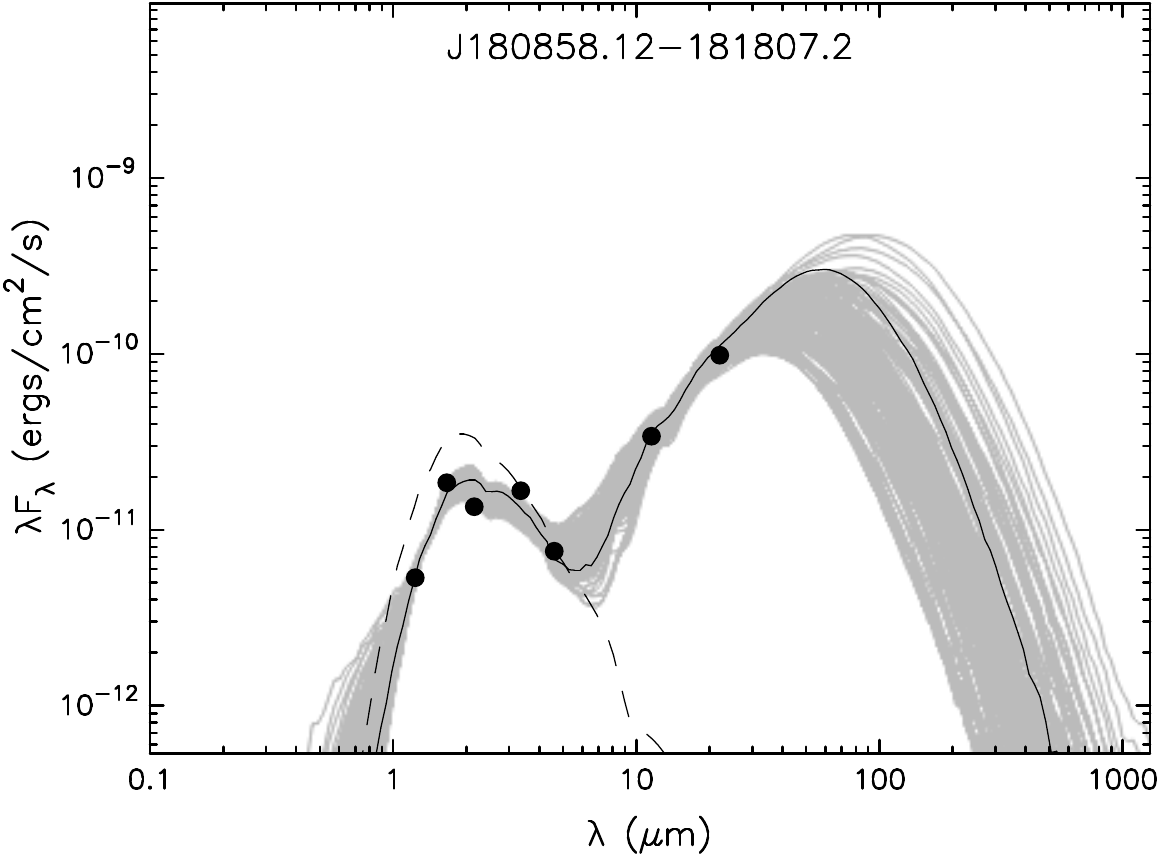}
\includegraphics[width=0.24\textwidth]{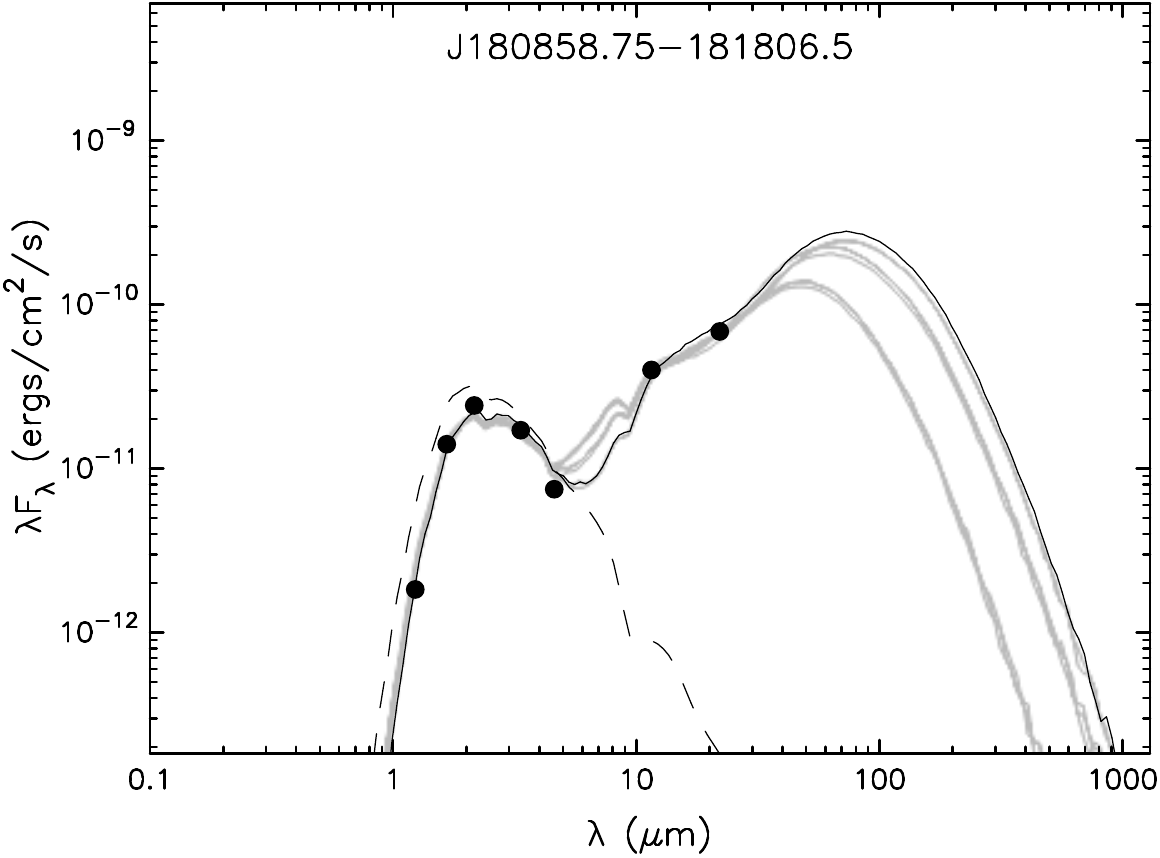}
\includegraphics[width=0.24\textwidth]{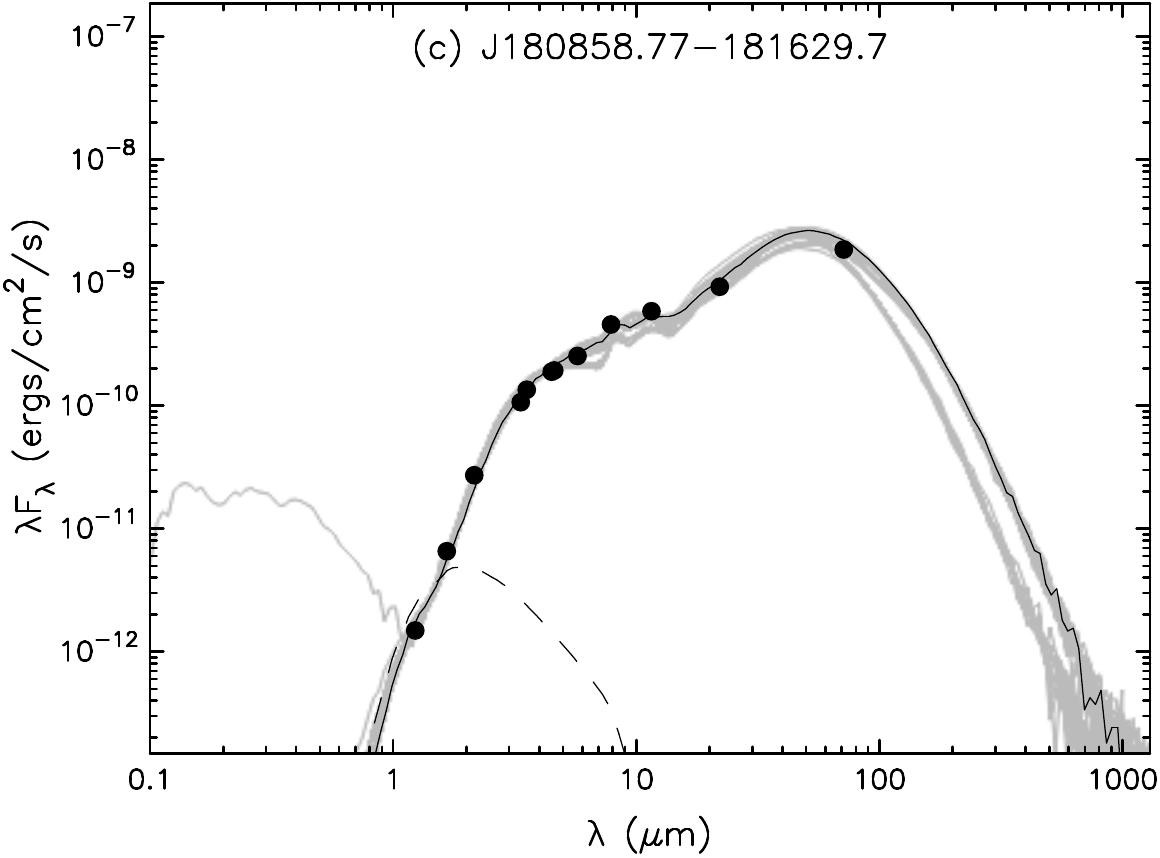}
\includegraphics[width=0.24\textwidth]{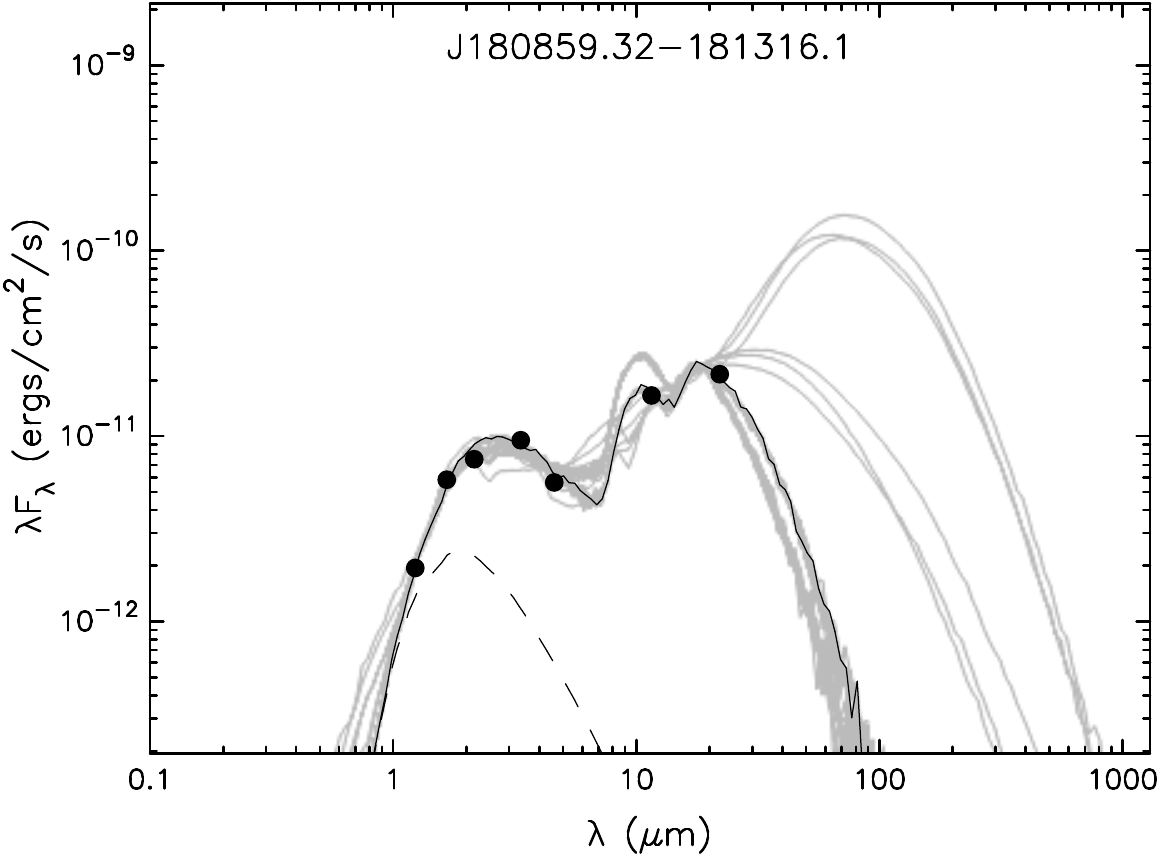}
\includegraphics[width=0.24\textwidth]{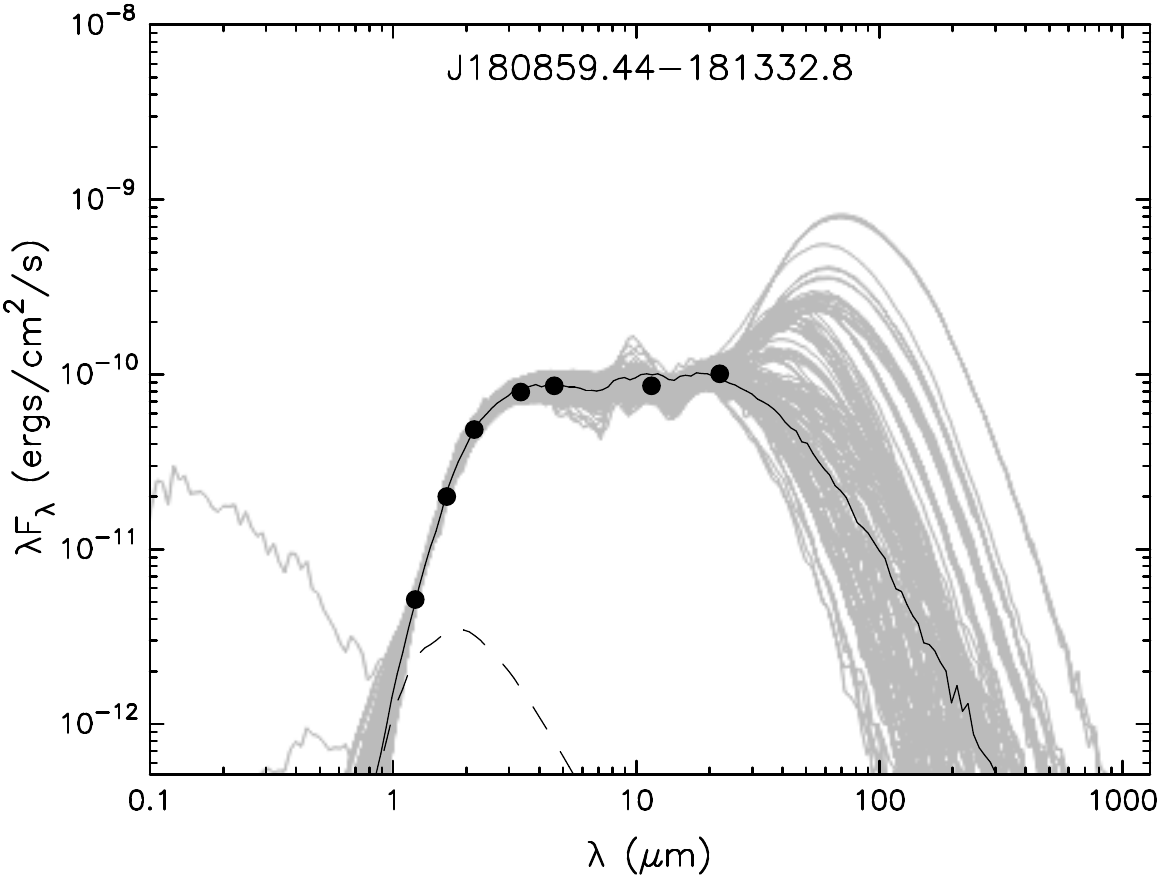}
\includegraphics[width=0.24\textwidth]{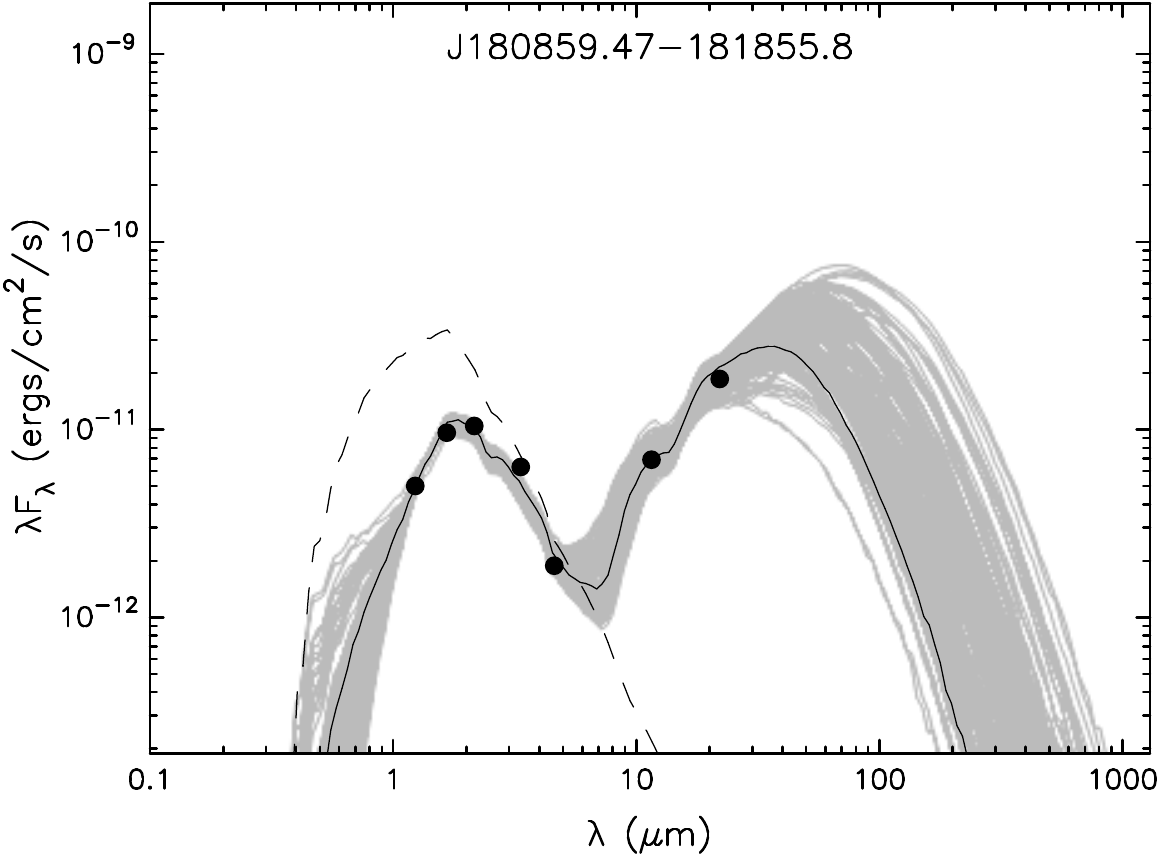}
\includegraphics[width=0.24\textwidth]{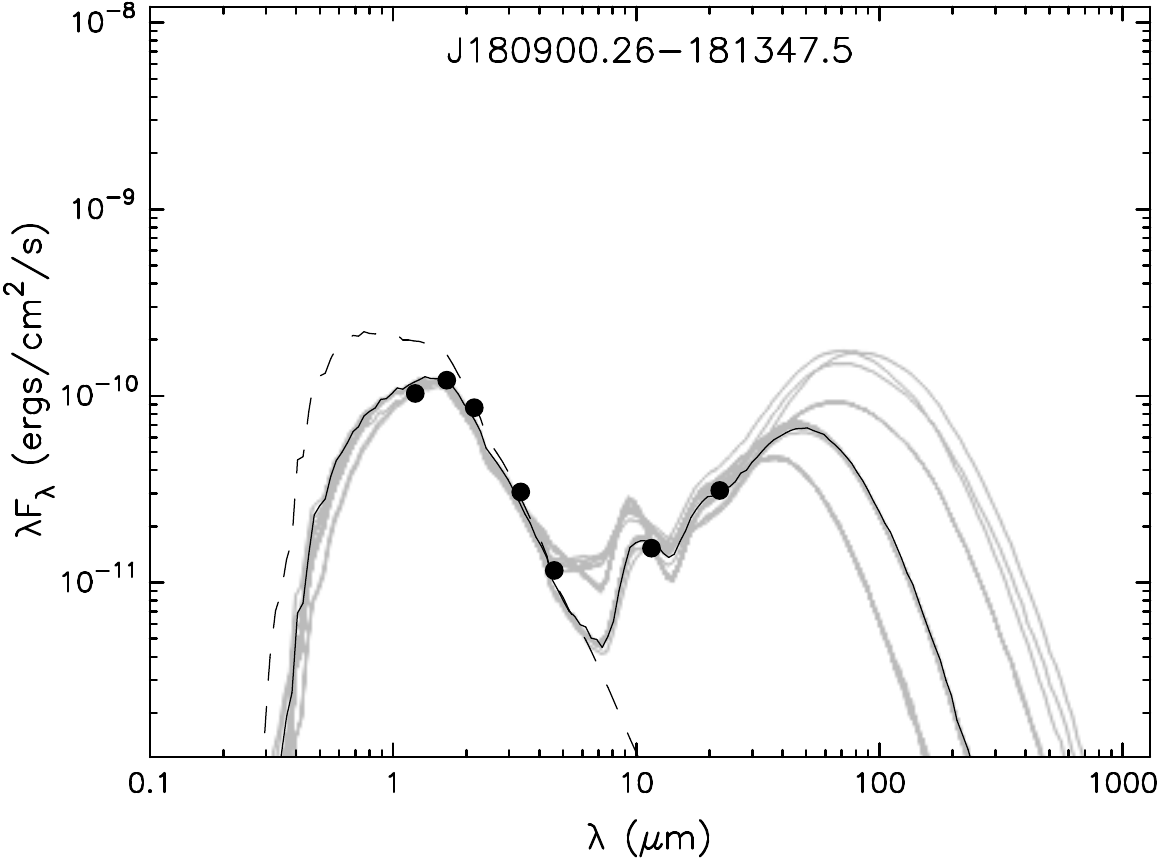}
\includegraphics[width=0.24\textwidth]{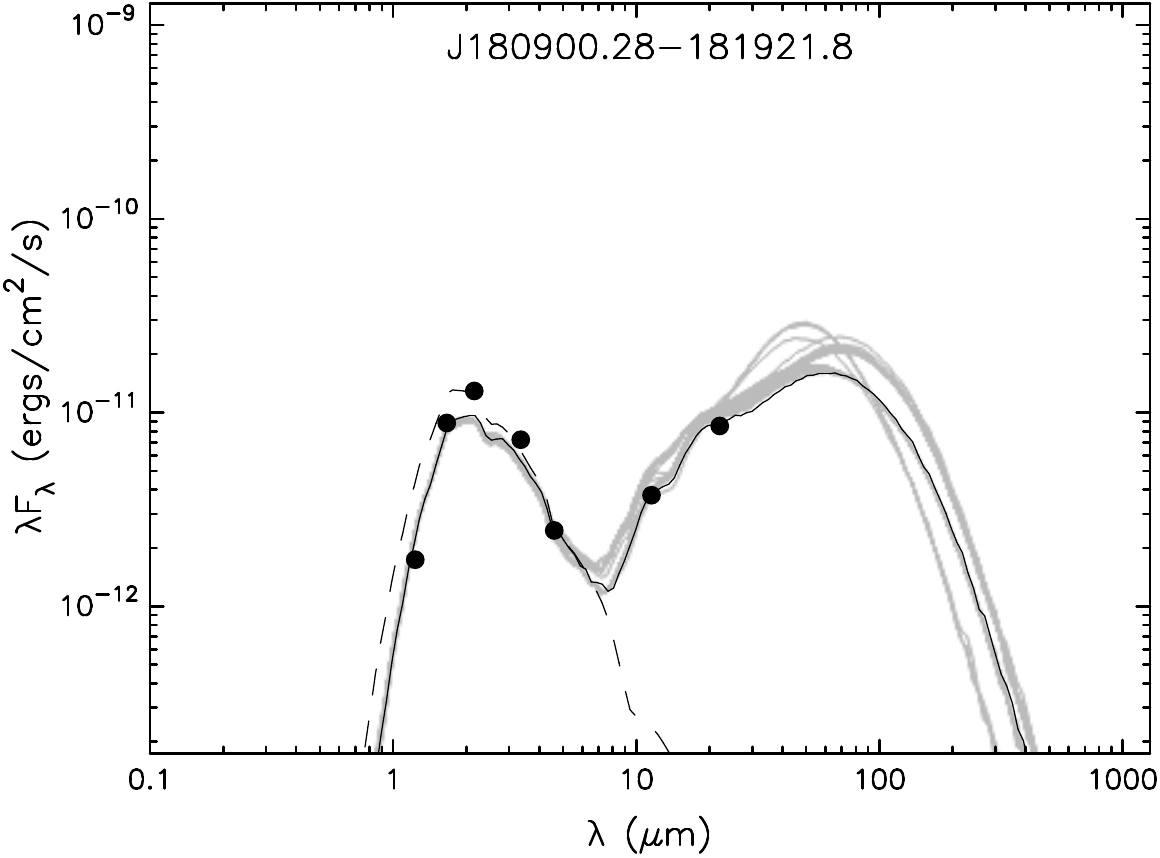}
\includegraphics[width=0.24\textwidth]{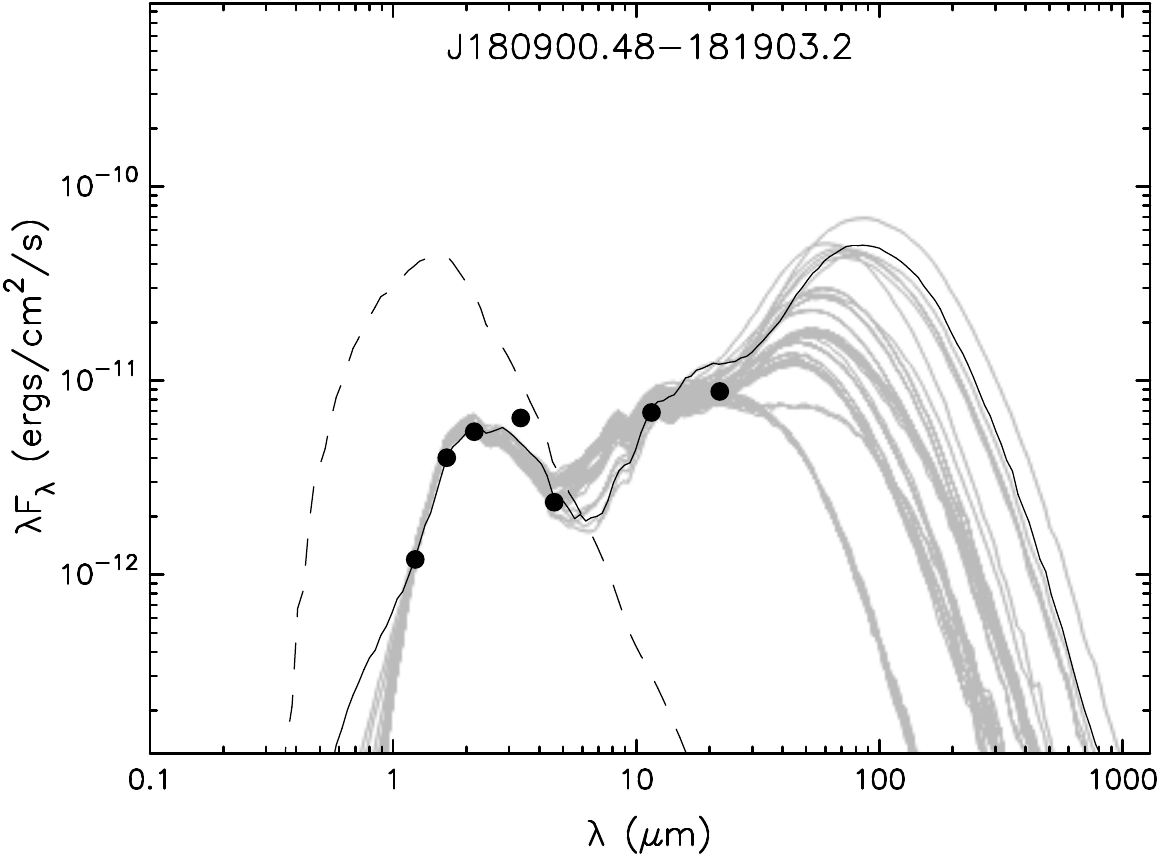}
\includegraphics[width=0.24\textwidth]{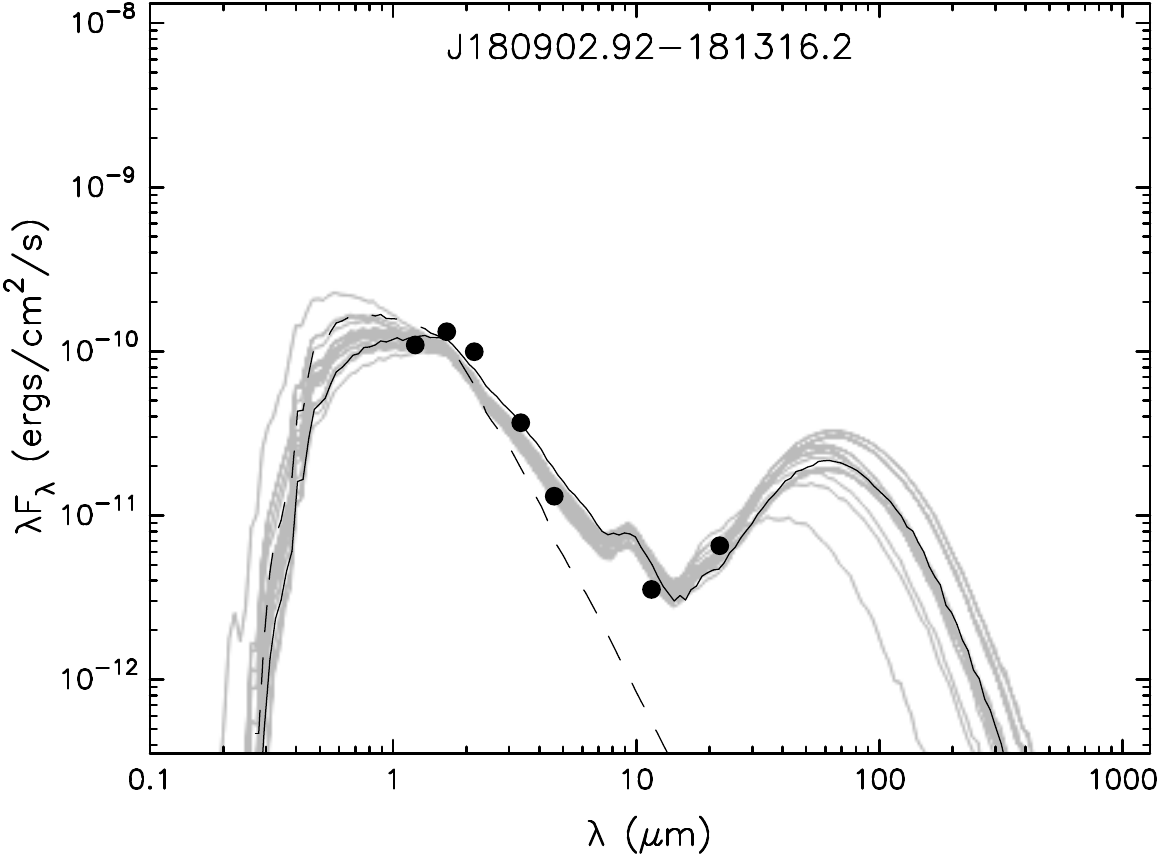}
\includegraphics[width=0.24\textwidth]{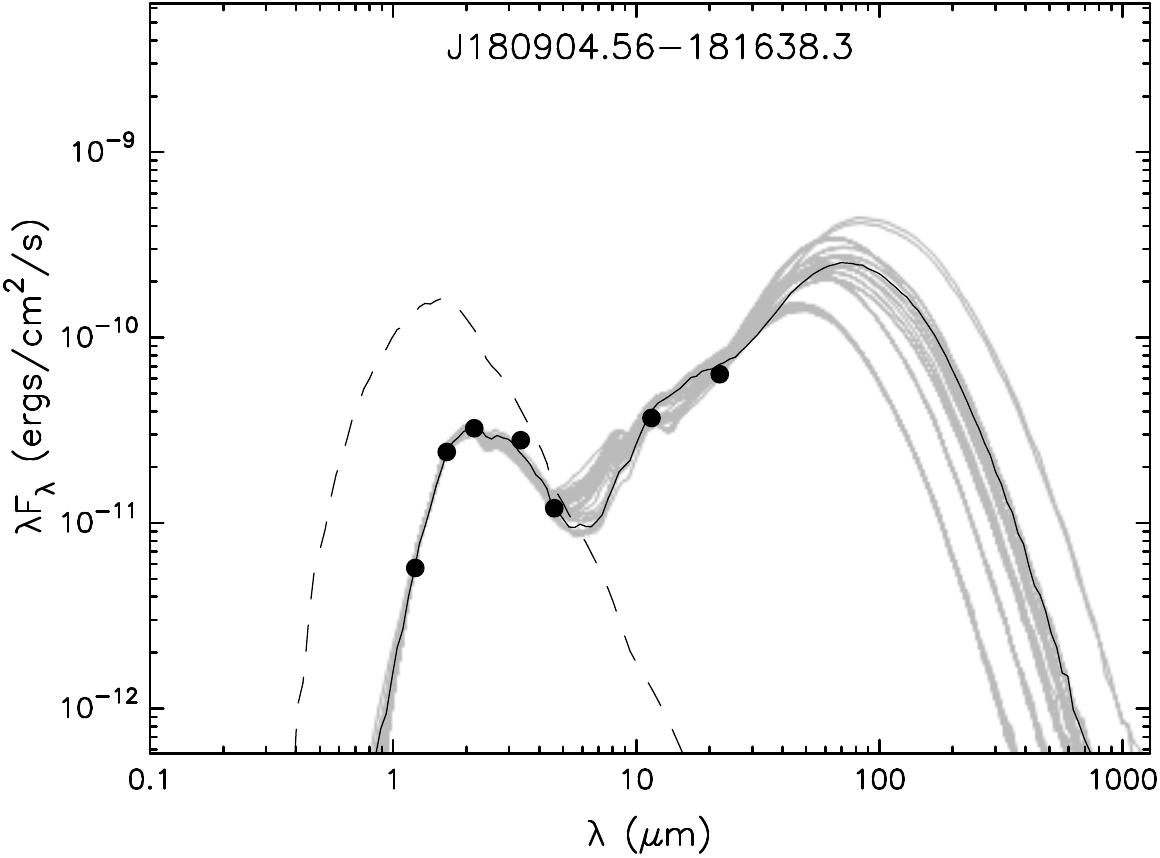}
\includegraphics[width=0.24\textwidth]{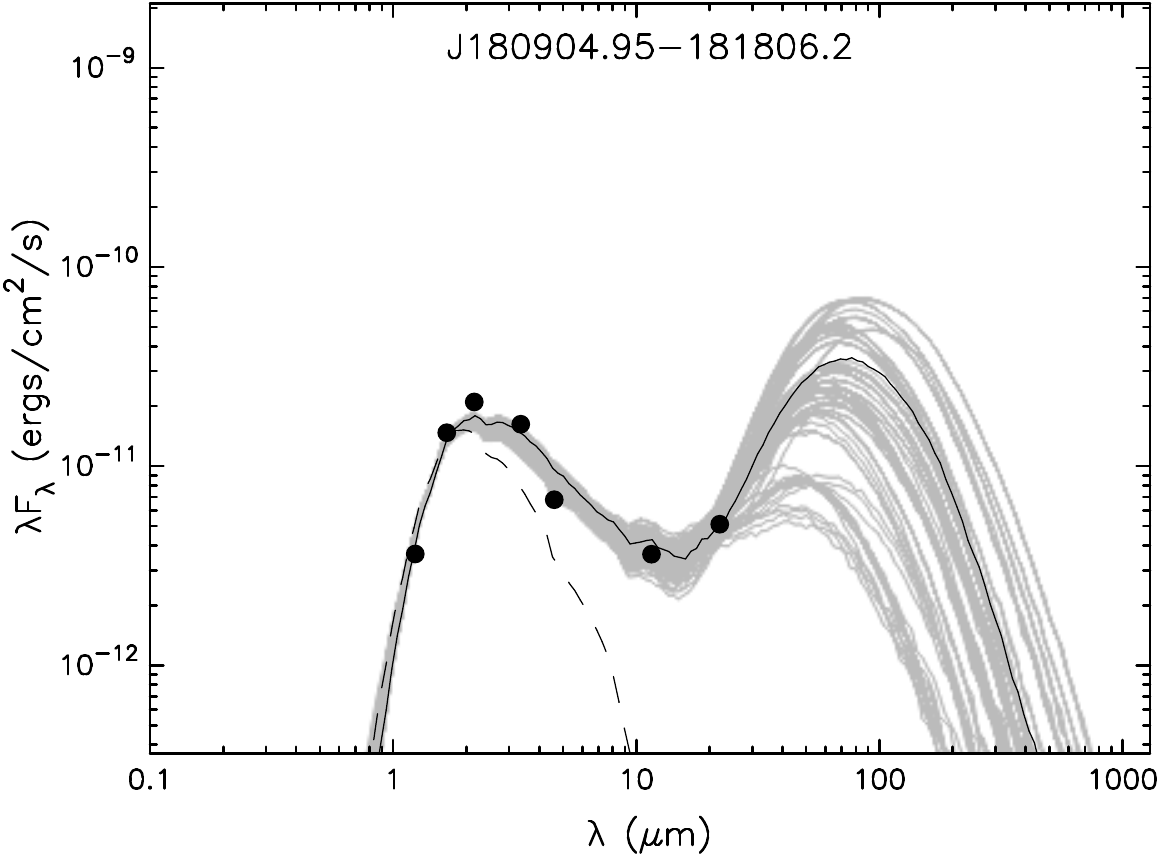}
\includegraphics[width=0.24\textwidth]{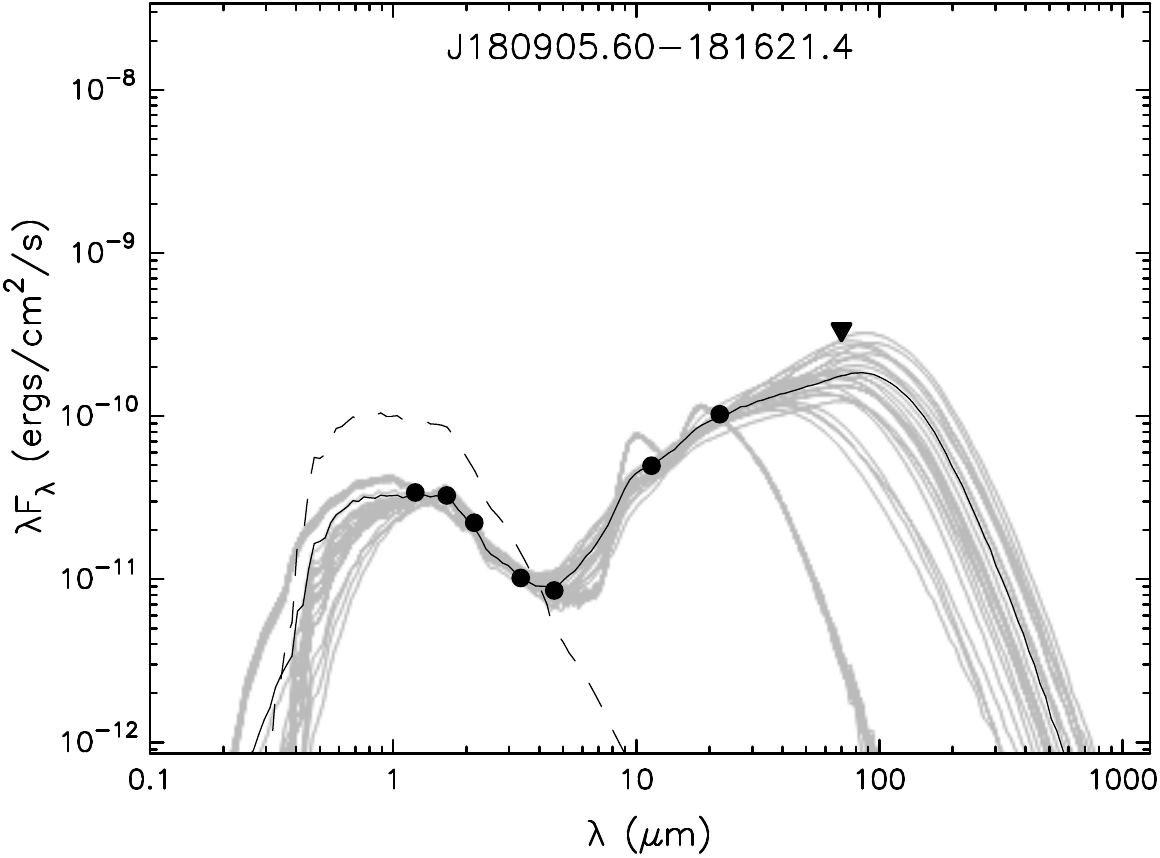}
\includegraphics[width=0.24\textwidth]{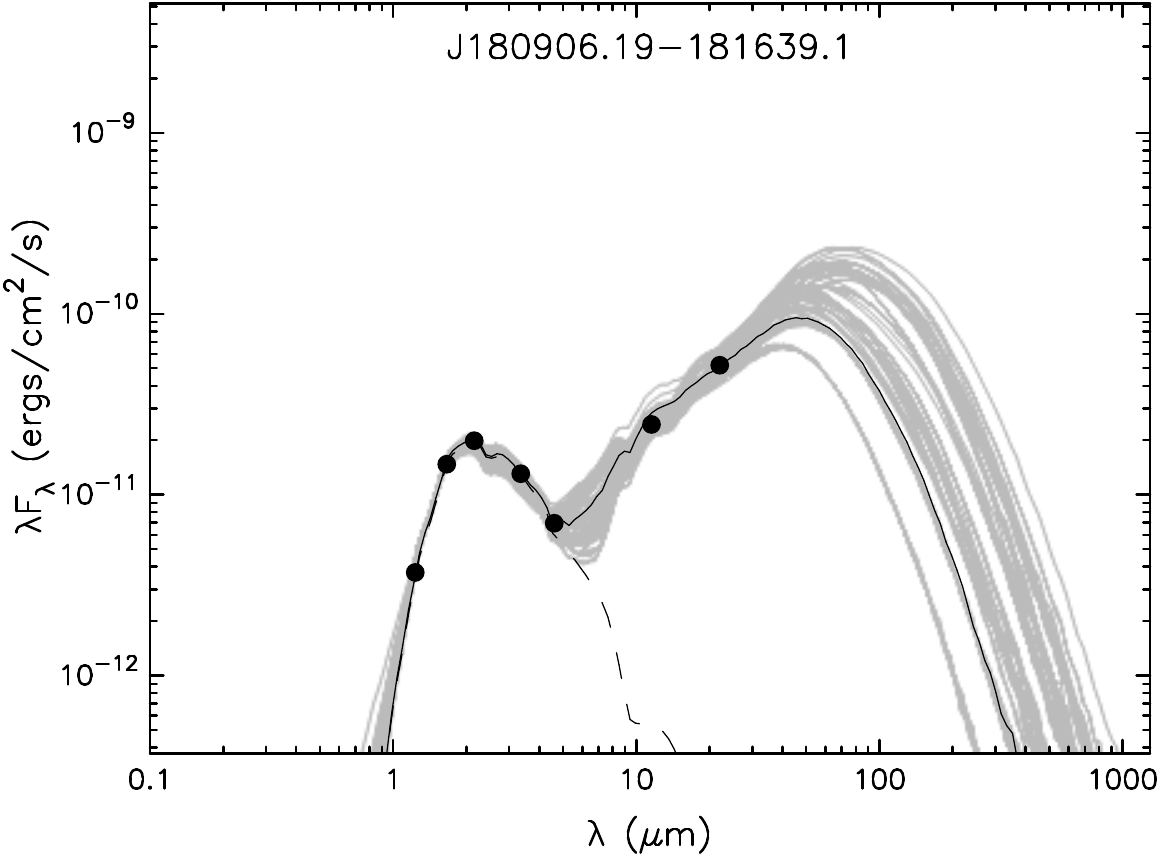}
\includegraphics[width=0.24\textwidth]{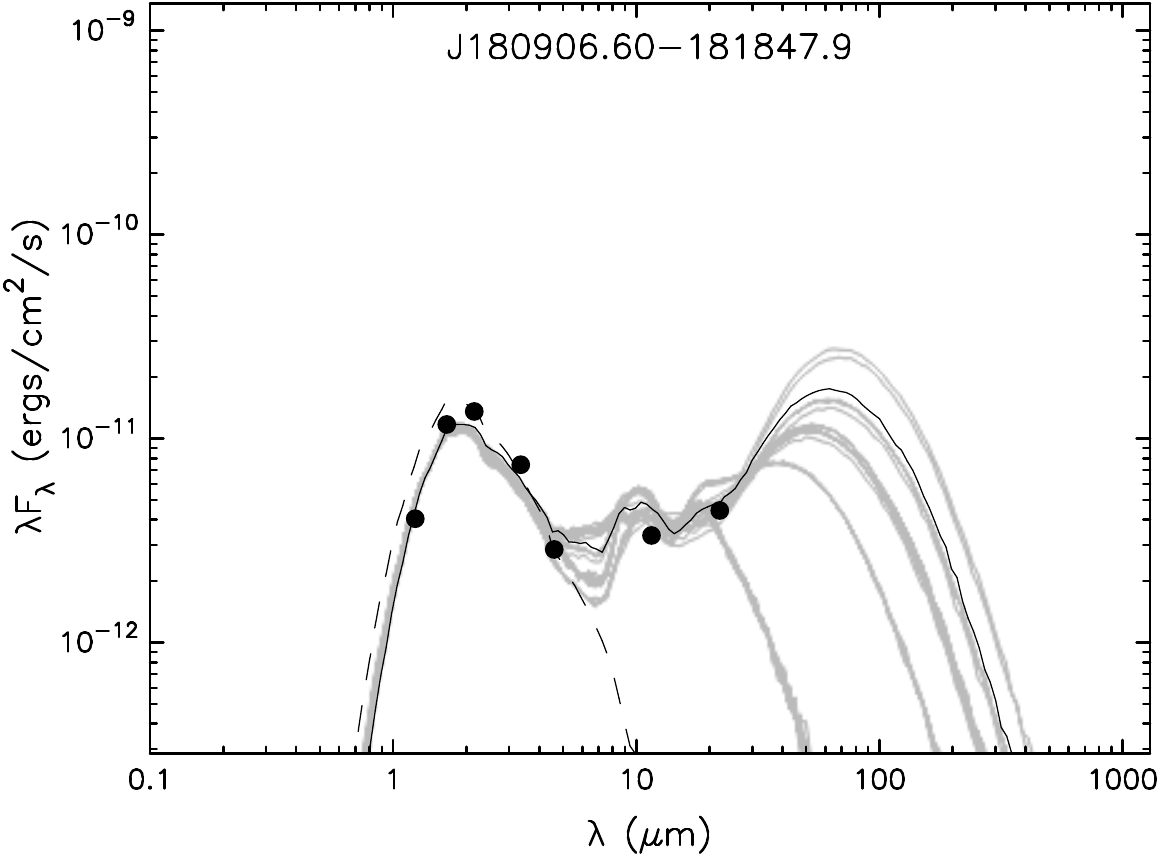}
\includegraphics[width=0.24\textwidth]{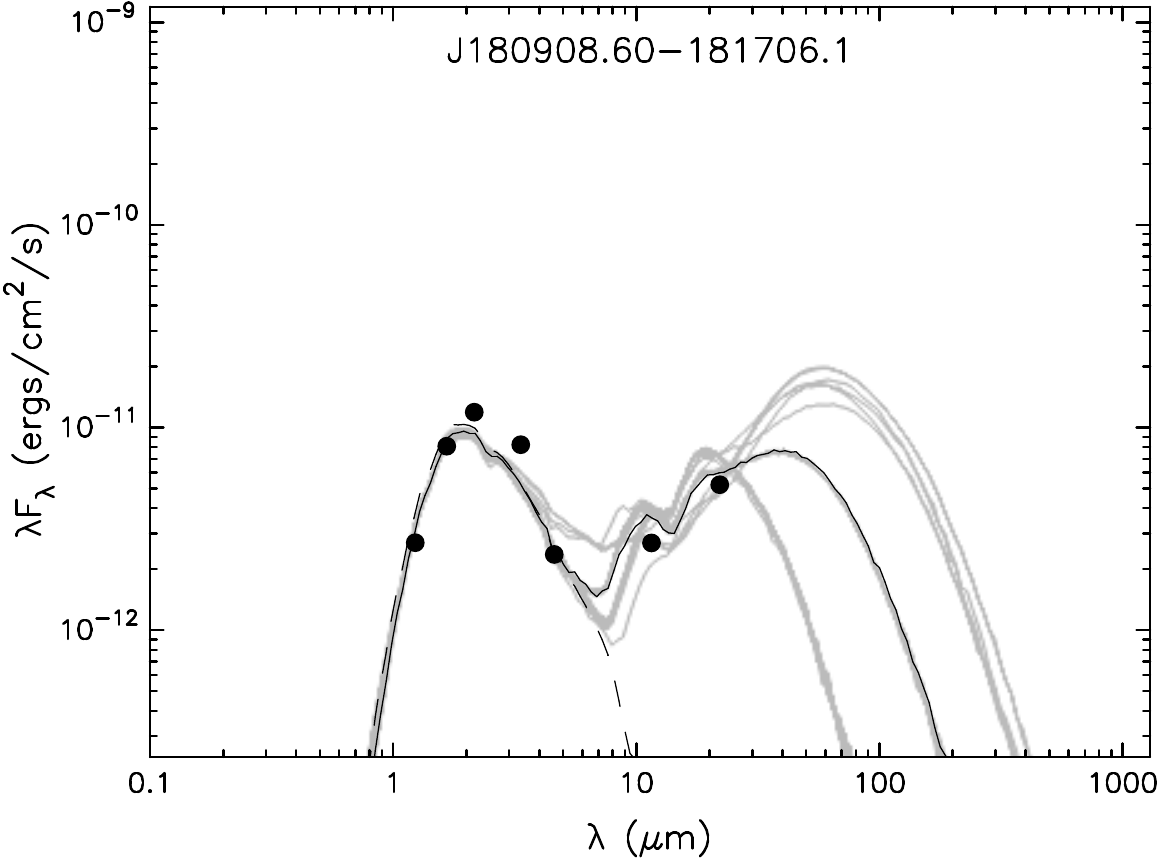}
\includegraphics[width=0.24\textwidth]{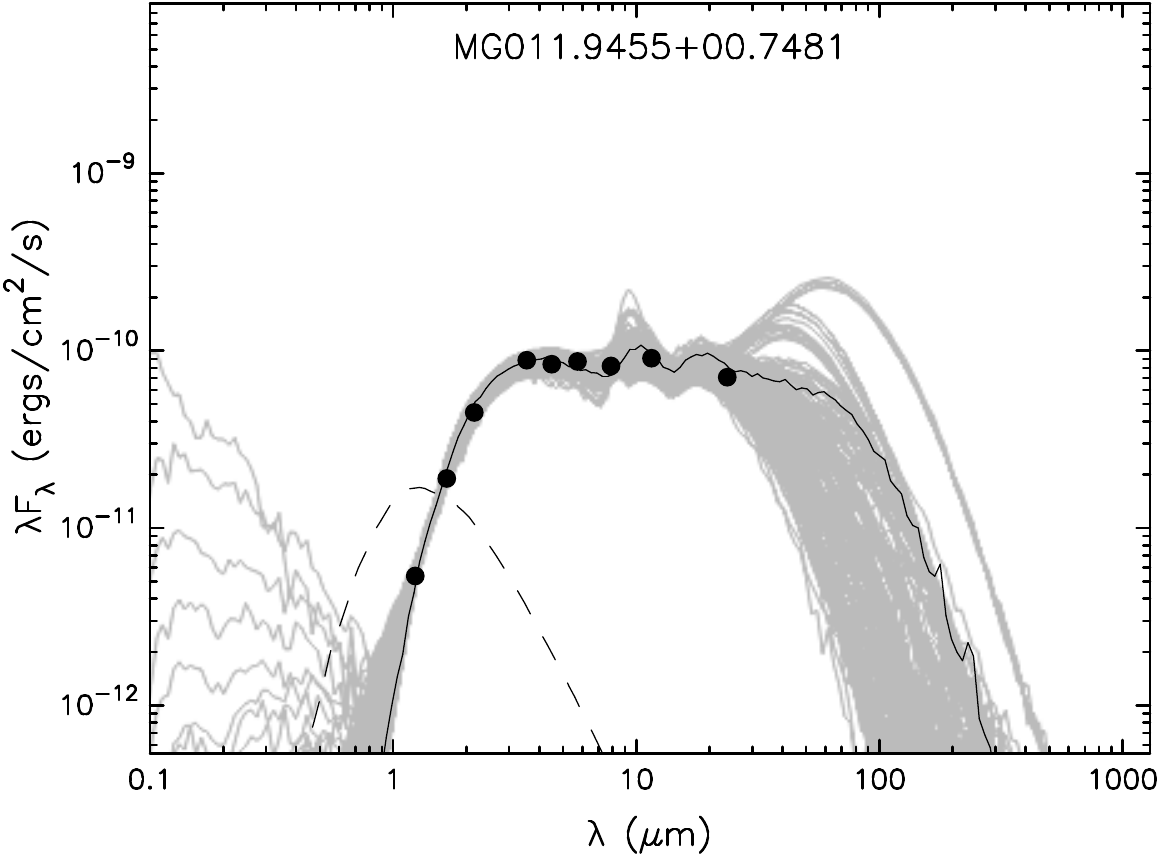}
\includegraphics[width=0.24\textwidth]{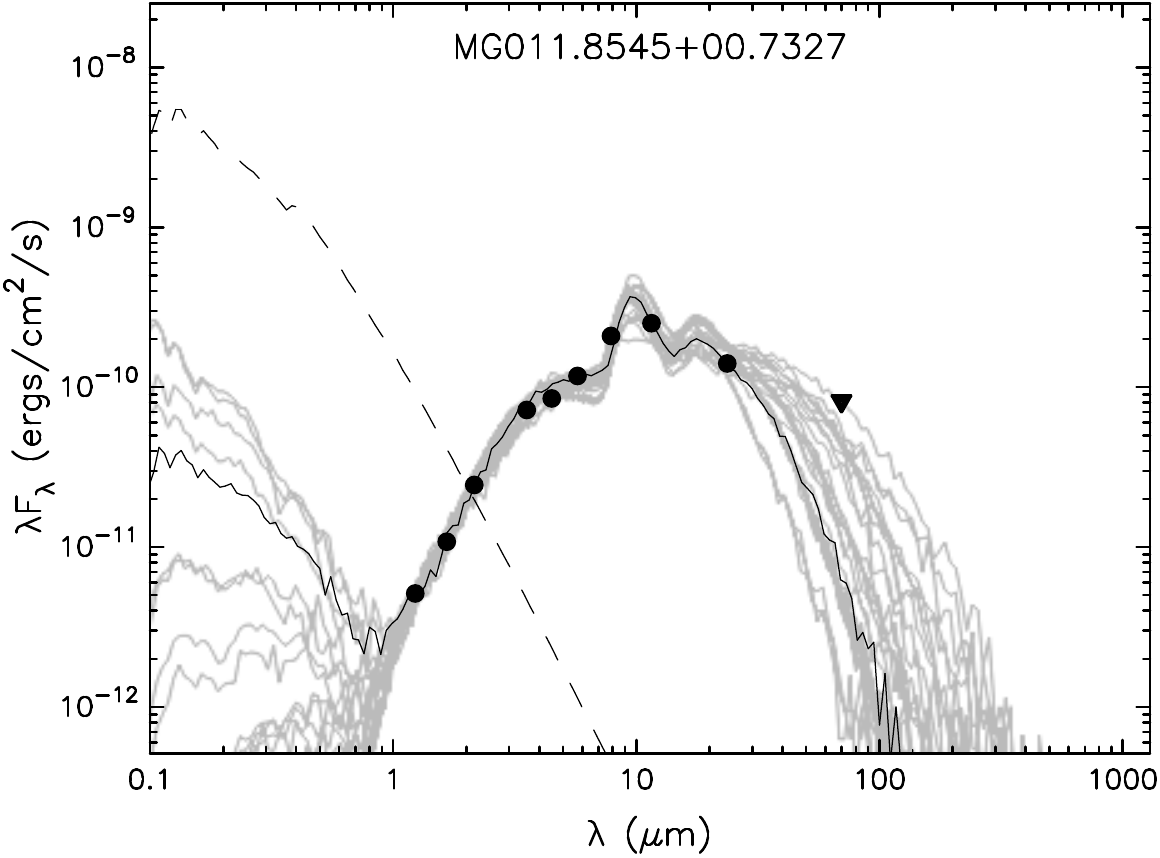}

   \caption{ -- continued.}
\label{sed-2}
\end{figure*}

\end{document}